# An evaluation for geometries, formation enthalpies, and dissociation energies of diatomic and triatomic (C, H, N, O), $NO_3$, and $HNO_3$ molecules from the PAW DFT method with PBE and optB88-vdW functionals (*AIP Advances*, in press, 2022)

Yong Han
Division of Chemical and Biological Sciences, Ames National Laboratory, Ames, Iowa 50011, USA.
Department of Physics and Astronomy, Iowa State University, Ames, Iowa 50011, USA.
*Email: y27h@ameslab.gov

## ABSTRACT

Structural geometries, formation enthalpies, and dissociation energies of all diatomic and triatomic molecules consisting of the four basic elements C, H, N, and/or O are calculated from the projector augmented wave (PAW) density functional theory (DFT) method with the PBE and optB88-vdW exchange-correlation functionals. The calculations are also extended to two larger molecules $NO_3$ and $HNO_3$, which consist of 4 and 5 atoms, respectively. In total, 82 molecules or isomers are considered in the calculations. The geometric parameters including 42 bond lengths and 15 bond angles of these molecules from the planewave DFT method are highly satisfactory, relative to available experimental data. The error analysis is also performed for 49 formation enthalpies and 138 dissociation energies (including 51 atomization energies as well as corresponding bond dissociation energies). The results are also compared with the previous data from various atomic-orbitals-based methods for molecules and from similar or different planewave DFT methods for various solids and other molecules. This provides an informative and instructive evaluation, especially for calculating the large-size material systems containing these small molecules as well as for further developing DFT methods.

## I. INTRODUCTION

Atomic-orbitals-based methods (AOBMs) [1, 2, 3] are generally regarded by the computational chemical community as more accurate numerically than the planewave density functional theory (DFT) methods [4, 5]. The reason is that, in earlier years, the pseudopotentials in the planewave DFT methods were not carefully generated when used in many planewave codes. However, since the projector augmented wave (PAW) method was suggested and used by Blöchl [6], a lot of effort has been made and the reliable PAW pseudopotentials have been generated [7] with constantly updated versions released, as implemented in the Vienna *Ab initio* Simulation Package (VASP) code [4]. Due to the high computational costs, AOBMs with higher accuracies are generally used to calculate relatively small systems like molecules. In contrast, the planewave DFT methods can be computationally much more efficient [4, 5] by using the supercell technology with the periodic boundary conditions to simulate any crystalline materials.

Due to the fundamental importance in various scientific areas including chemical physics, biophysics, environmental science, interstellar medium, etc., research on small molecules is constantly developing both theoretically and experimentally. Specifically, adsorption of small molecules on materials surfaces (including outer surfaces of materials or the surfaces of pores in materials, e.g., molecular sieves [8, 9]) is often considered for both theoretical studies and applications. The supercell for such a system by using a planewave DFT method often contains hundreds to thousands of atoms at least for obtaining reliable adsorption properties. In addition, *ab initio* molecular dynamics (AIMD) simulations [10] for such systems are also selectively implemented for, e.g., visualizing the diffusion paths of molecules on the surfaces, while AIMD simulations are even much more demanding than normal structural optimizations. Thus, using AOBMs with higher accuracies but high computational costs is impractical for such computations. Instead, the planewave DFT methods can be computationally practical due to the high efficiency. However, when a planewave DFT method is applied to a specific system, the reliability of the method must be first assessed because the accuracy of the DFT results can significantly depend on the electronic exchange-correlation energy functionals. To this end, we mention an example. It is well-known that Perdew-Burke-Ernzerhof (PBE) functional [11] cannot predict the interlayer spacing of graphite due to the absence of the dispersion corrections, e.g., the predicted lattice constant $c = 8.870$ Å along the [0001] direction of graphite from our previous PBE calculations [12] is notoriously much larger than the experimental value of 6.6720 Å [13]. In contrast, the value of 6.701 Å from our optB88-vdW [14] calculations with dispersion corrections can reproduce the above experimental $c$ value very well [12]. Thus, one should be particularly careful when selecting a functional with or without dispersion corrections to calculate the weakly bonded systems like graphite.

As a semilocal functional, the generalized gradient approximation (GGA) generally has comparably good accuracies for calculating ground state properties of neither weakly bonded nor strongly correlated systems. For example, recent extensive tests on the lattice constants, bulk moduli, and cohesive energies (or atomization energies) of 44 strongly and 17 weakly bonded solids from various local, semilocal, and nonlocal functionals (so-called "DFT Jacob's ladder") have been reported [15, 16]. For the weakly bonded systems, dispersion corrections need to be considered [17], while for strongly correlated systems, DFT+U corrections are usually needed [18]. In addition, more nonlocal functionals (upper rungs of the DFT Jacob's ladder) can have higher accuracies but higher computational costs than more local functionals (lower rungs). Therefore, to appropriately choose a functional before calculating a specific system, both accuracy and computational cost need to be balanced.

Recently, we have selected and applied the optB88-vdW functional [14], which typically considers dispersion correction including van der Walls (vdW) interactions, to various vdW materials with guest atoms [12, 19–30] and silica polymorphs with molecular groups consisting of C, H, and O [31]. These applications have already been proven very successful. The success is not surprising, given that these systems include weakly bonded interlayer spaces, while for the weakly bonded systems, the GGA plus dispersion corrections generally have satisfactory accuracies, as described above. In the near future, we will extensively involve the adsorption and diffusion properties of small molecules on the outer or inner pore surfaces of such weakly bonded materials. The elements that make up these small molecules will be C, H, N and/or O, which are also 4 most fundamental elements of organisms. Before extensively calculating these large-size systems, an evaluation even only for these small molecules in gas phase will be very instructive and necessary.

In this work, we use the PAW DFT method with the optB88-vdW functional to calculate the structural geometries, formation enthalpies, and dissociation energies of all diatomic and triatomic molecules consisting of C, H, N, and/or O, by considering 80 linear or triangular molecules or isomers. Then, we extend our calculations to two larger molecules $NO_3$ and $HNO_3$, containing 4 and 5 atoms, respectively, because these molecules will be the first candidates which will be involved in our ongoing projects for, e.g., separation of rare earth elements. As comparison, we also obtain the results using the most popular PBE GGA [11] without dispersion corrections.

The paper is organized as follows. The computational details are described in Sec. II. In Sec. III, we tabulate and discuss the structural geometries, formation enthalpies, dissociation energies, and corresponding spin states of all 82 molecular molecules or isomers consisting of C, H, N and/or O from our DFT calculations and previous experiments or AOBM calculations available in literature. In Sec. IV, we also discuss our results and make error analysis by comparing with other relevant DFT results in the literature. Sec. V provides a summary of this work. The formulation of formation enthalpies and dissociation energies is provided in Supplementary Material (SM) Sec. S1 and the original data for error analysis are also provided in SM.

## II. COMPUTATIONAL DETAILS

We use the VASP code [4] to perform all DFT calculations in this work, with the PAW pseudopotentials [7] developed by the VASP group and released in 2015. For the electron-electron exchange correlation part, as described in Sec. I, we use PBE and optB88-vdW functionals without and with dispersion corrections, respectively. In all DFT calculations, we take the energy cutoff to be 600 eV with sufficient accuracy (cf. the default energy cutoffs of 400 eV for C, 250 eV for H, 400 eV for N, and 400 eV for O in the PAW pseudopotential data files). Spin polarizations are always considered because the energy of a molecular configuration depends on the spin state. For 80 diatomic or triatomic molecules or isomers, the supercell is always taken to be a rectangular box with the size of 23 Å × 22 Å × 21 Å. For two larger molecules ($NO_3$ and $HNO_3$), the supercell is taken to be slightly larger with the size of 24.3 Å × 24.2 Å × 24.1 Å. These supercell sizes are sufficiently large so that the interactions between replicas can be ignored. The $k$ mesh is always taken to be $1 \times 1 \times 1$ which is sufficient because of the large supercell sizes. During energy minimization, the atoms in the supercell are fully relaxed after the initial configurations are judiciously selected based on previous experiments or AOBM calculations in the literature. The convergence of total energy is reached when the force exerted on each atom is less than 0.01 eV/Å.

## III. RESULTS AND DISCUSSION

In Table I, we list our PAW PBE and optB88-vdW results for all diatomic and triatomic molecules or isomers consisting of C, H, N, and/or O. As comparison, the corresponding AOBM and experimental data available in the previous literature are also listed. The theoretical and experimental data for two larger molecules $NO_3$ and $HNO_3$ are listed in Table II.

**TABLE I.** Theoretical and experimental data for 82 diatomic and triatomic molecules or isomers consisting of C, H, N, and/or O. Molecular formulas adopt the NIST notations [32] for convenient indexing, where molecules are listed in alphabetic order of element symbols and the numbers of elemental atoms in the molecule as subscripts 1, 2, 3, …, etc., but the conventional notations and names for molecules or isomers are also listed. All ball-and-stick geometric structures are from our PAW optB88-vdW calculations with the gray, white, blue, and red balls representing C, H, N, and O atoms, respectively. "PBE" and "optB88" are our PAW DFT calculations with PBE and optB88-vdW functionals, respectively. The results from other functionals in the previous literature (see Sec. IV) are not provided in this list, except for a few specific molecules or isomers. "AOBM" denotes the data from various atomic-orbitals-based methods in the literature. "Exp." denotes available experiment data from the literature. $M$ (in unit of Bohr magneton $\mu_{\mathrm{B}}$) is the spin magnetic moment of the molecule and taken to be the value with the lowest energy for a given configuration. $l_{\alpha\beta}$ (in Å) is the bond length (or interatomic distance or atomic spacing) between atom $\alpha$ and atom $\beta$. $\theta_{\alpha\beta\gamma}$ (in degree °) is the angle between $l_{\beta\alpha}$ and $l_{\beta\gamma}$. $\Delta H_{\mathrm{f}}$ (in eV) is the formation enthalpy from Eq. (S2). $D_{\mathrm{p1+p2+p3+\cdots}}$ (in eV) is the dissociation energy for products p1, p2, p3, … from Eq. (S4). The dissociation energy in the final column is also called the atomization energy from Eq. (S6). The data for $\Delta H_{\mathrm{f}}$ and $D_{\mathrm{p1+p2+p3+\cdots}}$ are obtained always at 0 K. Under the corresponding values in eV, the available experimental data from Active Thermochemical Tables (ATcT) [33] for $\Delta H_{\mathrm{f}}$ and $D_{\mathrm{p1+p2+p3+\cdots}}$ in units of kJ/mol with the uncertainties are also listed, as indicated by "±", corresponding to estimated 95% confidence limits [34, 35]. The species names from ATcT [33] are adopted partly.

| Formula: $C_1H_1$ | | | $M$ | | $\Delta H_{\mathrm{f}}$ | $D_{\mathrm{C+H}}$ |
|---|---|---|---|---|---|---|
| CH or HC (Methylidyne) | | | | $l_{\mathrm{CH}}$ | | |
| 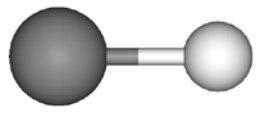 | PBE | | 1 | 1.1369 | 6.429 | 3.697 |
| | optB88 | | 1 | 1.1300 | 6.384 | 3.895 |
| | AOBM | [36] | | 1.1204 | | |
| | Exp. | [37] | | 1.1199 | | |
| | Exp. | [38] | | 1.119786 | | |

| | | | | | | | | | | |
|---|---|---|---|---|---|---|---|---|---|---|
| | Exp. | [33] | | | | 6.14432 | | | | 3.46791 |
| | | | | | | 592.837±0.097 | | | | 334.602±0.087 |
| Formula: $C_1H_1N_1$ | | | $M$ | | | $\Delta H_f$ | $D_{CH+N}$ | $D_{CN+H}$ | $D_{C+HN}$ | $D_{C+H+N}$ |
| Linear HCN or NCH (Hydrogen cyanide) | | | | $l_{CH}$ | $l_{CN}$ | | | | | |
| 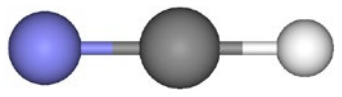 | PBE | | 0 | 1.0749 | 1.1611 | 1.240 | 10.386 | 5.544 | 10.220 | 14.083 |
| | optB88 | | 0 | 1.0699 | 1.1561 | 1.336 | 10.311 | 5.765 | 10.181 | 14.206 |
| | AOBM | [39] | | 1.067 | 1.160 | | | | | |
| | AOBM | [40] | | 1.0651-1.0826 | 1.1527-1.1758 | | | | | |
| | Exp. | [41] | | 1.06549 | 1.15321 | | | | | |
| | Exp. | [33] | | | | 1.34402 | 9.6775 | 5.4215 | 9.7473 | 13.14542 |
| | | | | | | 129.678±0.089 | 933.74±0.12 | 523.09±0.12 | 940.47±0.18 | 1268.340±0.085 |
| Linear HNC or CNH (Hydrogen isocyanide) | | | | $l_{NH}$ | $l_{NC}$ | | | | | |
| 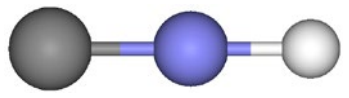 | PBE | | 0 | 1.0050 | 1.1774 | 1.857 | 9.769 | 4.927 | 9.603 | 13.466 |
| | optB88 | | 0 | 1.0019 | 1.1728 | 1.946 | 9.701 | 5.155 | 9.571 | 13.596 |
| | AOBM | [39] | | 0.996 | 1.175 | | | | | |
| | AOBM | [40] | | 0.9952-1.0063 | 1.1686-1.1895 | | | | | |
| | Exp. | [42] | | 0.9940 | 1.1689 | | | | | |
| | Exp. | [33] | | | | 1.9906 | 9.0310 | 4.7749 | 9.1008 | 12.4989 |
| | | | | | | 192.06±0.32 | 871.36±0.32 | 460.71±0.32 | 878.09±0.35 | 1205.96±0.31 |
| Linear CHN or NHC | | | | $l_{HC}$ | $l_{HN}$ | | | | | |
| | PBE | | 0 | 1.1043 | 1.0155 | 11.553 | 0.072 | -4.769 | -0.094 | 3.769 |

| | | | | | | | | | | | |
|---|---|---|---|---|---|---|---|---|---|---|---|
| 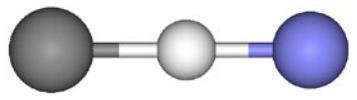 | optB88 | | 0 | 1.1054 | 1.0178 | | 11.633 | 0.014 | -4.532 | -0.116 | 3.909 |
| Triangular (cyclic) HCN or CHN, etc. | | | | $l_{CH}$ | $l_{CN}$ | $\theta_{HCN}$ | | | | | |
| 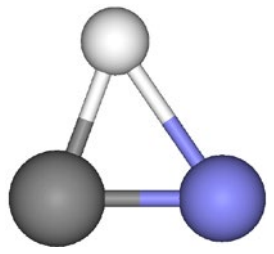 | PBE | | 0 | 1.2016 | 1.1961 | 70.16 | 3.225 | 8.400 | 3.559 | 8.234 | 12.097 |
| | optB88 | | 0 | 1.2082 | 1.1911 | 69.76 | 3.349 | 8.299 | 3.752 | 8.168 | 12.193 |
| | AOBM | [39] | | 1.186 | 1.195 | 71.67 | | | | | |
| | AOBM | [40] | | 1.1835-1.2017 | 1.1867-1.2089 | 71.623-71.827 | | | | | |
| Formula: $C_1H_1O_1$ | | | $M$ | | | | $\Delta H_f$ | $D_{CH+O}$ | $D_{CO+H}$ | $D_{C+HO}$ | $D_{C+H+O}$ |
| Linear HCO or OCH | | | | $l_{CH}$ | $l_{CO}$ | | | | | | |
| 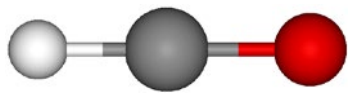 | PBE | | 1 | 1.0728 | 1.1944 | | 1.456 | 7.997 | 0.191 | 6.987 | 11.694 |
| | optB88 | | 1 | 1.0685 | 1.1928 | | 1.540 | 7.843 | 0.239 | 6.851 | 11.738 |
| Linear HOC or COH | | | | $l_{OH}$ | $l_{OC}$ | | | | | | |
| 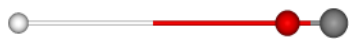 | PBE | | 1 | 6.6239 | 1.1431 | | 1.648 | 7.805 | 0.000 | 6.796 | 11.502 |
| | optB88 | | 1 | 6.6164 | 1.1394 | | 1.779 | 7.605 | 0.000 | 6.613 | 11.500 |
| Linear CHO or OHC | | | | $l_{HC}$ | $l_{HO}$ | | | | | | |
| 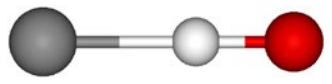 | PBE | | 1 | 1.3356 | 1.0108 | | 8.153 | 1.300 | -6.505 | 0.291 | 4.997 |
| | optB88 | | 1 | 1.5686 | 1.0051 | | 8.259 | 1.125 | -6.479 | 0.133 | 5.020 |
| Triangular HCO or OCH (Formyl) | | | | $l_{CH}$ | $l_{CO}$ | $\theta_{HCO}$ | | | | | |
| 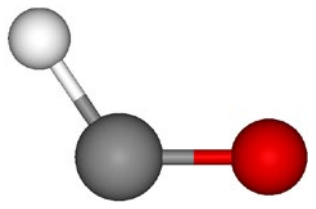 | PBE | | 1 | 1.1337 | 1.1882 | 123.91 | 0.465 | 8.988 | 1.183 | 7.979 | 12.685 |
| | optB88 | | 1 | 1.1294 | 1.1860 | 123.88 | 0.534 | 8.850 | 1.245 | 7.858 | 12.744 |
| | AOBM | [43] | | 1.112–1.1462 | 1.1729–1.1952 | 123.79–125.02 | 0.396–0.481 | | 0.607–0.824 | | 11.730–12.105 |
| | Exp. | [44] | | 1.1102 | 1.17115 | 127.426 | | | | | |
| | Exp. | [45] | | 1.125 | 1.175 | 124.95 | | | | | |
| | Exp. | [46] | | 1.1514 | 1.17708 | 123.01 | | | | | |
| | Exp. | [47] | | 1.1191 | 1.1754 | 124.43 | | | | | |
| | Exp. | [33] | | | | | 0.42888 | 8.2738 | 0.63066 | 7.3308 | 11.74169 |
| | | | | | | | 41.381±0.096 | 798.30±0.13 | 60.849±0.095 | 707.31±0.10 | 1132.901±0.097 |
| Triangular HOC or COH (Isoformyl) | | | | $l_{OH}$ | $l_{OC}$ | $\theta_{HOC}$ | | | | | |
| 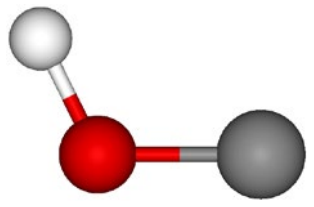 | PBE | | 1 | 1.0070 | 1.2755 | 115.98 | 2.304 | 7.149 | -0.656 | 6.140 | 10.846 |
| | optB88 | | 1 | 1.0058 | 1.2744 | 116.30 | 2.352 | 7.032 | -0.573 | 6.040 | 10.926 |
| | AOBM | [48] | | 0.980 | 1.299 | 111.7 | | | | | |
| | AOBM | [49] | | 0.97604 | 1.27300 | 112.956 | | | | | |
| | Exp. | [33] | | | | | 2.2516 | 6.4510 | -1.1921 | 5.5080 | 9.9190 |
| | | | | | | | 217.25±0.69 | 622.43±0.69 | -115.02±0.68 | 531.44±0.68 | 957.04±0.68 |
| Formula: $C_1H_2$ | | | $M$ | | | | $\Delta H_f$ | $D_{CH+H}$ | $D_{H_2+C}$ | | $D_{C+2H}$ |
| Linear HCH | | | | $l_{CH}$ | $l_{CH}$ | | | | | | |
| 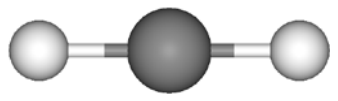 | PBE | | 2 | 1.0740 | 1.0730 | | 4.078 | 4.620 | 3.780 | | 8.317 |
| | optB88 | | 2 | 1.0700 | 1.0702 | | 4.368 | 4.510 | 3.416 | | 8.405 |
| Linear CHH or HHC | | | | $l_{HC}$ | $l_{HH}$ | | | | | | |
| 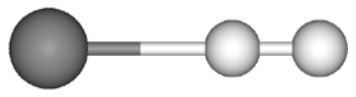 | PBE | | 2 | 1.4844 | 0.7812 | | 7.635 | 1.063 | 0.222 | | 4.760 |
| | optB88 | | 2 | 1.6136 | 0.7648 | | 7.704 | 1.174 | 0.080 | | 5.069 |
| Triangular HCH (Methylene) | | | | $l_{CH}$ | $l_{CH}$ | $\theta_{HCH}$ | | | | | |
| | PBE | | 2 | 1.0849 | 1.0849 | 135.06 | 3.890 | 4.808 | 3.967 | | 8.505 |
| 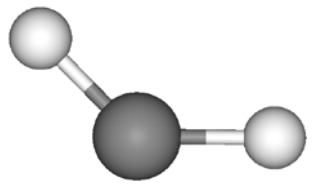 | optB88 | | 2 | 1.0816 | 1.0816 | 135.04 | 4.176 | 4.702 | 3.608 | | 8.597 |
| | AOBM | [50] | | 1.07481 | 1.07481 | 133.839 | | | | | |
| | Exp. | [51] | | 1.07530 | 1.07530 | 133.9308 | | | | | |
| | Exp. | [33] | | | | | 4.05299 | 4.3304 | 3.32019 | | 7.79826 |
| | | | | | | | 391.054±0.096 | 417.82±0.11 | 320.350±0.089 | | 752.418±0.089 |
| Formula: $C_1N_1$ | | | $M$ | | | | $\Delta H_f$ | | | | $D_{C+N}$ |
| CN or NC (Nitrilomethyl) | | | | $l_{CN}$ | | | | | | | |
| 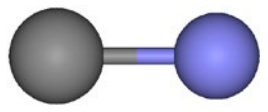 | PBE | | 1 | 1.1768 | | | 4.515 | | | | 8.539 |
| | optB88 | | 1 | 1.1733 | | | 4.606 | | | | 8.441 |
| | AOBM | [52] | | 1.18 | | | | | | | |
| | B3LYP | [53] | | 1.162 | | | | | | | |
| | Exp. | [37] | | 1.1718 | | | | | | | |
| | Exp. | [33] | | | | | 4.5264 | | | | 7.7240 |
| | | | | | | | 436.73±0.14 | | | | 745.25±0.13 |
| Formula: $C_1N_1O_1$ | | | $M$ | | | | $\Delta H_f$ | $D_{CN+O}$ | $D_{CO+N}$ | $D_{C+NO}$ | $D_{C+N+O}$ |
| Linear NCO or OCN (Cyanooxidanyl) | | | | $l_{CN}$ | $l_{CO}$ | | | | | | |
| 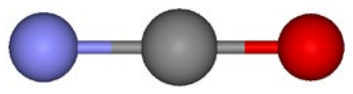 | PBE | | 1 | 1.2343 | 1.1932 | | 1.067 | 6.471 | 3.508 | 7.736 | 15.010 |
| | optB88 | | 1 | 1.2292 | 1.1914 | | 1.139 | 6.467 | 3.409 | 7.592 | 14.908 |
| | AOBM | [54] | | 1.241 | 1.182 | | | | | | |
| | AOBM | [55] | | 1.2330 | 1.1851 | | | | | | |
| | AOBM | [56] | | 1.223 | 1.177 | | | | | | |
| | Exp. | [57] | | 1.200 | 1.206 | | | | | | |
| | Exp. | [33] | | | | | 1.3156 | 5.7691 | 2.3820 | 6.9969 | 13.4931 |
| | | | | | | | 126.94±0.33 | 556.63±0.35 | 229.83±0.33 | 675.10±0.33 | 1301.89±0.33 |
| Linear NOC or CON | | | | $l_{OC}$ | $l_{ON}$ | | | | | | |
| 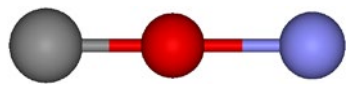 | PBE | | 1 | 1.2084 | 1.3038 | | 6.023 | 1.516 | -1.448 | 2.780 | 10.054 |
| | optB88 | | 1 | 1.2008 | 1.3225 | | 6.043 | 1.563 | -1.495 | 2.689 | 10.005 |
| | AOBM | [54] | | 1.211 | 1.303 | | | | | | |
| | AOBM | [55] | | 1.1948 | 1.3306 | | | | | | |
| | AOBM | [56] | | 1.184 | 1.325 | | | | | | |
| Linear CNO or ONC (Nitrosomethylidyne) | | | | $l_{NC}$ | $l_{NO}$ | | | | | | |
| 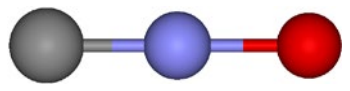 | PBE | | 1 | 1.2245 | 1.2237 | | 3.645 | 3.894 | 0.930 | 5.158 | 12.433 |
| | optB88 | | 1 | 1.2186 | 1.2269 | | 3.731 | 3.875 | 0.817 | 5.001 | 12.316 |
| | AOBM | [54] | | 1.226 | 1.221 | | | | | | |
| | AOBM | [55] | | 1.2168 | 1.2223 | | | | | | |
| | AOBM | [56] | | 1.210 | 1.216 | | | | | | |
| | Exp. | [33] | | | | | 4.035 | 3.049 | -0.338 | 4.277 | 10.774 |
| | | | | | | | 389.3±1.1 | 294.2±1.0 | -32.6±1.0 | 412.7±1.0 | 1039.5±1.0 |
| Triangular (cyclic) NCO or OCN (Oxazirinyl) | | | | $l_{CN}$ | $l_{CO}$ | $\theta_{NCO}$ | | | | | |
| 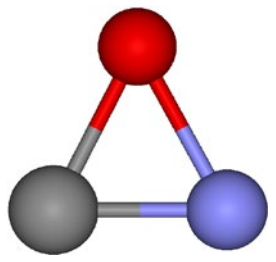 | PBE | | 1 | 1.3467 | 1.4224 | 62.62 | 4.455 | 3.084 | 0.121 | 4.349 | 11.623 |
| | optB88 | | 1 | 1.3396 | 1.4278 | 63.05 | 4.553 | 3.053 | -0.005 | 4.179 | 11.494 |
| | AOBM | [54] | | 1.349 | 1.467 | 59.54 | | | | | |
| | AOBM | [55] | | 1.3457 | 1.3937 | 63.92 | | | | | |
| | AOBM | [56] | | 1.344 | 1.380 | 63.70 | | | | | |

| | | | | | | | | | | |
|---|---|---|---|---|---|---|---|---|---|---|
| | Exp. | [33] | | | | | 4.665 | 2.420 | -0.967 | 10.14 |
| | | | | | | | 450.1±1.5 | 233.5±1.4 | -93.3±1.4 | 978.7±1.4 |
| Formula: $C_1N_2$ | | | $M$ | | | | $\Delta H_f$ | $D_{CN+N}$ | $D_{C+N_2}$ | $D_{C+2N}$ |
| Linear NCN (Methanetetraylbisamidogen) | | | | $l_{CN}$ | $l_{CN}$ | | | | | |
| 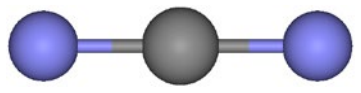 | PBE | | 2 | 1.2363 | 1.2363 | | 4.149 | 5.562 | 3.708 | 14.101 |
| | optB88 | | 2 | 1.2321 | 1.2321 | | 4.301 | 5.569 | 3.484 | 14.010 |
| | B3LYP | [58] | | 1.232 | 1.232 | | | | | |
| | B3LYP | [59] | | 1.225 | 1.225 | | | | | |
| | Exp. | [33] | | | | | 4.6768 | 4.7268 | 2.6964 | 12.4508 |
| | | | | | | | 451.24±0.44 | 456.07±0.44 | 260.16±0.43 | 1201.32±0.43 |
| Linear CNN or NNC (Diazomethylene) | | | | $l_{NC}$ | $l_{NN}$ | | | | | |
| 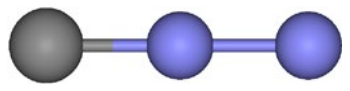 | PBE | | 2 | 1.2566 | 1.1999 | | 5.308 | 4.404 | 2.550 | 12.942 |
| | optB88 | | 2 | 1.2496 | 1.1931 | | 5.484 | 4.386 | 2.300 | 12.827 |
| | AOBM | [60] | | 1.2325 | 1.2158 | | | | | |
| | AOBM | [60] | | 1.2526 | 1.2241 | | | | | |
| | B3LYP | [58] | | 1.233 | 1.198 | | | | | |
| | Exp. | [33] | | | | | 6.004 | 3.399 | 1.369 | 11.123 |
| | | | | | | | 579.3±3.8 | 328.0±3.7 | 132.1±3.7 | 1073.2±3.7 |
| Triangular (cyclic) NCN (3H-Diazirin-3-ylidene) | | | | $l_{CN}$ | $l_{CN}$ | $\theta_{NCN}$ | | | | |
| 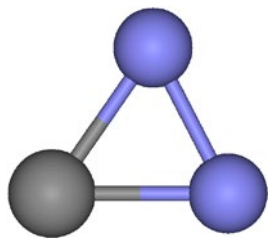 | PBE | | 0 | 1.4036 | 1.4036 | 54.98 | 5.733 | 3.978 | 2.124 | 12.517 |
| | optB88 | | 0 | 1.4023 | 1.4018 | 54.93 | 5.887 | 3.982 | 1.897 | 12.424 |
| | AOBM | [60] | | 1.4100 | 1.4100 | 54.9101 | | | | |
| | AOBM | [60] | | 1.3979 | 1.3979 | 55.0543 | | | | |
| | B3LYP | [61] | | 1.387 | 1.387 | 54.9 | | | | |
| | Exp. | [33] | | | | | 6.002 | 3.403 | 1.372 | 11.126 |
| | | | | | | | 579.1±2.7 | 328.3±2.7 | 132.4±2.7 | 1073.5±2.7 |
| Formula: $C_1O_1$ | | | $M$ | | | | $\Delta H_f$ | | | $D_{C+O}$ |
| CO or OC (Carbon monoxide) | | | | $l_{CO}$ | | | | | | |
| 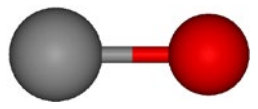 | PBE | | 0 | 1.1431 | | | -0.621 | | | 11.502 |
| | optB88 | | 0 | 1.1393 | | | -0.715 | | | 11.499 |
| | AOBM | [62] | | 1.1513 | | | | | | |
| | Exp. | [37] | | 1.128323 | | | | | | |
| | Exp. | [33] | | | | | -1.17950 | | | 11.11104 |
| | | | | | | | -113.804±0.026 | | | 1072.052±0.046 |
| Formula: $C_1O_2$ | | | $M$ | | | | $\Delta H_f$ | $D_{CO+O}$ | $D_{C+O_2}$ | $D_{C+2O}$ |
| Linear OCO (Carbon dioxide) | | | | $l_{CO}$ | $l_{CO}$ | | | | | |
| 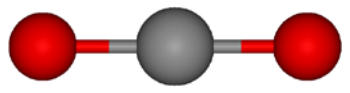 | PBE | | 0 | 1.1764 | 1.1764 | | -3.870 | 6.273 | 11.727 | 17.775 |
| | optB88 | | 0 | 1.1739 | 1.1738 | | -3.863 | 6.148 | 11.648 | 17.647 |
| | AOBM | [63] | | 1.15 | 1.15 | | | | | |
| | Exp. | [64] | | 1.162 | 1.162 | | | | | |
| | Exp. | [33] | | | | | -4.07430 | 5.45315 | 11.44748 | 16.56419 |
| | | | | | | | -393.110±0.015 | 526.149±0.024 | 1104.514±0.043 | 1598.201±0.043 |
| Linear COO or OOC (Dioxymethylidyne) | | | | $l_{OC}$ | $l_{OO}$ | | | | | |
| | PBE | | 0 | 1.1797 | 1.3236 | | 3.048 | -0.645 | 4.809 | 10.857 |
| 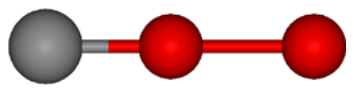 | optB88 | | 0 | 1.1739 | 1.3421 | | 2.975 | -0.690 | 4.809 | 10.809 |
| | Exp. | [33] | | | | | 3.107 | -1.729 | 4.266 | 9.383 |
| | | | | | | | 299.8±1.7 | -166.8±1.6 | 411.6±1.6 | 905.3±1.6 |
| Triangular OCO (Dioxiranylidene) | | | | $l_{CO}$ | $l_{CO}$ | $\theta_{OCO}$ | | | | |
| 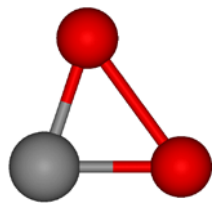 | PBE | | 0 | 1.3404 | 1.3403 | 71.21 | 2.198 | 0.205 | 5.660 | 11.707 |
| | optB88 | | 0 | 1.3397 | 1.3399 | 71.69 | 2.158 | 0.127 | 5.627 | 11.626 |
| | Exp. | [33] | | | | | 1.972 | -0.594 | 5.401 | 10.518 |
| | | | | | | | 190.3±1.4 | -57.3±1.4 | 521.1±1.4 | 1014.8±1.4 |
| Formula: $C_2$ | | | $M$ | | | | $\Delta H_f$ | | | $D_{2C}$ |
| CC (Ethynylene) | | | | $l_{CC}$ | | | | | | |
| 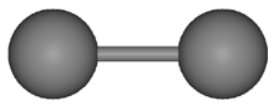 | PBE | | 2 | 1.3142 | | | 8.780 | | | 6.935 |
| | optB88 | | 2 | 1.3093 | | | 8.865 | | | 6.704 |
| | AOBM | [65] | | 1.248-<br>1.255 | | | | | | |
| | AOBM | [66] | | 1.247,<br>1.254 | | | | | | 6.072-6.371 |
| | Exp. | [37] | | 1.2425 | | | | | | |
| | Exp. | [33] | | | | | 8.49875 | | | 6.24762 |
| | | | | | | | 820.005±0.092 | | | 602.804±0.028 |
| Formula: $C_2H_1$ | | | $M$ | | | | $\Delta H_f$ | $D_{C+CH}$ | $D_{C_2+H}$ | $D_{2C+H}$ |
| Linear CCH (Ethynyl) | | | | $l_{CC}$ | $l_{CH}$ | | | | | |
| 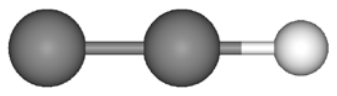 | PBE | | 1 | 1.2113 | 1.0720 | | 5.938 | 8.348 | 5.110 | 12.045 |
| | optB88 | | 1 | 1.2065 | 1.0675 | | 6.175 | 7.993 | 5.184 | 11.888 |
| | AOBM | [67] | | 1.223 | 1.075 | | | | | |
| | Exp. | [68] | | 1.21652 | 1.04653 | | | | | |
| | Exp. | [33] | | | | | 5.8430 | 7.6745 | 4.8948 | 11.1424 |
| | | | | | | | 563.76±0.15 | 740.48±0.15 | 472.28±0.13 | 1075.08±0.13 |
| Linear CHC | | | | $l_{HC}$ | $l_{HC}$ | | | | | |
| 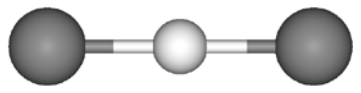 | PBE | | 1 | 1.1967 | 1.2014 | | 12.903 | 1.383 | -1.854 | 5.080 |
| | optB88 | | 1 | 1.1922 | 1.2018 | | 13.054 | 1.114 | -1.695 | 5.009 |
| Triangular (cyclic) CHC or CCH or HCC | | | | $l_{HC}$ | $l_{HC}$ | $\theta_{HCH}$ | | | | |
| 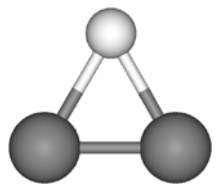 | PBE | | 1 | 1.2853 | 1.2789 | 59.94 | 6.750 | 7.536 | 4.299 | 11.233 |
| | optB88 | | 1 | 1.2897 | 1.2809 | 59.50 | 7.016 | 7.152 | 4.343 | 11.046 |
| Formula: $C_2N_1$ | | | $M$ | | | | $\Delta H_f$ | $D_{C+CN}$ | $D_{C_2+N}$ | $D_{2C+N}$ |
| Linear CCN or NCC (Cyanomethylidyne) | | | | $l_{CC}$ | $l_{CN}$ | | | | | |
| 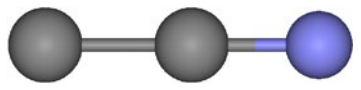 | PBE | | 1 | 1.3743 | 1.1972 | | 6.929 | 5.443 | 7.047 | 13.982 |
| | optB88 | | 1 | 1.3684 | 1.1920 | | 6.966 | 5.425 | 7.162 | 13.866 |
| | AOBM | [60] | | 1.4152 | 1.1993 | | | | | |
| | AOBM | [60] | | 1.3908 | 1.1811 | | | | | |
| | AOBM | [69] | | 1.4045 | 1.1889 | | | | | |
| | AOBM | [70] | | 1.3821 | 1.1847 | | | | | |
| | Exp. | [33] | | | | | 7.096 | 4.803 | 6.280 | 12.527 |
| | | | | | | | 684.7±2.1 | 463.4±2.0 | 605.9±2.0 | 1208.7±2.0 |
| Linear CNC (Isocyanomethylidyne) | | | | $l_{NC}$ | $l_{NC}$ | | | | | |
| 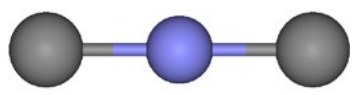 | PBE | | 1 | 1.2529 | 1.2531 | | 6.873 | 5.499 | 7.103 | 14.038 |
| | optB88 | | 1 | 1.2481 | 1.2482 | | 6.892 | 5.499 | 7.236 | 13.940 |
| | AOBM | [60] | | 1.2658 | 1.2658 | | | | | |
| | AOBM | [60] | | 1.2479 | 1.2479 | | | | | |
| | AOBM | [69] | | 1.2534 | 1.2534 | | | | | |
| | AOBM | [70] | | 1.2462 | 1.2462 | | | | | |
| | Exp. | [71] | | 1.245 | 1.245 | | | | | |
| | Exp. | [33] | | | | | 6.967 | 4.933 | 6.409 | 12.657 |
| | | | | | | | 672.2±1.7 | 476.0±1.7 | 618.4±1.7 | 1221.2±1.7 |
| Triangular (cyclic) CNC (2,3-didehydro-1H-azirin-1-yl) | | | | $l_{NC}$ | $l_{NC}$ | $\theta_{CNC}$ | | | | |
| 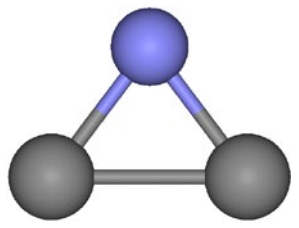 | PBE | | 1 | 1.3188 | 1.3188 | 72.73 | 7.375 | 4.997 | 6.601 | 13.536 |
| | optB88 | | 1 | 1.3129 | 1.3118 | 74.10 | 7.531 | 4.860 | 6.597 | 13.301 |
| | AOBM | [60] | | 1.3275 | 1.3275 | 74.5919 | | | | |
| | AOBM | [60] | | 1.3099 | 1.3099 | 75.7304 | | | | |
| | Exp. | [33] | | | | | 7.465 | 4.434 | 5.910 | 12.157 |
| | | | | | | | 720.3±2.1 | 427.8±2.0 | 570.2±2.0 | 1173.0±2.0 |
| Formula: $C_2O_1$ | | | $M$ | | | | $\Delta H_f$ | $D_{C+CO}$ | $D_{C_2+O}$ | $D_{2C+O}$ |
| Linear CCO or OCC (Dicarbon monoxide) | | | | $l_{CC}$ | $l_{CO}$ | | | | | |
| 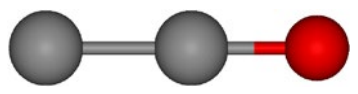 | PBE | | 2 | 1.3705 | 1.1785 | | 3.876 | 3.360 | 7.928 | 14.862 |
| | optB88 | | 2 | 1.3625 | 1.1770 | | 3.991 | 3.079 | 7.874 | 14.578 |
| | AOBM | [72] | | a=1.3699 | b=1.1627 | | | | | |
| | AOBM | [73] | | a=1.3718 | b=1.1638 | | | | | |
| | AOBM | [74] | | a=1.373 | b=1.169 | | | | | |
| | AOBM | [75] | | a=1.388 | b=1.149 | | | | | |
| | Exp. | [33] | | | | | 3.9086 | 2.2851 | 7.1486 | 13.3961 |
| | | | | | | | 377.12±0.83 | 220.48±0.82 | 689.73±0.82 | 1292.53±0.82 |
| Linear COC | | | | $l_{OC}$ | $l_{OC}$ | | | | | |
| 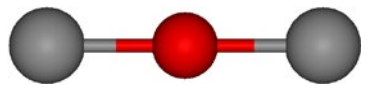 | PBE | | 2 | 1.2689 | 1.2690 | | 6.501 | 0.735 | 5.302 | 12.237 |
| | optB88 | | 2 | 1.2675 | 1.2669 | | 6.575 | 0.494 | 5.290 | 11.993 |
| | Exp. | [33] | | | | | 6.768 | -0.574 | 4.289 | 10.536 |
| | | | | | | | 653.0±1.5 | -55.4±1.5 | 413.8±1.5 | 1016.6±1.5 |
| Triangular CCO or OCC | | | | $l_{CC}$ | $l_{CO}$ | $\theta_{CCO}$ | | | | |
| 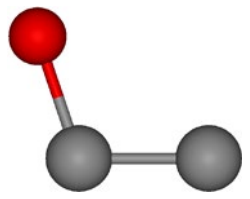 | PBE | | 2 | 1.3646 | 1.3540 | 109.03 | 5.957 | 1.280 | 5.847 | 12.782 |
| | optB88 | | 2 | 1.3598 | 1.3511 | 108.66 | 6.008 | 1.061 | 5.857 | 12.560 |
| Formula: $C_3$ | | | $M$ | | | | $\Delta H_f$ | $D_{C+C_2}$ | | $D_{3C}$ |
| Linear CCC (1,2-Propadiene-1,3-diylidene) | | | | $l_{CC}$ | $l_{CC}$ | | | | | |
| 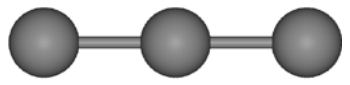 | PBE | | 0 | 1.3018 | 1.3019 | | 8.706 | 7.932 | | 14.866 |
| | optB88 | | 0 | 1.2948 | 1.2948 | | 8.658 | 7.991 | | 14.695 |
| | AOBM | [76] | | 1.29452 | 1.29452 | | | | | |
| | Exp. | [77] | | 1.29471 | 1.29471 | | | | | |
| | Exp. | [33] | | | | | 8.4422 | 7.4297 | | 13.6774 |
| | | | | | | | 814.55±0.53 | 716.86±0.51 | | 1319.67±0.51 |
| Triangular (quasilinear) CCC | | | | $l_{CC}$ | $l_{CC}$ | $\theta_{CCC}$ | | | | |

| | | | | | | | | | | | |
|---|---|---|---|---|---|---|---|---|---|---|---|
| 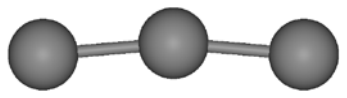 | PBE | | 0 | 1.3027 | 1.3028 | 149.25 | 8.700 | 7.937 | | | 14.872 |
| | optB88 | | 0 | 1.2949 | 1.2949 | 170.24 | 8.658 | 7.991 | | | 14.695 |
| | AOBM | [78] | | 1.289664 | 1.289664 | 161.6 | | | | | |
| Triangular (cyclic) CCC (2-Cyclopropyn-1-ylidene) | | | | $l_{CC}$ | $l_{CC}$ | $\theta_{CCC}$ | | | | | |
| 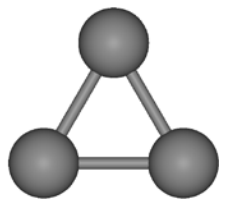 | PBE | | 2 | 1.3778 | 1.3777 | 59.99 | 9.212 | 7.425 | | | 14.360 |
| | optB88 | | 2 | 1.3746 | 1.3742 | 60.02 | 9.491 | 7.158 | | | 13.862 |
| | Exp. | [33] | | | | | 9.315 | 6.556 | | | 12.804 |
| | | | | | | | 898.8±1.6 | 632.6±1.5 | | | 1235.4±1.5 |
| Formula: $H_1N_1$ | | | $M$ | | | | $\Delta H_f$ | | | | $D_{H+N}$ |
| HN or NH (Imidogen) | | | | $l_{HN}$ | | | | | | | |
| 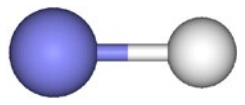 | PBE | | 2 | 1.0506 | | | 3.602 | | | | 3.863 |
| | optB88 | | 2 | 1.0485 | | | 3.732 | | | | 4.025 |
| | AOBM | [79] | | 1.035 | | | | | | | |
| | Exp. | [37] | | 1.0362 | | | | | | | |
| | Exp. | [33] | | | | | 3.7181 | | | | 3.3981 |
| | | | | | | | 358.74±0.16 | | | | 327.87±0.16 |
| Formula: $H_1N_1O_1$ | | | $M$ | | | | $\Delta H_f$ | $D_{HN+O}$ | $D_{H+NO}$ | $D_{HO+N}$ | $D_{H+N+O}$ |
| Linear HNO or ONH | | | | $l_{NH}$ | $l_{NO}$ | | | | | | |
| 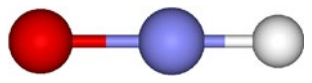 | PBE | | 2 | 1.0038 | 1.2291 | | 2.277 | 4.349 | 0.938 | 3.506 | 8.212 |
| | optB88 | | 2 | 1.0022 | 1.2314 | | 2.440 | 4.292 | 1.002 | 3.431 | 8.318 |
| Linear HON or NOH | | | | $l_{ON}$ | $l_{OH}$ | | | | | | |
| 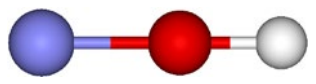 | PBE | | 2 | 1.3041 | 0.9619 | | 3.093 | 3.533 | 0.122 | 2.690 | 7.396 |
| | optB88 | | 2 | 1.3097 | 0.9612 | | 3.201 | 3.531 | 0.240 | 2.669 | 7.556 |
| Linear OHN or NHO | | | | $l_{HN}$ | $l_{HO}$ | | | | | | |
| 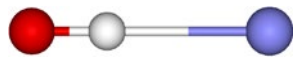 | PBE | | 4 | 2.1373 | 0.9920 | | 5.750 | 0.877 | -2.534 | 0.033 | 4.740 |
| | optB88 | | 4 | 2.1433 | 0.9905 | | 5.840 | 0.893 | -2.398 | 0.031 | 4.918 |
| Triangular HNO or ONH (Nitrosyl hydride) | | | | $l_{NH}$ | $l_{NO}$ | $\theta_{HNO}$ | | | | | |
| 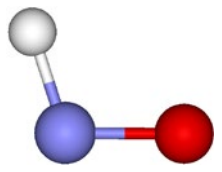 | PBE | | 0 | 1.0788 | 1.2168 | 108.49 | 0.960 | 5.666 | 2.255 | 4.822 | 9.529 |
| | optB88 | | 0 | 1.0757 | 1.2171 | 108.57 | 0.997 | 5.735 | 2.445 | 4.874 | 9.760 |
| | AOBM | [80] | | 1.0524 | 1.2085 | 108.08 | | | | | |
| | Exp. | [81] | | 1.0628 | 1.2116 | 108.058 | | | | | |
| | Exp. | [46] | | 1.09026 | 1.2090 | 108.047 | | | | | |
| | Exp. | [33] | | | | | 1.1395 | 5.1310 | 2.03894 | 4.1242 | 8.5351 |
| | | | | | | | 109.94±0.11 | 495.64±0.19 | 196.728±0.092 | 397.92±0.11 | 823.51±0.11 |
| Triangular NOH or HON (Hydroxyimidogen) | | | | $l_{ON}$ | $l_{OH}$ | $\theta_{NOH}$ | | | | | |
| 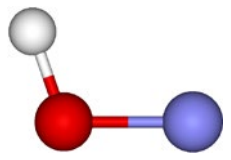 | PBE | | 2 | 1.3341 | 0.9849 | 109.04 | 1.922 | 4.705 | 1.294 | 3.861 | 8.568 |
| | optB88 | | 2 | 1.3380 | 0.9841 | 109.29 | 2.041 | 4.691 | 1.401 | 3.829 | 8.716 |
| | AOBM | [80] | | 1.3255 | 0.9676 | 107.47 | | | | | |
| | Exp. | [33] | | | | | 2.2580 | 4.0184 | 0.920 | 3.0056 | 7.4166 |
| | | | | | | | 217.86±0.77 | 387.72±0.79 | 88.81±0.77 | 290.00±0.77 | 715.59±0.77 |
| Formula: $H_1N_2$ | | | $M$ | | | | $\Delta H_f$ | $D_{HN+N}$ | $D_{H+N_2}$ | | $D_{H+2N}$ |
| Linear HNN or NNH | | | | $l_{NH}$ | $l_{NN}$ | | | | | | |
| | PBE | | 1 | 1.0045 | 1.1834 | | 3.124 | 5.674 | -0.855 | | 9.538 |

| | | | | | | | | | | |
|---|---|---|---|---|---|---|---|---|---|---|
| 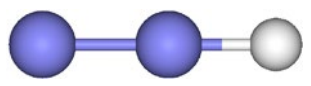 | optB88 | | 1 | 1.0022 | 1.1816 | | 3.285 | 5.711 | -0.791 | 9.736 |
| Linear NHN | | | | $l_{HN}$ | $l_{HN}$ | | | | | |
| 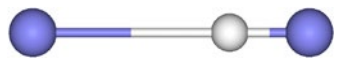 | PBE | | 1 | 1.0507 | 2.6013 | | 8.784 | 0.014 | -6.515 | 3.877 |
| | optB88 | | 1 | 1.0487 | 2.5896 | | 8.981 | 0.015 | -6.486 | 4.040 |
| Triangular HNN or NNH (Diazenyl) | | | | $l_{NH}$ | $l_{NN}$ | $\theta_{HNN}$ | | | | |
| 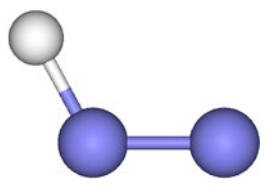 | PBE | | 1 | 1.0716 | 1.1806 | 118.36 | 2.016 | 6.783 | 0.253 | 10.646 |
| | optB88 | | 1 | 1.0698 | 1.1778 | 118.42 | 2.168 | 6.828 | 0.326 | 10.853 |
| | AOBM | [82] | | 1.045 | 1.157 | 118.0 | | | 0.438–1.184 | |
| | AOBM | [83] | | 1.031-<br>1.060 | 1.171-<br>1.181 | 113.2-<br>118.5 | | | | |
| | AOBM | [84] | | 1.062 | 1.197 | 116.3 | | | 0.323-0.898 | |
| | Exp. | [33] | | | | | 2.6133 | 5.9821 | -0.3743 | 9.3802 |
| | | | | | | | 252.14±0.45 | 577.18±0.47 | -36.11±0.45 | 905.05±0.45 |
| Formula: $H_1O_1$ | | | *M* | | | | $\Delta H_f$ | | | $D_{H+O}$ |
| OH or HO (Hydroxyl) | | | | $l_{HO}$ | | | | | | |
| 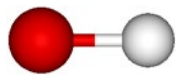 | PBE | | 1 | 0.9868 | | | 0.586 | | | 4.706 |
| | optB88 | | 1 | 0.9862 | | | 0.607 | | | 4.887 |
| | AOBM | [85] | | 0.971 | | | | | | |
| | AOBM | [86] | | | | | 0.3845 | | | |
| | Exp. | [37] | | 0.96966 | | | | | | |
| | Exp. | [33] | | | | | 0.38641 | | | 4.41098 |
| | | | | | | | 37.283±0.025 | | | 425.595±0.024 |
| Formula: $H_1O_2$ | | | $M$ | | | | $\Delta H_f$ | $D_{HO+O}$ | $D_{H+O_2}$ | $D_{H+2O}$ |
| Linear HOO or OOH | | | | $l_{OH}$ | $l_{OO}$ | | | | | |
| 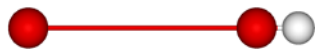 | PBE | | 3 | 0.9871 | 5.7400 | | 3.608 | 0.002 | -1.339 | 4.708 |
| | optB88 | | 3 | 0.9861 | 5.7397 | | 3.605 | 0.002 | -1.110 | 4.889 |
| Linear OHO | | | | $l_{HO}$ | $l_{HO}$ | | | | | |
| 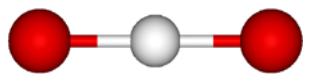 | PBE | | 3 | 1.1137 | 1.1195 | | 3.248 | 0.362 | -0.979 | 5.069 |
| | optB88 | | 3 | 1.1215 | 1.1152 | | 3.240 | 0.367 | -0.746 | 5.253 |
| Triangular HOO (Dioxidanyl) | | | | $l_{OH}$ | $l_{OO}$ | $\theta_{HOO}$ | | | | |
| 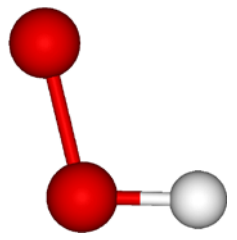 | PBE | | 1 | 0.9902 | 1.3451 | 105.01 | -0.004 | 3.614 | 2.273 | 8.320 |
| | optB88 | | 1 | 0.9888 | 1.3520 | 104.85 | -0.005 | 3.612 | 2.499 | 8.499 |
| | AOBM | [87] | | 1.00 | 1.35 | 104 | | | | |
| | AOBM | [87] | | 0.951 | 1.391 | 106 | | | | |
| | AOBM | [86] | | | | | 0.1535 | | | |
| | Exp. | [88] | | 0.9774 | 1.3339 | 104.15 | | | | |
| | Exp. | [89] | | 0.9707 | 1.33054 | 104.29 | | | | |
| | Exp. | [33] | | | | | 0.1569 | 2.7879 | 2.0822 | 7.1988 |
| | | | | | | | 15.14±0.15 | 268.99±0.15 | 200.90±0.15 | 694.58±0.15 |
| Formula: $H_2$ | | | $M$ | | | | $\Delta H_f$ | | | $D_{2H}$ |
| HH (Dihydrogen) | | | | $l_{HH}$ | | | | | | |
| 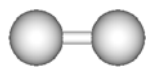 | PBE | | 0 | 0.7500 | | | 0 | | | 4.538 |
| | optB88 | | 0 | 0.7442 | | | 0 | | | 4.989 |
| | AOBM | [90] | | 0.7668 | | | | | | |
| | Exp. | [91] | | 0.74130 | | | | | | |
| | Exp. | [37] | | 0.74144 | | | | | | |
| | Exp. | [33] | | | | | | | | 4.478069884<br>432.0680600<br>±0.0000092 |
| Formula: $H_2N_1$ | | | $M$ | | | | $\Delta H_f$ | $D_{H+HN}$ | $D_{H_2+N}$ | $D_{2H+N}$ |
| Linear NHH or HHN | | | | $l_{HN}$ | $l_{HH}$ | | | | | |
| 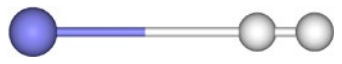 | PBE | | 3 | 2.8632 | 0.7506 | | 5.189 | 0.682 | 0.008 | 4.546 |
| | optB88 | | 3 | 2.9003 | 0.7443 | | 5.259 | 0.968 | 0.005 | 4.993 |
| Linear HNH | | | | $l_{NH}$ | $l_{NH}$ | | | | | |
| 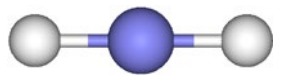 | PBE | | 1 | 0.9986 | 0.9986 | | 2.893 | 2.979 | 2.304 | 6.842 |
| | optB88 | | 1 | 0.9965 | 0.9965 | | 3.083 | 3.144 | 2.180 | 7.169 |
| Triangular HNH (Amidogen) | | | | $l_{NH}$ | $l_{NH}$ | $\theta_{HNH}$ | | | | |
| 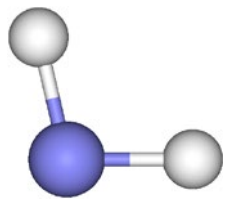 | PBE | | 1 | 1.0348 | 1.0347 | 102.89 | 1.555 | 4.317 | 3.642 | 8.180 |
| | optB88 | | 1 | 1.0374 | 1.0374 | 102.46 | 1.739 | 4.487 | 3.524 | 8.512 |
| | AOBM | [92] | | 1.023 | 1.023 | 102.9 | | | | |
| | Exp. | [64] | | 1.024 | 1.024 | 103.4 | | | | |
| | Exp. | [33] | | | | | 1.9580<br>188.92±0.11 | 3.9991<br>385.85±0.17 | 2.9192<br>281.66±0.11 | 7.3972<br>713.72±0.11 |
| Formula: $H_2O_1$ | | | $M$ | | | | $\Delta H_f$ | $D_{H+HO}$ | $D_{H_2+O}$ | $D_{2H+O}$ |
| Linear OHH or HHO | | | | $l_{HO}$ | $l_{HH}$ | | | | | |
| | PBE | | 2 | 0.9922 | 1.9384 | | 2.833 | 0.022 | 0.191 | 4.729 |
| 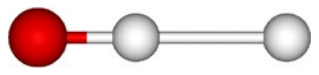 | optB88 | | 2 | 1.0039 | 1.5461 | | 3.055 | 0.047 | -0.055 | 4.934 |
| | Exp. | [33] | | | | | 2.566 | 0.059 | -0.007 | 4.470 |
| | | | | | | | 247.6±2.7 | 5.7±2.7 | -0.7±2.7 | 431.3±2.7 |
| Linear HOH | | | | $l_{OH}$ | $l_{OH}$ | | | | | |
| 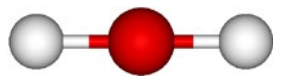 | PBE | | 0 | 0.9428 | 0.9428 | | -1.266 | 4.122 | 4.290 | 8.828 |
| | optB88 | | 0 | 0.9419 | 0.9420 | | -1.182 | 4.283 | 4.182 | 9.170 |
| Triangular1 HOH (Water) | | | | $l_{OH}$ | $l_{OH}$ | $\theta_{HOH}$ | | | | |
| 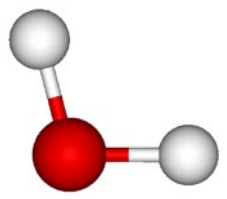 | PBE | | 0 | 0.9719 | 0.9718 | 104.51 | -2.520 | 5.375 | 5.544 | 10.081 |
| | optB88 | | 0 | 0.9710 | 0.9709 | 104.82 | -2.433 | 5.534 | 5.433 | 10.421 |
| | AOBM | [85] | | 0.960 | 0.960 | 103.4 | | | | |
| | Exp. | [93] | | 0.95781 | 0.95781 | 104.4776 | | | | |
| | Exp. | [33] | | | | | -2.47600 | 5.101446 | 5.03436 | 9.51243 |
| | | | | | | | -238.898±0.025 | 492.2147±0.0011 | 485.742±0.024 | 917.810±0.024 |
| Triangular2 HOH | | | | $l_{OH}$ | $l_{OH}$ | $\theta_{HOH}$ | | | | |
| 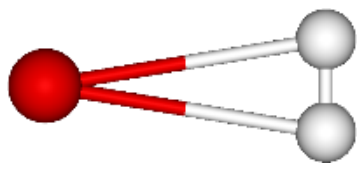 | PBE | | 0 | 2.3306 | 2.3306 | 18.66 | 2.991 | -0.136 | 0.033 | 4.571 |
| | optB88 | | 0 | 2.4567 | 2.4567 | 17.51 | 2.982 | 0.119 | 0.018 | 5.006 |
| | Exp. | [33] | | | | | 2.562 | 0.063 | -0.004 | 4.474 |
| | | | | | | | 247.2±2.0 | 6.1±1.9 | -0.4±1.9 | 431.7±1.9 |
| Formula: $H_3$ | | | $M$ | | | | $\Delta H_f$ | $D_{H+H_2}$ | | $D_{3H}$ |
| Linear1 HHH | | | | $l_{HH}$ | $l_{HH}$ | | | | | |
| 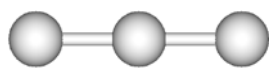 | PBE | | 0 | 0.9358 | 0.9358 | | 3.028 | -0.759 | | 3.779 |
| | optB88 | | 0 | 0.9330 | 0.9330 | | 3.109 | -0.614 | | 4.374 |

| | | | | | | | | | |
|---|---|---|---|---|---|---|---|---|---|
| | AOBM | [94] | | 0.90 | 0.90 | | | | |
| | AOBM | [95] | | 0.90-0.95 | 0.90-0.95 | | | | |
| Linear2 HHH (Trihydrogen) | | | | $l_{HH}$ | $l_{HH}$ | | | | |
| 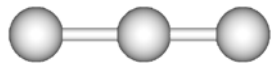 | PBE | | 1 | 0.8877 | 0.9968 | | 2.428 | -0.159 | 4.379 |
| | optB88 | | 1 | 0.8963 | 0.9773 | | 2.612 | -0.118 | 4.871 |
| | AOBM | [94] | | 0.828 | 1.040 | | | | |
| | AOBM | [95] | | 0.889 | 0.995 | | | | |
| | Exp. | [33] | | | | | 2.638 | -0.398 | 4.079 |
| | | | | | | | 254.5±1.2 | -38.4±1.1 | 393.6±1.1 |
| Linear3 HHH | | | | $l_{HH}$ | $l_{HH}$ | | | | |
| 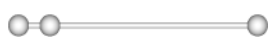 | PBE | | 1 | 0.7500 | 4.9214 | | 2.269 | 0.000 | 4.538 |
| | optB88 | | 1 | 0.7443 | 4.9213 | | 2.494 | 0.001 | 4.989 |
| | AOBM | [94] | | 0.74 | 4.92 | | | | |
| | Exp. | [33] | | | | | 2.238 | 0.002 | 4.479 |
| | | | | | | | 215.9±1.1 | 0.2±1.1 | 432.2±1.1 |
| Triangular (cyclic) HHH | | | | $l_{HH}$ | $l_{HH}$ | $\theta_{HHH}$ | | | |
| 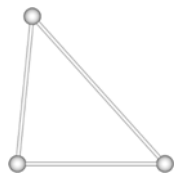 | PBE | | 3 | 3.6979 | 3.8235 | 83.45 | 6.805 | -4.536 | 0.002 |
| | optB88 | | 3 | 4.1644 | 4.2243 | 84.30 | 7.483 | -4.988 | 0.000 |
| Formula: $N_1O_1$ | | | $M$ | | | | $\Delta H_f$ | | $D_{N+O}$ |
| NO or ON (Nitric oxide) | | | | $l_{NO}$ | | | | | |
| 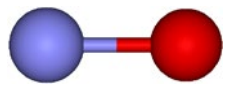 | PBE | | 1 | 1.1687 | | | 0.946 | | 7.274 |
| | optB88 | | 1 | 1.1676 | | | 0.948 | | 7.316 |

| | | | | | | | | | | |
|---|---|---|---|---|---|---|---|---|---|---|
| | AOBM | [96] | | 1.169 | | | | | | |
| | AOBM | [96] | | 1.154 | | | | | | |
| | Exp. | [37] | | 1.15077 | | | | | | |
| | Exp. | [33] | | | | | 0.93940 | | | 6.49617 |
| | | | | | | | 90.638±0.066 | | | 626.785±0.062 |
| Formula: $N_1O_2$ | | | $M$ | | | | $\Delta H_f$ | $D_{NO+O}$ | $D_{N+O_2}$ | $D_{N+2O}$ |
| Linear NOO or OON | | | | $l_{ON}$ | $l_{OO}$ | | | | | |
| 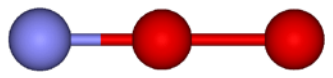 | PBE | | 1 | 1.2215 | 1.3137 | | 4.623 | -0.653 | 0.574 | 6.622 |
| | optB88 | | 1 | 1.2221 | 1.3260 | | 4.606 | -0.658 | 0.658 | 6.657 |
| Triangular NOO (Peroxyimidogen) | | | | | | | | | | |
| 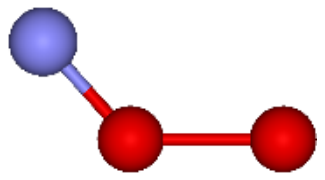 | PBE | | 1 | 1.2019 | 1.3714 | | 3.389 | 0.581 | 1.808 | 7.855 |
| | optB88 | | 1 | 1.1960 | 1.3948 | | 3.378 | 0.569 | 1.885 | 7.885 |
| | Exp. | [33] | | | | | 4.202 | -0.704 | 0.676 | 5.793 |
| | | | | | | | 405.4±1.9 | -67.9±1.9 | 65.2±1.9 | 558.9±1.9 |
| Linear ONO | | | | $l_{NO}$ | $l_{NO}$ | | | | | |
| 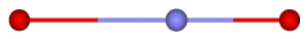 | PBE | | 7 | 3.2516 | 4.5098 | | 11.228 | -7.258 | -6.031 | 0.016 |
| | optB88 | | 7 | 3.2515 | 4.5096 | | 11.247 | -7.300 | -5.984 | 0.016 |
| Triangular ONO (Nitrogen dioxide) | | | | $l_{NO}$ | $l_{NO}$ | $\theta_{ONO}$ | | | | |
| 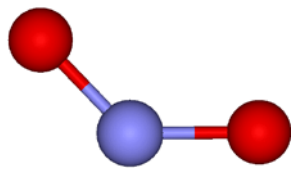 | PBE | | 1 | 1.2118 | 1.2119 | 133.86 | -0.195 | 4.165 | 5.391 | 11.439 |
| | optB88 | | 1 | 1.2128 | 1.2129 | 133.40 | -0.110 | 4.057 | 5.373 | 11.372 |
| | AOBM | [96] | | 1.216 | 1.216 | 133.7 | | | | |
| | AOBM | [96] | | 1.202 | 1.202 | 134.0 | | | | |
| | Exp. | [64] | | 1.193 | 1.193 | 134.1 | | | | |
| | Exp. | [33] | | | | | 0.38221<br>36.878±0.066 | 3.1155438<br>300.60428±0.00031 | 4.49500<br>433.702±0.062 | 9.61171<br>927.389±0.062 |
| Formula: $N_2$ | | | $M$ | | | | $\Delta H_f$ | | | $D_{2N}$ |
| NN (Dinitrogen) | | | | $l_{NN}$ | | | | | | |
| 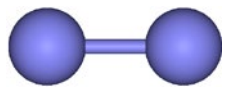 | PBE | | 0 | 1.1129 | | | 0 | | | 10.393 |
| | optB88 | | 0 | 1.1087 | | | 0 | | | 10.527 |
| | AOBM | [84] | | 1.102-<br>1.109 | | | | | | 9.19-9.50 |
| | Exp. | [37] | | 1.09768 | | | | | | |
| | Exp. | [33] | | | | | | | | 9.75443<br>941.159±0.046 |
| Formula: $N_2O_1$ | | | $M$ | | | | $\Delta H_f$ | $D_{N+NO}$ | $D_{N_2+O}$ | $D_{2N+O}$ |
| Linear NNO or ONN (Nitrous oxide) | | | | $l_{NN}$ | $l_{NO}$ | | | | | |
| 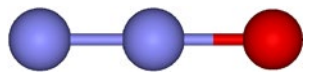 | PBE | | 0 | 1.1431 | 1.1977 | | 0.168 | 5.974 | 2.856 | 13.248 |
| | optB88 | | 0 | 1.1394 | 1.1996 | | 0.308 | 5.903 | 2.692 | 13.219 |
| | AOBM | [97] | | 1.13 | 1.21 | | | | 3.36 | |
| | AOBM | [97] | | 1.12 | 1.20 | | | | 3.63 | |
| | Exp. | [98] | | 1.127292 | 1.185089 | | | | | |
| | Exp. | [33] | | | | | 0.89157<br>86.023±0.096 | 4.9250<br>475.19±0.12 | 1.66678<br>160.820±0.096 | 11.4212<br>1101.98±0.11 |
| Linear NON | | | | $l_{ON}$ | $l_{ON}$ | | | | | |
| | PBE | | 0 | 1.2010 | 1.2012 | | 4.722 | 1.420 | -1.698 | 8.695 |
| 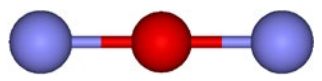 | optB88 | | 0 | 1.2028 | 1.2027 | | 4.828 | 1.382 | -1.829 | 8.698 |
| | Exp. | [33] | | | | | 5.654 | 0.163 | -3.096 | 6.659 |
| | | | | | | | 545.5±1.7 | 15.7±1.7 | -298.7±1.7 | 642.5±1.7 |
| Triangular (cyclic) NON (Oxadiazirene) | | | | $l_{\mathrm{ON}}$ | $l_{\mathrm{ON}}$ | $\theta_{\mathrm{NON}}$ | | | | |
| 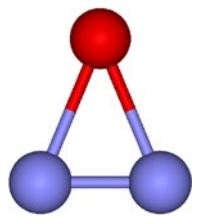 | PBE | | 0 | 1.5344 | 1.5344 | 45.98 | 3.255 | 2.888 | -0.231 | 10.162 |
| | optB88 | | 0 | 1.5507 | 1.5507 | 45.16 | 3.294 | 2.917 | -0.294 | 10.232 |
| | Exp. | [33] | | | | | 3.645 | 2.171 | -1.087 | 8.668 |
| | | | | | | | 351.7±1.7 | 209.5±1.6 | -104.9±1.6 | 836.3±1.6 |
| Formula: $N_3$ | | | $M$ | | | | $\Delta H_{\mathrm{f}}$ | $D_{\mathrm{N+N_2}}$ | | $D_{\mathrm{3N}}$ |
| Linear NNN (Azido radical) | | | | $l_{\mathrm{NN}}$ | $l_{\mathrm{NN}}$ | | | | | |
| 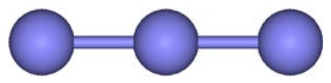 | PBE | | 1 | 1.1888 | 1.1888 | | 3.612 | 1.584 | | 11.977 |
| | optB88 | | 1 | 1.1866 | 1.1866 | | 3.797 | 1.466 | | 11.993 |
| | AOBM | [99] | | 1.1570 | 1.1570 | | | | | |
| | AOBM | [99] | | 1.1538 | 1.1538 | | | | | |
| | Exp. | [100] | | 1.1815 | 1.1815 | | | | | |
| | Exp. | [101] | | 1.18115 | 1.18115 | | | | | |
| | Exp. | [33] | | | | | 4.6871 | 0.1901 | | 9.9445 |
| | | | | | | | 452.24±0.60 | 18.34±0.59 | | 959.50±0.5 |
| Triangular1 (cyclic) NNN | | | | $l_{\mathrm{NN}}$ | $l_{\mathrm{NN}}$ | $\theta_{\mathrm{NNN}}$ | | | | |
| 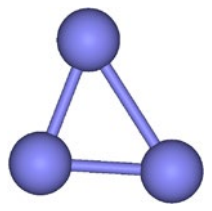 | PBE | | 1 | 1.3092 | 1.3092 | 71.38 | 5.270 | -0.074 | | 10.319 |
| | optB88 | | 1 | 1.3078 | 1.3077 | 72.02 | 5.521 | -0.258 | | 10.269 |

| | | | | | | | | | |
|---|---|---|---|---|---|---|---|---|---|
| Triangular2 (cyclic) NNN (1H-Triazirin-1-yl) | | | | $l_{NN}$ | $l_{NN}$ | $\theta_{NNN}$ | | | |
| 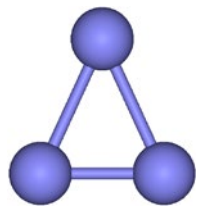 | PBE | | 1 | 1.4654 | 1.4654 | 50.05 | 5.271 | -0.075 | 10.318 |
| | optB88 | | 1 | 1.4749 | 1.4749 | 49.48 | 5.509 | -0.246 | 10.281 |
| | B3LYP | [102] | | 1.453 | 1.453 | 49.69 | | | |
| | Exp. | [33] | | | | | 6.07243 | -1.195 | 8.55985 |
| | | | | | | | 585.9±2.0 | -115.3±1.9 | 825.9±1.9 |
| Triangular3 (cyclic) NNN | | | | $l_{NN}$ | $l_{NN}$ | $\theta_{NNN}$ | | | |
| 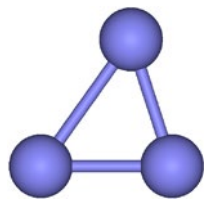 | PBE | | 1 | 1.5222 | 1.3596 | 70.62 | 5.271 | -0.074 | 10.319 |
| | optB88 | | 1 | 1.5381 | 1.3166 | 72.01 | 5.521 | -0.258 | 10.269 |
| Formula: $O_2$ | | | $M$ | | | | $\Delta H_f$ | | $D_{2O}$ |
| OO (Dioxygen) | | | | $l_{OO}$ | | | | | |
| 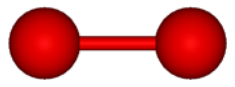 | PBE | | 2 | 1.2327 | | | 0 | | 6.048 |
| | optB88 | | 2 | 1.2360 | | | 0 | | 6.000 |
| | AOBM | [96] | | 1.229 | | | | | |
| | AOBM | [96] | | 1.210 | | | | | |
| | PBE | [103] | | 1.2343 | | | | | |
| | Exp. | [104] | | 1.2074 | | | | | |
| | Exp. | [37] | | 1.20752 | | | | | |
| | Exp. | [33] | | | | | | | 5.116708 |
| | | | | | | | | | 493.6873±0.0041 |
| Formula: $O_3$ | | | $M$ | | | | $\Delta H_f$ | $D_{O+O_2}$ | $D_{3O}$ |
| Linear OOO | | | | $l_{OO}$ | $l_{OO}$ | | | | |
| 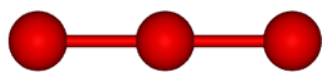 | PBE | | 2 | 1.3098 | 1.3097 | | 3.778 | -0.754 | 5.294 |
| | optB88 | | 2 | 1.3176 | 1.3176 | | 3.688 | -0.688 | 5.312 |
| Triangular OOO (Ozone) | | | | $l_{OO}$ | $l_{OO}$ | $\theta_{OOO}$ | | | |
| 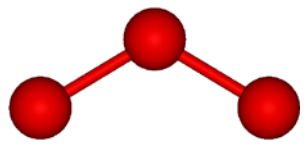 | PBE | | 0 | 1.2844 | 1.2840 | 118.19 | 1.287 | 1.737 | 7.784 |
| | optB88 | | 0 | 1.2894 | 1.2894 | 117.99 | 1.218 | 1.782 | 7.782 |
| | AOBM | [105] | | 1.299 | 1.299 | 116.0 | | | |
| | AOBM | [106] | | 1.277 | 1.277 | 116.75 | | | |
| | AOBM | [96] | | 1.296 | 1.296 | 116.5 | | | |
| | AOBM | [96] | | 1.276 | 1.276 | 116.9 | | | |
| | Exp. | [107] | | 1.278 | 1.278 | 116.82 | | | |
| | Exp. | [108] | | 1.278 | 1.278 | 116.75 | | | |
| | Exp. | [109] | | 1.2717 | 1.2717 | 116.78 | | | |
| | Exp. | [110] | | 1.27276 | 1.27276 | 116.7542 | | | |
| | Exp. | [33] | | | | | 1.49666 | 1.06168 | 6.1784 |
| | | | | | | | 144.406±0.039 | 102.437±0.039 | 596.125±0.039 |
| Triangular (cyclic) OOO | | | | $l_{HH}$ | $l_{HH}$ | $\theta_{HHH}$ | | | |
| 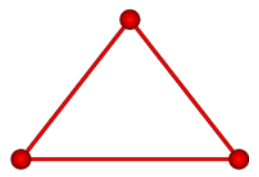 | PBE | | 6 | 5.1778 | 5.1778 | 76.46 | 9.067 | -6.043 | 0.004 |
| | optB88 | | 6 | 5.2392 | 5.2392 | 75.48 | 8.999 | -5.999 | 0.001 |

## A. Diatomic molecules

The energy of a diatomic molecule is a function of interatomic distance (or atomic spacing), i.e., the potential energy surface (PES) is a one-dimensional (1D) curve. The bond length $l$ between two atoms is defined as the atomic spacing at their equilibrium positions generally corresponding to the sole energy minimum of the PES.

Let us first see 3 one-element diatomic molecules HH or $H_2$, NN or $N_2$, and OO or $O_2$, which are best known in the air in our daily lives. For $H_2$, the bond lengths 0.7500 Å and 0.7442 Å from our PBE and optB88-vdW calculations reproduce very well the experimental values 0.74144 Å [37] and 0.74130 Å [91]. These values from our PAW DFT calculations are even better than 0.7668 Å from previous Gaussian basis *ab initio* calculations [90]; the dissociation energies 4.458 eV and 4.989 eV from our PBE and optB88-vdW calculations are fairly good when compared with the experimental value 4.4781 eV [33]. For $N_2$, the bond lengths 1.1129 Å (PBE) and 1.1087 Å (optB88-vdW) also reproduce very well the experimental value 1.09768 Å [37] and very close to the values 1.102 to 1.109 Å from previous AOBM calculations [84]; the dissociation energies 10.393 eV (PBE) and 10.527 eV (optB88-vdW) are fairly good when compared with the experimental value 9.754 eV [33] and the values 9.19 to 9.50 eV from previous AOBM calculations [84]. For $O_2$, the bond lengths 1.2327 Å (PBE) and 1.2360 Å (optB88-vdW) still reproduce very well the experimental values 1.20752 Å [37] and 1.2074 Å [104], also very close to the values 1.210 to 1.229 Å from previous AOBM calculations [96]; the dissociation energies 6.048 eV (PBE) and 6.000 eV (optB88-vdW) are comparably good with the experimental value 5.117 eV [33].

For the C dimer CC or $C_2$, the bond lengths 1.3142 Å and 1.3093 Å from our PBE and optB88-vdW calculations somewhat overestimate the experimental value 1.2425 Å [37] than the values 1.247 to 1.255 Å from previous AOBM calculations [65, 66]. As will analyzed below, the errors for $C_2$ are the upper limits for all bond lengths obtained from our PAW DFT calculations in this work. However, the formation enthalpies 8.780 eV (PBE) and 8.865 eV (optB88-vdW) match well the experimental values 8.499 eV [33] with very low relative errors (see the error analysis in Sec. IV C). The very positive formation enthalpies indicate that forming a $C_2$ molecule from the bulk graphite is extremely unfavorable. The dissociation energies 6.935 eV (PBE) and 6.704 eV (optB88-vdW) are also comparably good relative to the experimental value 6.248 eV [33] and the AOBM values of 6.072 to 6.371 eV [66]. The very positive dissociation energies indicate that the bond strength in a $C_2$ molecule is covalently strong.

For CH or HC, the bond lengths 1.1369 Å and 1.1300 Å from our PBE and optB88-vdW calculations have good agreements with the experimental values 1.1199 Å [37] and 1.119786 Å [38], as well as 1.1204 Å from previous AOBM calculation [36]. The formation enthalpies are 6.429 eV (PBE) and 6.384 eV (optB88-vdW), good in line with the experimental value 6.144 eV [33]. The dissociation energies 3.697 eV (PBE) and 3.895 eV (optB88-vdW) are also comparably good relative to the experimental values 3.468 eV [33].

For CN or NC, the bond lengths 1.1768 Å and 1.1733 Å from our PBE and optB88-vdW calculations have excellent agreements with the experimental value 1.1718 Å [37], an AOBM value 1.18 Å [52], and the value 1.162 Å from a hybrid functional B3LYP [53]. The formation enthalpies 4.515 eV (PBE) and 4.606 eV (optB88-vdW) are also in excellent agreements with the experimental value 4.526 eV [33]. The dissociation energies 8.539 eV (PBE) and 8.441 eV (optB88-vdW) are comparable with the experimental value 7.724 eV [33].

For CO or OC, the bond lengths 1.1431 Å and 1.1393 Å from our PBE and optB88-vdW calculations have good agreements with the experimental value 1.128323 Å [37] and the AOBM value 1.1513 Å [62]. Our PBE and optB88-vdW calculations predict the negative formation enthalpies −0.621 eV and −0.715 eV, consistent with the experimental value −1.1795 eV [33]. The negative formation enthalpy indicates that forming CO from gas phase $O_2$ and bulk graphite is favorable thermodynamically. The dissociation energies 11.502 eV (PBE) and 11.499 eV (optB88-vdW) are in good agreements with the experimental value 11.111 eV [33]. The dissociation energy of CO is largest among those of all 10 diatomic molecules in this section, indicating the strongest bond strength.

For HN or NH, the bond lengths 1.0506 Å and 1.0485 Å from our PBE and optB88-vdW calculations are in good agreements with the experimental value 1.0362 Å [37] and the AOBM value 1.035 Å [79]. The formation enthalpies 3.602 eV (PBE) and 3.732 eV (optB88-vdW) are in excellent agreement with the experimental value 3.718 eV [33]. The dissociation energies 3.863 eV (PBE) and 4.025 eV (optB88-vdW) are fairly good relative to the experimental value 3.398 eV [33].

For HO or OH, the bond lengths 0.9868 Å and 0.9862 Å from our PBE and optB88-vdW calculations are in good agreements with the experimental value 0.96966 Å [37] and the AOBM value 0.971 Å [85]. The formation enthalpies 0.586 eV (PBE) and 0.607 eV (optB88-vdW) are comparable with the experimental value 0.386 eV [33] and the AOBM value 0.3845 eV [86]. The dissociation energies 4.706 eV (PBE) and 4.887 eV (optB88-vdW) are also comparable with the experimental value 4.411 eV [33].

For NO or ON, the bond lengths 1.1687 Å and 1.1676 Å from our PBE and optB88-vdW calculations have good agreements with the experimental value 1.15077 Å [37] and the AOBM values 1.154 Å and 1.169 Å [96]. The formation enthalpies 0.946 eV (PBE) and 0.948 eV (optB88-vdW) are in excellent agreements with the experimental value 0.939 eV [33]. The dissociation energies 7.274 eV (PBE) and 7.316 eV (optB88-vdW) are comparable with the experimental value 6.496 eV [33].

For the above 10 diatomic molecules, we also rank them in order of bond lengths, formation enthalpies, and dissociation energies. From smallest (about 0.74 Å) to largest (about 1.25 Å) bond length, the order from our PBE and optB88-vdW calculations or from experiments is always consistently HH < HO < HN < NN < CH < CO < NO < CN < OO < CC. From lowest (about −1 eV) to highest (about 9 eV) formation enthalpy, the order from our PBE and optB88-vdW calculations or from experiments is always consistently CO < (HH = NN = OO) < HO < NO < HN < CN < CH < CC, where the formation enthalpy of HH, NN, or OO is zero (i.e., the energy reference point). CO is the sole one with a negative formation enthalpy among the 10 diatomic molecules, indicating that it is most stable thermodynamically relative to its reference systems. From lowest (about 3 eV) to highest (about 11 eV) dissociation energy (i.e., from weakest to strongest bond strength), the order from our optB88-vdW calculations or from experiments is consistently CH ≈ HN < HO < HH < OO < CC < NO < CN < NN < CO, but there is a switch in the order of HO and HH from our PBE calculations.

## B. Triatomic molecules

The energy of a triatomic molecule is a function of three interatomic distances (or equivalently two interatomic distances and the angle between them), and therefore the PES is a 3D curve and much more complicated than the 1D curve for a diatomic molecule. In general, the PES of a triatomic molecule has multiple energy minima corresponding to different structural geometries (or spatial arrangements of atoms) which are commonly called the locally stable isomers, and the global energy minimum corresponds to the most stable isomer. For clarity, we refer to all isomers having the same number of atoms in each element but distinct structural geometries as *one isomer type*. For triatomic molecules consisting of C, H, N, and/or O, there are 20 isomer types in total. In principle, all locally stable isomers of each isomer type can be found by calculating and searching its full 3D PES curve. However, such work is demanding. In this paper, we only focus on several typical isomers for each type, including linear and triangular isomers, especially for those with available data from previous experiments or AOBM calculations.

In Table I, we show the optimized structural geometries of all isomers from our optB88-vdW calculations (the corresponding structural geometries from our PBE calculations are very close to those from the optB88-vdW calculations and therefore not shown). All bond lengths, formation enthalpies, dissociation energies, and spin magnetic moments from our PBE and optB88-vdW calculations, as well as from previous experiments and AOBM calculations, are provided in Table I. In this section, we selectively discuss the 20 types of triatomic molecular isomers. For the isomer types, we adopt the NIST notations [32] for convenient indexing with the formulas in alphabetic order of element symbols and the numbers of elemental atoms in the molecule as subscripts 1, 2, 3, …, etc. In Table I, the conventional notations and names for isomers are also listed.

For each isomer type, we consider various possible linear and triangular isomers by relaxing the judiciously selected initial configurations, e.g., based on previous experiments or AOBM calculations in the literature. For a linear isomer, the name $\alpha\beta\gamma$ (or equivalently $\gamma\beta\alpha$) indicates that atom $\beta$ is in the middle position with atoms $\alpha$ and $\gamma$ on the two sides, where $\alpha$, $\beta$, and $\gamma$ can be C, H, N, and/or O; correspondingly, the two bond lengths are denoted by $l_{\beta\alpha}$ and $l_{\beta\gamma}$. The triangular isomers can be obtuse (or quasilinear) and acute (or cyclic). The order of atoms in the name of a triangular isomer also indicates the optimized configuration, e.g., for the triangular isomer name $\alpha\beta\gamma$ or $\gamma\beta\alpha$, atom $\beta$ is in the middle position with atoms $\alpha$ and $\gamma$ on the two sides; correspondingly, the two bond lengths are denoted by $l_{\beta\alpha}$ and $l_{\beta\gamma}$ with the bond angle denoted as $\theta_{\alpha\beta\gamma}$ or $\theta_{\gamma\beta\alpha}$. Below, we also use notations $\alpha\beta\gamma$ (L), $\alpha\beta\gamma$ (Q), and $\alpha\beta\gamma$ (C) to indicate that the isomer $\alpha\beta\gamma$ is linear, quasilinear triangular, and cyclic triangular, respectively.

$C_1H_1N_1$. We obtain 1 triangular isomer and 3 linear (HCN, HNC, and CHN) isomers. The triangular isomer is cyclic, and no any stable or metastable quasilinear isomer are found. A lower (higher) formation enthalpy indicates a more (less) stable isomer. From more to less stable, the order is HCN (L) > HNC (L) > HCN (C) > CHN (L) with the formation enthalpies of about 1.3 eV, 1.9 eV, 3.3 eV, and 11.6 eV, respectively. Our DFT (PBE and optB88-vdW) values for bond lengths, bond angles, formation enthalpies, and dissociation energies are in good agreements with available experimental and AOBM values (see Table I). Generally, the experimental measurements for less stable isomers with significantly higher formation enthalpies are often more difficult, and therefore the experimental data for the linear CHN and cyclic isomers are still unavailable in the literature. Note that, for experimental observations, a "less stable" isomer means not only a relatively higher formation enthalpy (i.e., thermodynamically less stable), but also a low energy barrier (i.e., kinetically less stable) from that isomer to another with a lower formation enthalpy, while the latter isomer is "more stable" both thermodynamically and kinetically. In this work, we do not consider the energy barriers between isomers. Therefore, in all statements below, the word “stable” always means “thermodynamically stable”, i.e., we omit the word “thermodynamically”.

$C_1H_1O_1$. We obtain 2 quasilinear (HCO and HOC) and 3 linear (HCO, HOC, and CHO) isomers. From more to less stable, the order is HCO (Q) > HCO (L) > HOC (Q) > HOC (L) > CHO (L) with the formation enthalpies of about 0.5 eV, 1.5 eV, 1.7 eV, 2.3 eV, and 8.3 eV, respectively. Our DFT values for bond lengths, bond angles, formation enthalpies, and dissociation energies are in good agreements with available experimental and AOBM values for HCO (Q) and HOC (Q) (see Table I). In addition, our DFT values of $D_{\mathrm{CO+H}}$ for HOC (L) are zero, indicating no bonding between H and CO, i.e., HOC (L) is not one single “real” molecule but an isolated CO molecule plus an isolated H atom, as also indicated by the large H-O distance beyond 6.6 Å (see Table I). In other words, no energy minimum is found for linear HOC.

$C_1H_2$. We obtain 1 quasilinear (HCH) isomer and 2 linear (HCH and CHH) isomers. The most stable isomer is HCH (Q) with the formation enthalpies 3.890 eV (PBE) and 4.176 eV (optB88-vdW), cf. the experimental value 4.053 eV [33]. The C-H bond lengths 1.0849 Å (PBE) and 1.0816 Å (optB88-vdW) as well as the bond angle 135.06° (PBE) and 135.04° (optB88-vdW) are in excellent agreements with the experimental bond length 1.07530 Å of and bond angle of 133.9308° [51] as well as the AOBM bond length 1.07481 Å of and bond angle of 133.839° [50]. In contrast, the isomer HCH (L) is slightly unstable with the formation enthalpies 4.078 eV (PBE) and 4.368 eV (optB88-vdW), while the isomer CHH (L) is extremely unstable with the significantly higher formation enthalpies 7.635 eV (PBE) and 7.704 eV (optB88-vdW). For dissociation energies of HCH (Q), our DFT values are also in good agreements with available experimental values (see Table I).

$C_1N_1O_1$. We obtain 1 triangular isomer and 3 linear (CNO, CON, and NCO) isomers. The triangular isomer is cyclic, and no any stable or metastable quasilinear isomer are found. From more to less stable, the order is NCO (L) > CNO (L) > NCO (C) > CON (L) with the formation enthalpies of about 1 eV, 4 eV, 4.5 eV, and 6 eV, respectively. Our DFT values for bond lengths, bond angles, and formation enthalpies are in good agreements with available experimental and AOBM values (see Table I).

$C_1N_2$. We obtain 1 triangular isomer and 2 linear (NCN and CNN) isomers. The triangular isomer is cyclic, and no any stable or metastable quasilinear isomer are found. The most stable isomer is NCN (L) with the formation enthalpies 4.149 eV (PBE) and 4.301 eV (optB88-vdW), cf. the experimental value 4.677 eV [33]. CNN (L) has the formation enthalpies 5.308 eV (PBE) and 5.484 eV (optB88-vdW), while NCN (C) has the highest formation enthalpies 5.733 eV (PBE) and 5.887 eV (optB88-vdW). Our DFT values for bond lengths and bond angles for these isomers are in good agreements with available B3LYP and AOBM values (see Table I). No experimental data for geometric parameters are available for this isomer type. The difficulty of experimental measurements for these isomers is likely because of their instabilities with the high formation enthalpies beyond 4 eV and likely low energy barriers for dissociations.

$C_1O_2$. The most stable isomer from our DFT calculations is linear OCO (L) (i.e., conventionally $CO_2$) with the very negative formation enthalpies −3.870 eV (PBE) and −3.863 eV (optB88-vdW), cf. the experimental value −4.074 eV [33], indicating its high stability. We also obtain another linear isomer COO (L) and a triangular cyclic isomer, both of which have very positive formation enthalpies and therefore relatively very unstable. For bond length $l_{\mathrm{CO}}$ and other formation enthalpies or dissociation energies, our DFT values are also comparably good relative to available experimental and AOBM values (see Table I).

$C_2H_1$. The most stable isomer from our DFT calculations is linear CCH (L) with the very positive formation enthalpies 5.938 eV (PBE) and 6.175 eV (optB88-vdW), cf. the experimental value 5.843 eV [33], indicating its high instability. For bond lengths $l_{\mathrm{CC}}$ and $l_{\mathrm{CH}}$, as well as dissociation energies of CCH (L), our DFT values have good agreements with available experimental and AOBM values (see Table I). A triangular cyclic isomer is obtained with the higher formation enthalpies 6.750 eV (PBE) and 7.016 eV (optB88-vdW). We also obtain another linear isomer CHC (L) with very high formation enthalpies 12.903 eV (PBE) and 13.054 eV (optB88-vdW).

$C_2N_1$. The most stable isomer from our DFT calculations is linear CNC (L) with the very positive formation enthalpies 6.873 eV (PBE) and 6.892 eV (optB88-vdW), cf. the experimental value 6.967 eV [33], indicating its high instability. We also obtain another linear isomer CCN (L) with slightly higher formation enthalpies 6.929 eV (PBE) and 6.966 eV (optB88-vdW). The high instability makes experimental measurements difficult for its structural parameters [71]. A triangular cyclic isomer is also obtained with the further higher formation enthalpies 7.375 eV (PBE) and 7.531 eV (optB88-vdW). For bond lengths, formation enthalpies, and dissociation energies of all these isomers, our DFT values have good agreements with available experimental and AOBM values (see Table I).

$C_2O_1$. The most stable isomer from our DFT calculations is linear CCO (L) with the formation enthalpies 3.876 eV (PBE) and 3.991 eV (optB88-vdW), cf. the experimental values 3.909 eV [33]. For bond lengths $l_{\mathrm{CC}}$ and $l_{\mathrm{CO}}$, as well as dissociation energies, our DFT values are comparably good relative to available experimental and AOBM values (see Table I). A triangular isomer CCO (Q) with bond angles 109.034° (PBE) and 108.665° (optB88-vdW) (see Table I) is obtained with the much higher formation enthalpies 5.957 eV (PBE) and 6.008 eV (optB88-vdW). We also obtain another linear isomer COC (L) with high formation enthalpies 6.501 eV (PBE) and 6.575 eV (optB88-vdW). The formation enthalpies and dissociation energies of these isomers are also in good agreements with the experimental values listed in Table I.

$C_3$. Our DFT results show that the PES from quasilinear CCC (Q) to linear CCC (L) is almost flat in the region with the bond angles close to 180°. From our PBE calculations, the change in formation enthalpy is only 0.0058 eV (from 8.6997 eV to 8.7055 eV) with the bond length from 1.3027 Å to 1.3018 Å and the bond angle from 149.255° to 180°. From our optB88-vdW calculations, the change in formation enthalpy is only −0.0002 eV (from 8.6579 eV to 8.6577 eV) with the bond length from 1.2949 Å to 1.2948 Å and the bond angle from 170.243° to 180°. This energy degenerate behavior from CCC (Q) to linear CCC (L) is consistent with previous AOBM calculations [78, 111]. These DFT bond lengths, formation enthalpies, and dissociation energies are in good agreements with previous experimental and AOBM data (see Table I). We also obtain a cyclic isomer CCC (C) with formation enthalpies 9.212 eV (PBE) and 9.491 eV (optB88-vdW) which are higher (i.e., the isomer is less stable) than the above values for CCC (Q) or CCC (L). Also note that the spin magnetic momentum $M = 2\ \mu_{\mathrm{B}}$ of CCC (C) is different from $M = 0\ \mu_{\mathrm{B}}$ for CCC (Q) or CCC (L). Note that the formation enthalpy and dissociation energies for the cyclic isomer are also in good agreements with the experimental values in Table I.

$H_1N_1O_1$. We obtain 2 quasilinear (HNO and HON) and 3 linear (HNO, HON, and NHO) isomers. From more to less stable, the order is HNO (Q) > HON (Q) > HNO (L) > HON (L) > NHO (L) with the formation enthalpies of about 1.0 eV, 2.0 eV, 2.4 eV, 3.2 eV, and 5.8 eV, respectively. The bond lengths, bond angles, formation enthalpies, and dissociation energies from our DFT calculations are in good agreements with available experimental and AOBM values (see Table I).

$H_1N_2$. The most stable isomer from our DFT calculations is quasilinear HNN (Q) with the formation enthalpies 2.016 eV (PBE) and 2.168 eV (optB88-vdW), cf. the experimental value 2.613 eV [33]. The values 0.253 eV (PBE) and 0.326 eV (optB88-vdW) for dissociation energy $D_{\mathrm{H+N_2}}$ of HNN(Q) are comparable to the AOBM values 0.438 eV to 1.184 eV [82] and 0.323 eV to 0.898 eV [84], cf. the relatively slightly negative value −0.374 eV from experiments [33]. Other PBE and optB88-vdW dissociation energies are also comparable with the experimental values in Table I. The bond lengths and bond angles from our DFT calculations are in good agreements with available AOBM values (see Table I), while the experimental data in the literature are unavailable for these geometric parameters. We also obtain the linear isomer HNN (L) with higher formation enthalpies 3.124 eV (PBE) and 3.285 eV (optB88-vdW), as well as the linear isomer NHN (L) with much higher formation enthalpies 8.784 eV (PBE) and 8.981 eV (optB88-vdW). In addition, our DFT values of $D_{\mathrm{HN+N}}$ for NHN (L) are 0.014 eV (PBE) and 0.015 eV (optB88-vdW), indicating a weak bond between N and HN with the bond length of about 2.6 Å (see Table I), which approaches the range of vdW interactions.

$H_1O_2$. The most stable isomer from our DFT calculations is triangular HOO (Q) with the slightly negative formation enthalpies −0.004 eV (PBE) and −0.005 eV (optB88-vdW), cf. the positive but relatively low experimental value 0.157 eV [33] and AOBM value 0.1535 eV [86]. All dissociation energies from PBE and optB88-vdW calculations for HOO(Q) are comparable to the experimental values [33] in Table I. The bond lengths and bond angles from our DFT calculations are in good agreements with available experimental values (also see Table I). For the linear isomer HOO (L), we obtain the significantly higher formation enthalpies 3.608 eV (PBE) and 3.605 eV (optB88-vdW), as well as $D_{\mathrm{HO+O}} = 0.002$ eV (PBE and optb88-vdW), indicating negligible bonding between O and HO, i.e., they are almost isolated, as indicated by the large O-O distance beyond 5.7 Å (see Table I). In addition, we also obtain the linear isomer OHO (L) with formation enthalpies 3.248 eV (PBE) and 3.240 eV (optB88-vdW).

$H_2N_1$. The most stable isomer from our DFT calculations is triangular HNH (Q) with the formation enthalpies 1.555 eV (PBE) and 1.739 eV (optB88-vdW), cf. the experimental value 1.958 eV [33]. All dissociation energies from PBE and optB88-vdW calculations for HNH(Q) are comparable to the experimental values [33] in Table I. The bond lengths and bond angles from our DFT calculations for HNH (Q) are in good agreements with available experimental and AOBM values (also see Table I). We also obtain the linear isomer HNH (L) with significantly higher formation enthalpies 2.893 eV (PBE) and 3.083 eV (optB88-vdW). For the linear isomer HHN (L), we obtain further significantly higher formation enthalpies 5.189 eV (PBE) and 5.259 eV (optB88-vdW), as well as $D_{\mathrm{H_2+N}} = 0.008$ eV (PBE) and 0.005 eV (optb88-vdW), indicating negligible bonding between N and HH, i.e., both N and HH (or $H_2$) are nearly isolated, as indicated by the large H-N distance beyond 2.8 Å (see Table I again).

$H_2O_1$. The most stable isomer from our DFT calculations is triangular HOH (Q) with the negative formation enthalpies −2.520 eV (PBE) and −2.433 eV (optB88-vdW) which match well the experimental value −2.476 eV [33]. The bond lengths and bond angles from our DFT calculations for HOH (Q) are in excellent agreements with available experimental and AOBM values (see Table I). We also obtain a triangular HOH, which is actually a complex with weak vdW interaction between O and HH, as indicated by $D_{\mathrm{H+HO}} \approx 0$ (see Triangular2 HOH in Table I). For the linear isomer OHH (L), we obtain positive formation enthalpies 2.833 eV (PBE) and 3.055 eV (optB88-vdW), as well as $D_{\mathrm{H+HO}} = 0.022$ eV (PBE) and 0.047 eV (optb88-vdW), again indicating weak bonding between H and HO. The formation enthalpies and dissociation energies from the DFT calculaitons for these isomers have good consistency with the experimental values in Table I. Additionally, we also obtain the linear isomer HOH (L) with higher but still negative formation enthalpies −1.266 eV (PBE) and −1.182 eV (optB88-vdW).

$H_3$. The PES of linear HHH are available from the early ABOM work [94, 95]. From the PES, there are no stable linear geometries for 3 bonded H atoms. To assess the consistency of our DFT results with these ABOM results, we first select 2 locally equilibrium linear HHH with short bond lengths around 1 Å (see linear1 and linear2 HHH in Table I) and always obtain the negative $D_{\mathrm{H+H_2}}$, which indeed indicates the instabilities of these two linear geometries. We also obtain a linear geometry with two H-H distances of about 0.75 Å and 4.72 Å (see linear3 HHH in Table I) and lowest formation enthalpies 2.269 eV (PBE) and 2.494 eV (optB88-vdW), as well as $D_{\mathrm{H+H_2}} = 0.000$ eV (PBE) and 0.001 eV (optb88-vdW), indicating no bonding between H and HH, i.e., a system of an isolated H atom plus an isolated $H_2$ molecule is more stable than any linearly fully bonded HHH. For the consistency of formation enthalpies and other dissociation energies of these isomers from the DFT calculations with available experimental data, see Table I. In early work [112, 113, 114], equilateral triangular states with the side lengths of about 0.85 Å for HHH are predicted. However, we cannot find such fully bonded triangular HHH from our DFT calculations. Instead, we find that three H atoms for any initially triangular HHH always become isolated after full relaxation (see, e.g., a triangular cyclic HHH in Table I), as indicated by the corresponding $D_{3\mathrm{H}} = 0.002$ eV (PBE) and 0.000 eV (optb88-vdW). This seems to imply that the fully bonded stable $H_3$ structures do not exist at all, thus consistent with our above DFT predictions. This instability is also consistent with the very negative experimental values of $D_{\mathrm{H+H_2}} = -5.491$ eV and $D_{3\mathrm{H}} = -1.013$ eV [33] for an equilateral triangular state.

$N_1O_2$. The most stable isomer from our DFT calculations is triangular ONO (Q) (i.e., conventionally $NO_2$) with the negative formation enthalpies $-0.195$ eV (PBE) and $-0.110$ eV (optB88-vdW), cf. the positive but quite low experimental value 0.382 eV [33]. For bond length $l_{\mathrm{NO}}$ and bond angle $\theta_{\mathrm{ONO}}$, our DFT values are in excellent agreements with available experimental and AOBM values (see Table I). The dissociation energies from the DFT calculations are comparable with the experimental values (see Table I). For the linear isomer ONO (L), we obtain much higher DFT formation enthalpies of about 11.2 eV, as well as $D_{\mathrm{N+2O}} = 0.016$ eV, indicating weak bonding between three atoms with large bond lengths beyond 3.2 Å. We also obtain the linear isomer NOO (L) with DFT formation enthalpies of about 4.6 eV. Again, the relatively high formation enthalpies imply high instabilities of the linear isomers. We also obtain a triangular NOO (Q) (see Table I) with a relatively small positive $D_{\mathrm{NO+O}} \approx 0.6$ eV, indicating weak bonding between O and NO, while the corresponding experimental value $-0.704$ eV [33] is slightly negative, indicating the instability of this isomer. We also tried to obtain a cyclic ONO, but no stable configuration was found. This is also consistent with the negative value of $D_{\mathrm{NO+O}} \approx -0.182$ eV (again indicating the instability of the isomer) from experiments [33].

$N_2O_1$. The most stable isomer from our DFT calculations is linear NNO (L) with the formation enthalpies 0.168 eV (PBE) and 0.308 eV (optB88-vdW), cf. the experimental value 0.892 eV [33]. The bond lengths from our DFT calculations are in excellent agreements with available experimental and AOBM values (see Table I). For the linear isomer NON (L), we obtain much higher DFT formation enthalpies of about 4.8 eV. We also obtain an isosceles triangular isomer NON (C) with DFT formation enthalpies of about 3.3 eV (see triangular cyclic NON in Table I). Both NON (L) and NON (C) have the relatively high formation enthalpies implying their instabilities relative to NNO (L). For the dissociation energies of these isomers, the DFT values also have good agreements with the experimental values and AOBM values, as listed in Table I.

$N_3$. The most stable isomer from our DFT calculations is the linearly symmetric NNN (L) with the formation enthalpies 3.612 eV (PBE) and 3.797 eV (optB88-vdW), cf. the experimental value 4.687 eV [33]. The bond lengths 1.1888 Å (PBE) and 1.1866 Å (optB88-vdW) from our DFT calculations are in excellent agreements with the experimental values 1.1815 Å [100] and 1.18115 Å [101], even better than the AOBM values 1.1570 Å and 1.1538 Å [99]. We also obtain three triangular isomers with slightly different geometries (see triangular1, triangular2, triangular3 for cyclic NNN in Table I) and formation enthalpies (about 5.5 eV). For the triangular2, the bond lengths and bond angle are consistent with the previous B3LYP values [102], as well as the formation enthalpies with experimental values (see again triangular2 for cyclic NNN in Table I). For the consistency of dissociation energies between the DFT and experimental values, see Table I.

$O_3$. The most stable isomer from our DFT calculations is quasilinear OOO (Q) with the formation enthalpies 1.287 eV (PBE) and 1.218 eV (optB88-vdW), cf. the experimental value 1.497 eV [33]. The bond lengths from our DFT calculations are in good agreements with available experimental and AOBM values (see Table I). The dissociation energy $D_{O+O_2}$ values of OOO (Q) are 1.737 eV (PBE) and 1.782 eV (optB88-vdW), cf., the experimental value 1.06 eV [33]. The dissociation energy $D_{3O}$ values of OOO (Q) are 7.784 eV (PBE) and 7.782 eV (optB88-vdW), cf., the experimental value 6.178 eV [33]. For the linear isomer OOO (L), we obtain higher DFT formation enthalpies of about 3.7 eV, and thus it is relatively unstable. We also try to calculate cyclic geometries and obtain a configuration with 3 almost isolated O atoms (see triangular cyclic OOO in Table I), for which the formation enthalpies are about 9 eV high and the $D_{3O}$ values are nearly zero, implying no stable bonded cyclic OOO isomers. This is also consistent with a negative value of $D_{O+O_2} = -0.268$ eV from experiments for a cyclic OOO isomer [33].

### C. $NO_3$ and $HNO_3$ molecules

As described in Sec. III B, the PES of a triatomic molecule can be much more complicated than the 1D curve for a diatomic molecule. A PES for a 4-atom (e.g., $NO_3$) or 5-atom (e.g., $HNO_3$) molecule will be even more complicated, and therefore obtaining the full PES with all spatial arrangements of atoms is impractical computationally. In this section, we choose the initial configurations based on previous experiments and perform our DFT structure optimizations for $NO_3$ and $HNO_3$ molecules.

#### *1*. $NO_3$

From experiments [115], the $NO_3$ has a planar structure of $D_{3h}$ symmetry (see Table II) in the ground electronic state. It is interesting to first note that there is a long dispute on the symmetry breaking of $NO_3$ from various quantum chemical investigations [116, 117, 118, 119] (also see more references cited in Ref. [124]), until Eisfeld *et al.* use the CASSCF approach (which is an AOBM) to recognize that the $C_{2v}$ geometry previously obtained from different AOBM calculations is likely artificially erroneous [119, 120]. Thus, it is very necessary and interesting to examine the geometric symmetry of a $NO_3$ molecule from the PAW DFT method.

As listed in Table II, the optimized geometry of $NO_3$ from our DFT calculations is of $D_{3h}$ symmetry and no stable $C_{2v}$ geometry is found, therefore consistent with experiments and the above CASSCF calculations from Eisfeld *et al*. The bond lengths 1.251 Å (PBE) and 1.252 Å (optB88-vdW) are in good agreements with the experimental value 1.240 Å [115], as well as previous ABOM values 1.2518 Å [119], 1.24 Å [121], and 1.246 Å [122].

The formation enthalpies −0.213 eV (PBE) and −0.136 eV (optB88-vdW) are negative, in contrast to the positive experimental value 0.823 eV [33]. The dissociation energy $D_{NO_2+O}$ values 3.042 eV (PBE) and 3.027 eV (optB88-vdW) are comparable to the experimental value 2.118 eV [33]. For a comparison of other dissociation energies, see Table II. For the error analysis and discussion relevant to these data, see Sec. IV C.

### *2.* $HNO_3$

By fully relaxing different initial geometries, e.g., based on three planar (2D) models and one 3D model assumed by Maxwell *et al.* [123], the most stable configuration for a $HNO_3$ molecule is a planar structure from our DFT calculations, e.g., the optimized geometry from our optB88-vdW calculations is shown in Table II. All bond lengths and bond angles from our PBE and optB88-vdW calculations are in excellent agreements with previous experimental data [123, 124] and AOBM values [125] (see Table II). The formation enthalpies −2.107 eV (PBE) and −2.005 eV (optB88-vdW) are negative, in good agreements with the experimental value −1.290 eV [33]. Again, for a comparison of other dissociation energies, see Table II; for the error analysis and discussion relevant to these data, see Sec. IV C.

**TABLE II.** Theoretical and experimental data for $NO_3$ and $HNO_3$ molecules. For the notation details, see the caption of Table I.

| Formula: $N_1O_3$ | | | $M$ | | | $\Delta H_f$ | $D_{NO_2+O}$ | $D_{NO+O_2}$ | $D_{3O+N}$ |
|---|---|---|---|---|---|---|---|---|---|
| **Planar $NO_3$** (nitrogen trioxide, etc.) | | | | $l_{NO_I} = l_{NO_{II}} = l_{NO_{III}}$ | $\theta_{O_INO_{II}} = \theta_{O_{II}NO_{III}} = \theta_{O_{III}NO_I}$ | | | | |
| 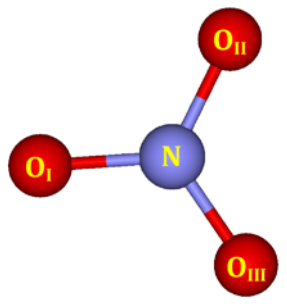 | PBE | | 1 | 1.251 | 120.0 | -0.213 | 3.042 | 1.159 | 14.481 |
| | optB88 | | 1 | 1.252 | 120.0 | -0.136 | 3.027 | 1.084 | 14.399 |
| | AOBM | [121] | | 1.24 | 120 | | | | |
| | AOBM | [122] | | 1.246 | 120 | | | | |
| | AOBM | [119] | | 1.2518 | 120 | | | | |
| | Exp. | [115] | | 1.240 | 120 | | | | |
| | Exp. | [33] | | | | 0.8229 | 2.1176 | 0.1165 | 11.7294 |
| | | | | | | 79.40±0.19 | 204.32±0.17 | 11.24±0.17 | 1131.71±0.18 |
| Formula: $H_1N_1O_3$ | | | $M$ | | | $\Delta H_f$ | $D_{H+NO_3}$ | $D_{HO+NO_2}$ | $D_{H+N+3O}$ |
| **Planar $HNO_3$** (nitric acid, etc.) | | | | | | | | | |
| 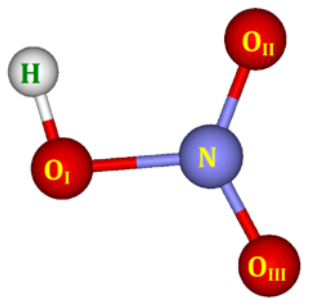 | PBE | | 0 | $l_{NO_I} = 1.44571$ | $\theta_{O_INO_{II}} = 115.4725$ | -2.107 | 4.163 | 2.498 | 18.644 |
| | | | | $l_{NO_{II}} = 1.22310$ | $\theta_{O_INO_{III}} = 113.5584$ | | | | |
| | | | | $l_{NO_{III}} = 1.20878$ | $\theta_{O_{II}NO_{III}} = 130.9692$ | | | | |
| | | | | $l_{HO_I} = 0.98328$ | $\theta_{HO_IN} = 101.9683$ | | | | |
| | optB88 | | 0 | $l_{NO_I} = 1.45489$ | $\theta_{O_INO_{II}} = 115.4328$ | -2.005 | 4.363 | 2.502 | 18.762 |
| | | | | $l_{NO_{II}} = 1.22329$ | $\theta_{O_INO_{III}} = 113.4929$ | | | | |
| | | | | $l_{NO_{III}} = 1.20875$ | $\theta_{O_{II}NO_{III}} = 131.0743$ | | | | |
| | | | | $l_{HO_I} = 0.98258$ | $\theta_{HO_IN} = 102.0232$ | | | | |
| | AOBM | [125] | | $l_{NO_I} = 1.39960$ | $\theta_{O_INO_{II}} = 115.7199$ | | | | |

| | | | | | | | |
|---|---|---|---|---|---|---|---|
| | | $l_{\mathrm{NO_{II}}} = 1.21032$ | $\theta_{\mathrm{O_I NO_{III}}} = 114.0088$ | | | | |
| | | $l_{\mathrm{NO_{III}}} = 1.19531$ | $\theta_{\mathrm{O_{II} NO_{III}}} = 130.2713$ | | | | |
| | | $l_{\mathrm{HO_I}} = 0.96985$ | $\theta_{\mathrm{HO_I N}} = 102.2040$ | | | | |
| Exp. | [123] | $l_{\mathrm{NO_I}} = 1.41 \pm 0.02$ | $\theta_{\mathrm{O_I NO_{II}}} = 115 \pm 2.5$ | | | | |
| | | $l_{\mathrm{NO_{II}}} = 1.22 \pm 0.02$ | $\theta_{\mathrm{O_I NO_{III}}} = 115 \pm 2.5$ | | | | |
| | | $l_{\mathrm{NO_{III}}} = 1.22 \pm 0.02$ | $\theta_{\mathrm{O_{II} NO_{III}}} = 130 \pm 5$ | | | | |
| Exp. | [124] | $l_{\mathrm{NO_I}} = 1.406$ | $\theta_{\mathrm{O_I NO_{II}}} = 115.88$ | | | | |
| | | $l_{\mathrm{NO_{II}}} = 1.211$ | $\theta_{\mathrm{O_I NO_{III}}} = 113.85$ | | | | |
| | | $l_{\mathrm{NO_{III}}} = 1.199$ | $\theta_{\mathrm{O_{II} NO_{III}}} = 130.27$ | | | | |
| | | $l_{\mathrm{HO_I}} = 0.964$ | $\theta_{\mathrm{HO_I N}} = 102.15$ | | | | |
| Exp. | [33] | | | -1.2900 | 4.3520 | 2.0587 | 16.0813 |
| | | | | -124.47±0.18 | 419.90±0.23 | 198.63±0.17 | 1551.61±0.17 |

## IV. Error analysis relative to experimental data

To assess the reliability of a DFT method by rating the quality or accuracy of the obtained DFT values, the corresponding error analysis is always needful. Before performing error analysis, it is necessary to make a review for the definitions of the errors often used for rating DFT values in the literature. For the $i$th value $Q_i$ in a dataset of DFT values from $i = 1$ to $n$ (where $n$ is often called the sample size), the absolute error (AE) is defined as

$$\Delta Q_i \equiv Q_i - Q_{0i}, \tag{1}$$

where the reference $Q_{0i}$ is often taken to be the corresponding experimental value. The quantity "$Q$" can be, e.g., bond length $l$, bond angle $\theta$, formation enthalpy $\Delta H_{\mathrm{f}}$, or dissociation energy $D_{\mathrm{p1+p2+p3+\cdots}}$. In this work, we always select the appropriate experimental data as the reference values $Q_{0i}$. Then, $\Delta Q_i > 0$ indicates overestimation of the experimental value and $\Delta Q_i < 0$ indicates underestimation. The mean absolute error (MAE) over the $n$ DFT values is defined as

$$\Delta Q_{\mathrm{MAE}} \equiv \frac{1}{n}\sum_{i=1}^{n} |\Delta Q_i|, \tag{2}$$

which describes the mean deviation in the magnitudes of DFT values. Note that the mean error (ME)

$$\Delta Q_{\mathrm{ME}} \equiv \frac{1}{n}\sum_{i=1}^{n} \Delta Q_i \tag{3}$$

is generally not very appropriate for assessing DFT values because of the cancellation in positive and negative $\Delta Q_i$ values, although ME values are often given in the literature.

The relative error (RE) or percentage error (PE) is defined as

$$\delta Q_i \equiv \frac{Q_i - Q_{0i}}{Q_{0i}} \times 100\%. \tag{4}$$

The mean absolute relative error (MARE) or mean absolute percentage error (MAPE) is defined as

$$\delta Q_{\mathrm{MAPE}} \equiv \frac{1}{n}\sum_{i=1}^{n} |\delta Q_i\,|, \tag{5}$$

and the mean relative error (MRE) or mean percentage error (MPE) is defined as

$$\delta Q_{\mathrm{MPE}} \equiv \frac{1}{n}\sum_{i=1}^{n} \delta Q_i. \tag{6}$$

Here, we need to mention that in the literature, the RE (or PE), MARE (or MAPE), and MRE (or MPE) are often given to assess DFT values, but sometimes using these errors is not very instructive (even misleading) especially for relatively small magnitudes, e.g., for the magnitudes ($\lesssim$ 0.2 eV) of atomization energies of inert gas crystals [126], which are very sensitive to their AEs, as will be mentioned below.

For the above dataset $\{Q_i\}$, we also give the corresponding standard deviation

$$s = \sqrt{\frac{1}{n}\sum_{i=1}^{n} |\Delta Q_i|^2} \text{ or } s = \sqrt{\frac{1}{n}\sum_{i=1}^{n} |\delta Q_i|^2}. \tag{7}$$

Note that the standard deviation $s$ reflects the discreteness of the dataset $\{|\Delta Q_i|\}$ or $\{|\delta Q_i|\}$. Larger (smaller) $s$ indicates that the dataset is more (less) discrete.

It must be mentioned that a quantity often has a significant uncertainty from different experimental measurements. In Tables I and II, the experimental data for formation enthalpies and dissociation energies from ATcT are listed with uncertainties corresponding to estimated 95% confidence limits [34, 35]. To describe the uncertainty for a dataset $\{Q_i\}$, one can also give a corresponding confidence limit by

$$u_{1-\alpha} = \pm t_{\alpha,n} s, \qquad (8)$$

where $s$ is obtained from Eq. (7) and $t_{\alpha,n}$ with the confidence level $1-\alpha$ is an $n$-dependent factor related to Student's $t$-distribution. For a 95% confidence interval, $\alpha$ in Eq. (8) is equal to $1 - 0.95 = 0.05$ and then $t_{\alpha,n}$ can be determined from the degree of freedom $n$. The 95% confidence limit described by Eq. (8) gives an estimate of what kind of accuracy could be expected when the same kind of quantity (e.g., the formation enthalpy) for an additional molecule or isomer (not contained in the dataset $\{Q_i\}$) is calculated. It should be noted that Eq. (8) has a different form from the well-known confidence intervals for evaluating the uncertainty of a new measurement after $n$ measurements of a quantity for a given molecule.

In the error analysis below, the selected reference experimental values based on Tables I and II for bond lengths, bond angles, formation enthalpies, and dissociation energies are provided in Tables S1, S2, S3, and S4 in the supplementary material, respectively. In the selection of experimental data for formation enthalpies or dissociation energies, we have considered the datasets used previously in assessment of PBE GGA [127, 128, 129, 130, 131, 132, 133, 134, 135]. Fortunately, the recently updated ATcT values [33] have been already available for most of formation enthalpies and dissociation energies of molecules or isomers in this work and therefore the ATcT values are prioritized, as listed in Tables I and II, as well as in Tables S3 and S4 with more significant figures.

## A. Bond lengths

By selecting 42 bond lengths based on the available experimental data in Tables I and II, the AEs ($\Delta l$) and PEs ($\delta l$) of the corresponding PBE and optB88-vdW values are plotted in Fig. 1. Among these bond lengths, the AE or |AE| maxima are $\Delta l_{\mathrm{max}} = |\Delta l|_{\mathrm{max}} = 0.072$ Å (PBE) and 0.067 Å (optB88-vdW) for $l_{\mathrm{CC}}$ of CC or $C_2$ dimer. The AE minima are $\Delta l_{\mathrm{min}} = -0.013$ Å (PBE) for $l_{\mathrm{CO}}$ of linear NCO and $-0.015$ Å (optB88-vdW) for $l_{\mathrm{NH}}$ of triangular HNO. The |AE| minima are $|\Delta l|_{\mathrm{min}} = 0.005$ Å (PBE) for $l_{\mathrm{CN}}$ of CN and 0.000 Å (optB88-vdW) for $l_{\mathrm{CC}}$ of linear CCC or $C_3$ trimer.

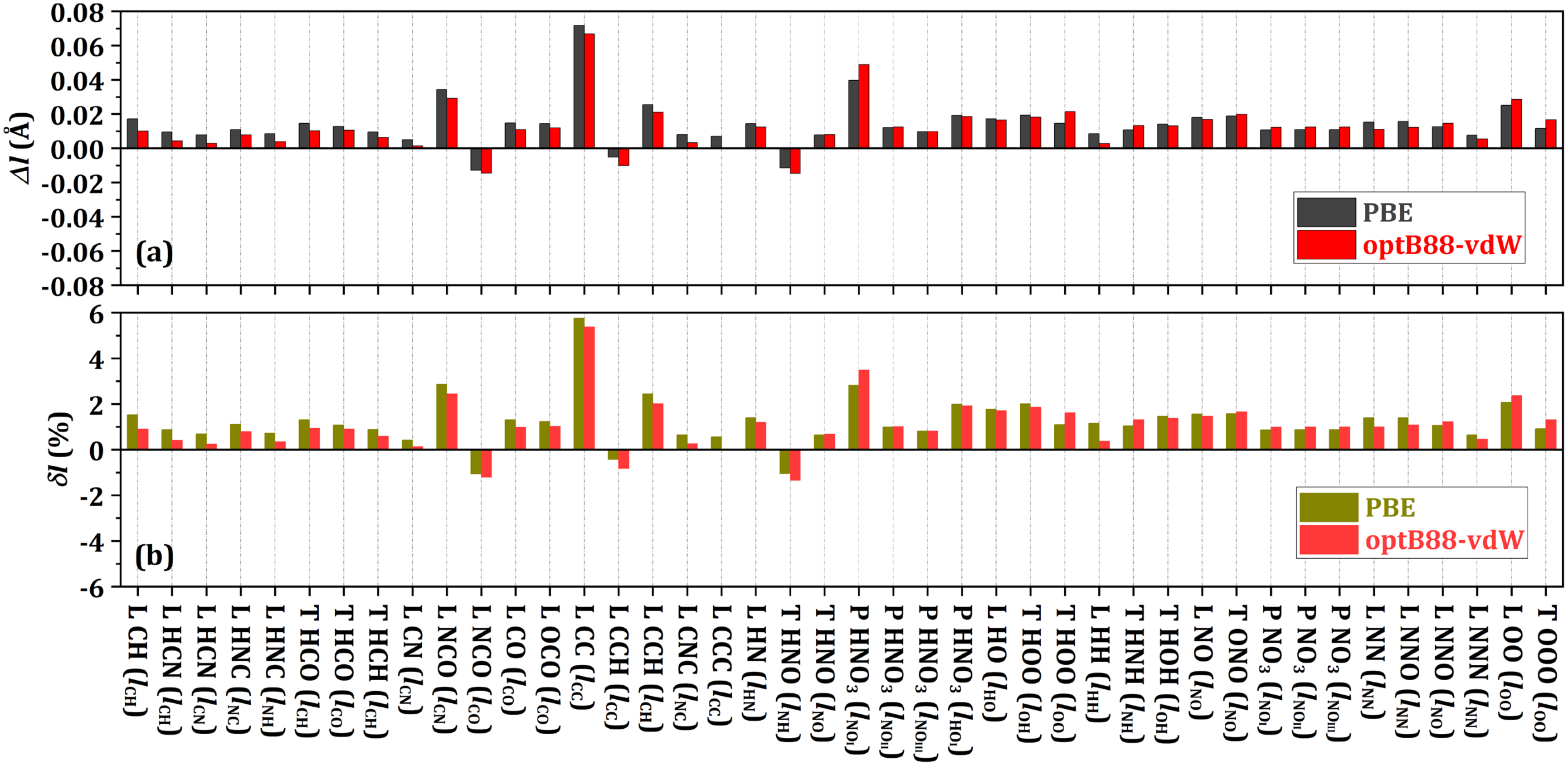

**FIG. 1.** (a) AEs and (b) PEs of 42 bond lengths. On the horizontal axis, "L" denotes "linear", "T" denotes "triangular", and "P" denotes "planar". For original data, see Table S1.

The PE or |PE| maxima are $\delta l_{\max} = |\delta l|_{\max} = 5.768\%$ (PBE) and 5.374% (optB88-vdW) for $l_{CC}$ of CC or $C_2$ dimer. The PE minima are $\delta l_{\min} = -1.061\%$ (PBE) for $l_{CO}$ of linear NCO and $-1.339\%$ (optB88-vdW) for $l_{NH}$ of triangular HNO. The |PE| minima are $|\delta l|_{\min} = 0.427\%$ (PBE) for $l_{CN}$ of CN and 0.005% (optB88-vdW) for $l_{CC}$ of linear CCC or $C_3$ trimer.

The MAEs (as well as the standard deviations and uncertainties) of these 42 bond lengths are $\Delta l_{MAE} = 0.015$ ($s = 0.019, u_{0.95} = \pm 0.038$) Å (PBE) and 0.014 ($s = 0.019, u_{0.95} = \pm 0.037$) Å (optB88-vdW). Correspondingly, the MAPEs (as well as $s$ and $u_{0.95}$ values) are $\delta l_{\text{MAPE}} = 1.345\%$ ($s = 1.616\%, u_{0.95} = \pm 3.262\%$) (PBE) and 1.228% ($s = 1.542\%, u_{0.95} = \pm 3.111\%$) (optB88-vdW). All above errors are also summarized in Table III.

For comparison, below we list the MAEs and MAPEs (as well as $s$ and $u_{0.95}$ values) for equilibrium lattice constants of solids or equilibrium bond lengths of molecules from similar or different PBE or optB88-vdW calculations in the literature. The $s$ and $u_{0.95}$ values are obtained by using the original data in these references.

$\Delta l_{\text{MAE}} = 0.060$ ($s = 0.069, u_{0.95} = \pm 0.145$) Å and $\delta l_{\text{MAPE}} = 1.297\%$ ($s = 1.458\%, u_{0.95} = \pm 3.063\%$) for equilibrium lattice constants of 18 tested solids from the planewave PBE calculations of Wu *et al.* [130].

$\Delta l_{\text{MAE}} = 0.071$ ($s = 0.084, u_{0.95} = \pm 0.172$)Å and $\delta l_{\text{MAPE}} = 1.411\%$ ($s = 1.591\%, u_{0.95} = \pm 3.248\%$) for equilibrium lattice constants of 30 tested solids from the PAW PBE calculations of Schimka *et al.* [131].

$\Delta l_{\text{MAE}} = 0.067$ ($s = 0.079, u_{0.95} = \pm 0.163$) Å and $\delta l_{\text{MAPE}} = 1.403\%$ ($s = 1.614\%, u_{0.95} = \pm 3.338\%$) for equilibrium lattice constants of 23 tested solids from the PAW PBE calculations of Klimeš *et al.* [132].

$\Delta l_{\text{MAE}} = 0.066$ ($s = 0.074, u_{0.95} = \pm 0.154$) Å and $\delta l_{\text{MAPE}} = 1.361\%$ ($s = 1.460\%, u_{0.95} = \pm 3.020\%$) for equilibrium lattice constants of 23 tested solids from the PAW optB88-vdW calculations of Klimeš *et al.* [132].

$\Delta l_{\text{MAE}} = 0.037$ ($s = 0.059, u_{0.95} = \pm 0.122$) Å and $\delta l_{\text{MAPE}} = 1.367\%$ ($s = 2.278\%, u_{0.95} = \pm 4.674\%$)% for shortest interatomic distances of 27 transition metals from the PAW PBE calculations of Janthon *et al.* [133].

$\Delta l_{\text{MAE}} = 0.061$ ($s = 0.075, u_{0.95} = \pm 0.151$) Å and $\delta l_{\text{MAPE}} = 1.246\%$ ($s = 1.448\%, u_{0.95} = \pm 2.919\%$) for equilibrium lattice constants of 44 tested solids from the PBE calculations of Tran *et al.* by using the augmented planewave plus local orbitals (APW+lo) method [16].

$\Delta l_{\text{MAE}} = 0.062$ ($s = 0.075, u_{0.95} = \pm 0.151$) Å and $\delta l_{\text{MAPE}} = 1.252\%$ ($s = 1.418\%, u_{0.95} = \pm 2.857\%$) for equilibrium lattice constants of 44 tested solids from the APW+lo optB88-vdW calculations of Tran *et al.* [16].

From the above list, the PAW PBE or optB88-vdW results of 42 bond lengths from our DFT calculations are highly satisfactory with a smaller MAE (as well as smaller $s$ and $u_{0.95}$ values) and a similar MAPE relative to the values for solids from various PBE or optB88-vdW calculations in the literature.

In addition, based on the data for equilibrium bond lengths of 16 dimers from the PAW PBE calculations of Paier *et al.* [134], $\Delta l_{\text{MAE}} = 0.013$ ($s = 0.018, u_{0.95} = \pm 0.038$) Å and $\delta l_{\text{MAPE}} = 0.832\%$ ($s = 0.992\%, u_{0.95} = \pm 2.103\%$), which are very close to the MAE and MAPE values (as well as the corresponding $s$ and $u_{0.95}$ values) for the above 42 bond lengths from our PAW PBE or optB88-vdW calculations. Based on the data for equilibrium bond lengths of 19 covalent molecules (22 molecular isomers) from the APW+lo PBE calculations of Tran *et al.* [135], $\Delta l_{\text{MAE}} = 0.009$ ($s = 0.010, u_{0.95} = \pm 0.021$) Å and $\delta l_{\text{MAPE}} =$ 0.716% ($s = 0.801\%, u_{0.95} = \pm 1.661\%$), which are better than but close to the MAE and MAPE values (as well as the corresponding $s$ and $u_{0.95}$ values) for the above 42 bond lengths from our PAW PBE or optB88-vdW calculations.

**TABLE III.** The MAEs ($\Delta Q_{\mathrm{MAE}}$) and MAPEs ($\delta Q_{\mathrm{MAPE}}$) with standard errors ($s$) and $u_{0.95}$ values of 42 bond lengths $Q = l$, 15 bond angles $Q = \theta$, 49 formation energies $Q = \Delta H_{\mathrm{f}}$ or $\Delta H_{\mathrm{f}}^{*}$, and 138 dissociation energies $Q = D$ or $D^{*}$ from our PAW PBE and optB88-vdW calculations, corresponding to Tables S1, S2, S3, and S4, respectively. For a given quantity $Q$, the minima ($\Delta Q_{\min}$ and $\delta Q_{\min}$) and maxima ($\Delta Q_{\max}$ and $\delta Q_{\max}$) of AEs and PEs as well as the minima ($|\Delta Q|_{\min}$ and $|\delta Q|_{\min}$) and maxima ($|\Delta Q|_{\max}$ and $|\delta Q|_{\max}$) of absolute values of AEs and PEs are also listed.

| | | PBE | optB88 |
|---|---|---|---|
| | $\Delta l_{\mathrm{MAE}}$ | 0.015 Å | 0.014 Å |
| **42 bond lengths** | | $s$ = 0.019 Å | $s$ = 0.019 Å |
| | | $u_{0.95} = \pm 0.038$ Å | $u_{0.95} = \pm 0.037$ Å |
| | $\|\Delta l\|_{\min}$, $\|\Delta l\|_{\max}$ (Å) | 0.005, 0.072 | 0.000, 0.067 |
| | $\Delta l_{\min}$, $\Delta l_{\max}$ (Å) | -0.013, 0.072 | -0.015, 0.067 |
| | $\delta l_{\mathrm{MAPE}}$ | 1.345% | 1.228% |
| | | $s$ = 1.616% | $s$ = 1.542% |
| | | $u_{0.95} = \pm 3.262\%$ | $u_{0.95} = \pm 3.111\%$ |
| | $\|\delta l\|_{\min}$, $\|\delta l\|_{\max}$ (%) | 0.427, 5.768 | 0.005, 5.374 |
| | $\delta l_{\min}$, $\delta l_{\max}$ (%) | -1.061, 5.768 | -1.339, 5.374 |
| | $\Delta\theta_{\mathrm{MAE}}$ | 0.446° | 0.517° |
| **15 bond angles** | | $s$ = 0.601° | $s$ = 0.639° |
| | | $u_{0.95} = \pm 1.282°$ | $u_{0.95} = \pm 1.362°$ |
| | $\|\Delta\theta\|_{\min}$, $\|\Delta\theta\|_{\max}$ (°) | 0.004, 1.437 | 0.002, 1.236 |
| | $\Delta\theta_{\min}$, $\Delta\theta_{\max}$ (°) | -0.518, 1.437 | -0.937, 1.236 |
| | $\delta\theta_{\mathrm{MAPE}}$ | 0.379% | 0.440% |
| | | $s$ = 0.506% | $s$ = 0.540% |
| | | $u_{0.95} = \pm 1.078\%$ | $u_{0.95} = \pm 1.151\%$ |
| | $\|\delta\theta\|_{\min}$, $\|\delta\theta\|_{\max}$ (%) | 0.003, 1.231 | 0.001, 1.058 |
| | $\delta\theta_{\min}$, $\delta\theta_{\max}$ (%) | -0.490, 1.231 | -0.906, 1.058 |
| | $\Delta(\Delta H_{\mathrm{f}})_{\mathrm{MAE}}$ | 0.325 eV = 31.3 kJ/mol | 0.287 eV = 27.7 kJ/mol |
| **49 formation enthalpies** | | $s$ = 0.431 eV = 41.6 kJ/mol | $s$ = 0.376 eV = 36.3 kJ/mol |
| | | $u_{0.95} = \pm 0.866$ eV = $\pm 83.5$ kJ/mol | $u_{0.95} = \pm 0.755$ eV = $\pm 72.9$ kJ/mol |
| | $\|\Delta(\Delta H_{\mathrm{f}})\|_{\min}$, $\|\Delta(\Delta H_{\mathrm{f}})\|_{\max}$ (eV) | 0.007, 1.075 | 0.008, 0.959 |
| | $\Delta(\Delta H_{\mathrm{f}})_{\min}$, $\Delta(\Delta H_{\mathrm{f}})_{\max}$ (eV) | -1.075, 0.558 | -0.959, 0.489 |
| | $\Delta(\Delta H_{\mathrm{f}}^{*})_{\mathrm{MAE}}$ | 0.109 eV/atom | 0.097 eV/atom |
| | | $s$ = 0.142 eV/atom | $s$ = 0.124 eV/atom |
| | | $u_{0.95} = \pm 0.285$ eV/atom | $u_{0.95} = \pm 0.249$ eV/atom |
| | $\|\Delta(\Delta H_{\mathrm{f}}^{*})\|_{\min}$, $\|\Delta(\Delta H_{\mathrm{f}}^{*})\|_{\max}$ (eV/atom) | 0.003, 0.358 | 0.003, 0.297 |
| | $\Delta(\Delta H_{\mathrm{f}}^{*})_{\min}$, $\Delta(\Delta H_{\mathrm{f}}^{*})_{\max}$ (eV/atom) | -0.358, 0.279 | -0.297, 0.232 |
| | $\delta(\Delta H_\mathrm{f})_\mathrm{MAPE}$ | 19.741% | 18.001% |
| | | $s =$ 37.621% | $s =$ 34.138% |
| | | $u_{0.95} = \pm$75.603% | $u_{0.95} = \pm$68.604% |
| | $\lvert\delta(\Delta H_\mathrm{f})\rvert_\mathrm{min}$, $\lvert\delta(\Delta H_\mathrm{f})\rvert_\mathrm{max}$ (%) | 0.252, 150.986 | 0.387, 128.677 |
| | $\delta(\Delta H_\mathrm{f})_\mathrm{min}$, $\delta(\Delta H_\mathrm{f})_\mathrm{max}$ (%) | -150.986, 63.313 | -128.677, 57.163 |
| | $\Delta D_\mathrm{MAE}$ | 0.876 eV = 84.5 kJ/mol | 0.860 eV = 83.0 kJ/mol |
| **138 dissociation energies** | | $s$ = 1.017 eV = 98.1 kJ/mol | $s$ = 0.986 eV = 95.1 kJ/mol |
| | | $u_{0.95} = \pm$2.010 eV = $\pm$194.0 kJ/mol | $u_{0.95} = \pm$1.949 eV = $\pm$188.1 kJ/mol |
| | $\lvert\Delta D\rvert_\mathrm{min}$, $\lvert\Delta D\rvert_\mathrm{max}$ (eV) | 0.031, 2.752 | 0.011, 2.680 |
| | $\Delta D_\mathrm{min}$, $\Delta D_\mathrm{max}$ (eV) | -0.199, 2.752 | -0.048, 2.680 |
| | $\Delta D^*_\mathrm{MAE}$ | 0.292 eV/atom | 0.288 eV/atom |
| | | $s =$ 0.334 eV/atom | $s =$ 0.324 eV/atom |
| | | $u_{0.95} = \pm$0.660 eV/atom | $u_{0.95} = \pm$0.641 eV/atom |
| | $\lvert\Delta D^*\rvert_\mathrm{min}$, $\lvert\Delta D^*\rvert_\mathrm{max}$ (eV/atom) | 0.010, 0.688 | 0.002, 0.697 |
| | $\Delta D^*_\mathrm{min}$, $\Delta D^*_\mathrm{max}$ (eV/atom) | -0.066, 0.688 | -0.016, 0.697 |
| | $\delta D_\mathrm{MAPE}$ | 73.674% | 52.357% |
| | | $s =$ 278.412% | $s =$ 141.844% |
| | | $u_{0.95} = \pm$550.505% | $u_{0.95} = \pm$280.468% |
| | $\lvert\delta D\rvert_\mathrm{min}$, $\lvert\delta D\rvert_\mathrm{max}$ (%) | 1.309, 2730.320 | 0.247, 830.539 |
| | $\delta D_\mathrm{min}$, $\delta D_\mathrm{max}$ (%) | -2730.320, 859.014 | -524.383, 830.539 |

## B. Bond angles

We select 15 bond angles based on the available experimental data in Tables I and II. The AEs ($\Delta\theta$) and PEs ($\delta\theta$) of the corresponding PBE and optB88-vdW values are plotted in Fig. 2. Among these bond angles, the AE or |AE| maxima are $\Delta\theta_{\mathrm{max}} = |\Delta\theta|_{\mathrm{max}} = 1.437°$ (PBE) and 1.236° (optB88-vdW) for $\theta_{\mathrm{OOO}}$ of triangular OOO or $O_3$ trimer. The AE minima are $\Delta\theta_{\mathrm{min}} = -0.518°$ (PBE) for $\theta_{\mathrm{HCO}}$ of triangular HCO and $-0.937°$ (optB88-vdW) for $\theta_{\mathrm{HNH}}$ of triangular HNH. The |AE| minima are $|\Delta\theta|_{\mathrm{min}} = 0.004°$ (PBE) for $\theta_{\mathrm{O_I NO_{III}}}$ of planar $NO_3$ and 0.002° (optB88-vdW) for $\theta_{\mathrm{O_I NO_{II}}}$ of planar $NO_3$.

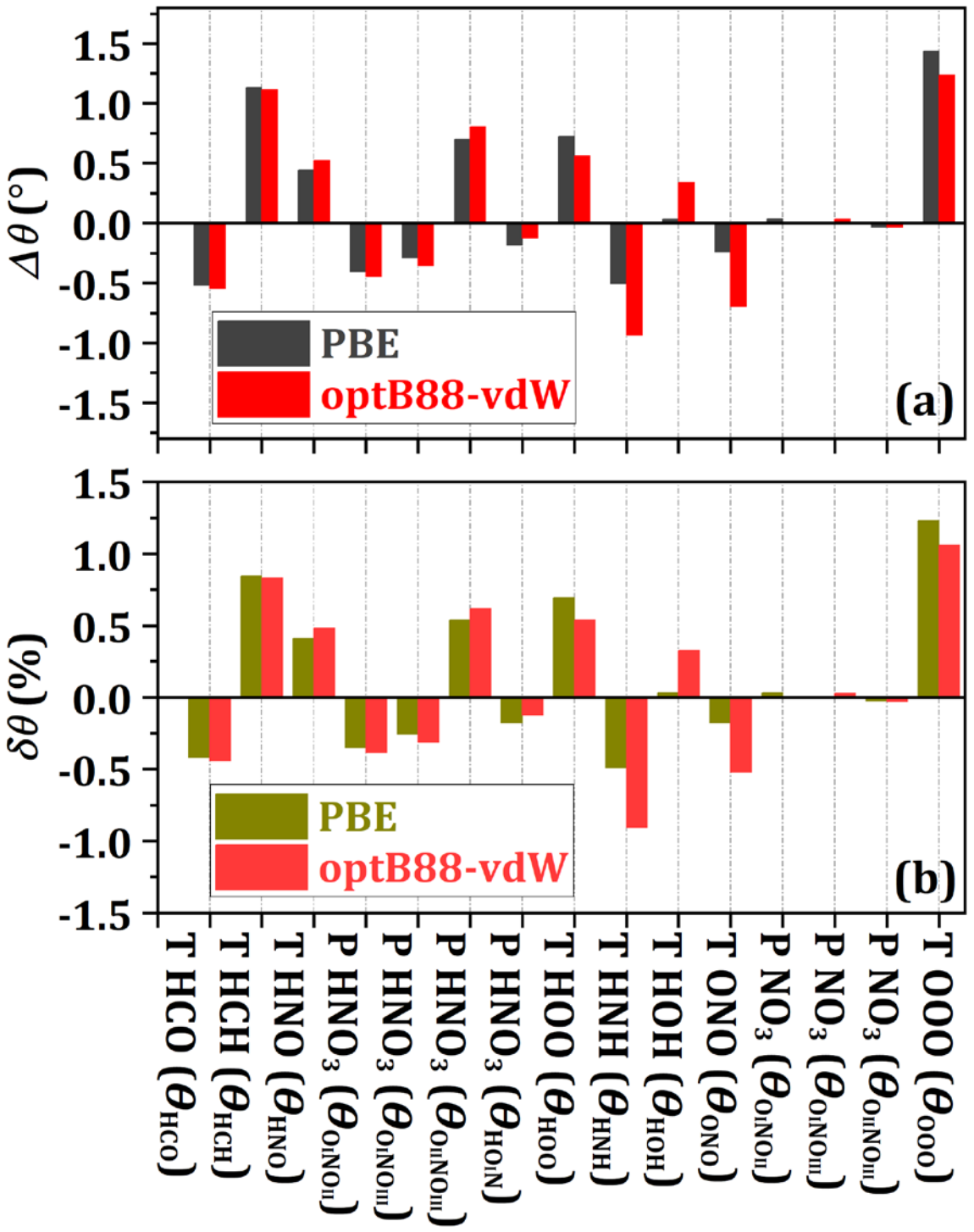

**FIG. 2.** (a) AEs and (b) PEs of 15 bond angles. On the horizontal axis, "T" denotes "triangular" and "P" denotes "planar". For original data, see Table S2.

The PE or |PE| maxima are $\delta\theta_{\max} = |\delta\theta|_{\max} = 1.231\%$ (PBE) and 1.058% (optB88-vdW) for $\theta_{\rm OOO}$ of triangular OOO or $O_3$ trimer. The PE minima are $\delta\theta_{\min} = -0.490\%$ (PBE) and $-0.906\%$ (optB88-vdW) for $\theta_{\rm HNH}$ of triangular HNH. The |PE| minima are $|\delta\theta|_{\min} = 0.003\%$ (PBE) for $\theta_{\rm O_INO_{III}}$ of planar $NO_3$ and 0.001% (optB88-vdW) for $\theta_{\rm O_INO_{II}}$ of planar $NO_3$.

The MAEs (as well as $s$ and $u_{0.95}$ values) of these 15 bond angles are $\Delta\theta_{\rm MAE} = 0.446°$ ($s = 0.601°, u_{0.95} = \pm1.281°$) (PBE) and 0.517° ($s = 0.639°, u_{0.95} = \pm1.362°$) (optB88-vdW). Correspondingly, the MAPEs (as well as $s$ and $u_{0.95}$ values) are $\delta\theta_{\rm MAPE} = 0.379\%$ ($s = 0.506\%, u_{0.95} = \pm1.078\%$) (PBE) and 0.440% ($s = 0.540\%, u_{0.95} = \pm1.151\%$) (optB88-vdW). All these errors for bond angles are also listed in Table III.

### C. Formation enthalpies and dissociation energies

Based on the available experimental data in Tables I and II, we select 49 formation enthalpies ($\Delta H_{\rm f}$ or $\Delta H_{\rm f}^*$) and 138 dissociation energies ($D$ or $D^*$). For definitions of the total formation enthalpy $\Delta H_{\rm f}$ and the per-atom formation enthalpy $\Delta H_{\rm f}^*$, as well as the total dissociation energy $D$ and the per-atom dissociation energy $D^*$, see Sec. S1. The AEs [$\Delta(\Delta H_{\rm f}^*)$ and $\Delta(\Delta H_{\rm f})$] and PEs [$\delta(\Delta H_{\rm f}) = \delta(\Delta H_{\rm f}^*)$] of the corresponding PBE and optB88-vdW values are plotted in Fig. 3. The AEs ($\Delta D^*$ and $\Delta D$) and PEs ($\delta D = \delta D^*$) of the corresponding PBE and optB88-vdW values are plotted in Fig. 4.

For AE or MAE analysis, using $\Delta H_{\rm f}^*$ and $D^*$ is more reasonable than directly using $\Delta H_{\rm f}$ and $D$ because the number of atoms in a molecule can vary (for this work, the number can be 2, 3, 4, or 5), especially for an error comparison between DFT datasets with different samples, e.g., between molecules with varying number of atoms and crystals with different number of atoms in their primitive cells. The per-atom cohesive energies or atomization energies have been widely used for error analysis in the DFT functional tests [16, 127, 129, 130, 132], as analyzed below. Note that the per-atom formation enthalpies are also used in the previous literature, e.g., for comparison between experimental and DFT reaction energies of compounds [136]. Thus, we mainly focus on $\Delta(\Delta H_{\rm f}^*)$ or $\Delta D^*$ [not $\Delta(\Delta H_{\rm f})$ or $\Delta D$] for AE or MAE analysis in this work, although all errors are included in Table III. In addition, using $\Delta H_{\rm f}^*$ and $\Delta H_{\rm f}$ or using $D^*$ and $D$ is no distinction for PE or MAPE analysis, according to their definitions in Sec. S1 and Eqs. (1) to (6).

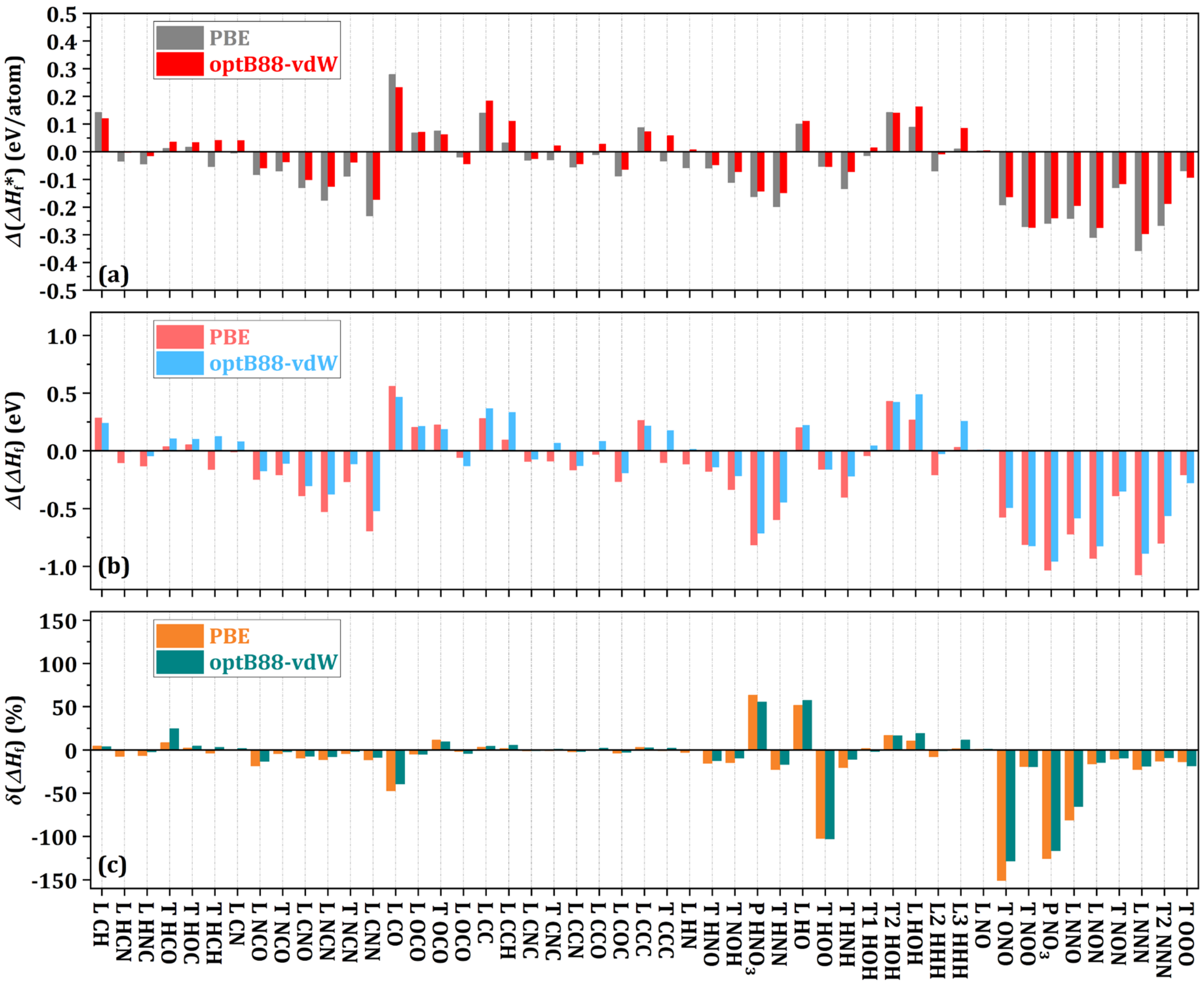

**FIG. 3.** (a, b) AEs and (c) PEs of 49 formation enthalpies. On the horizontal axis, "L" denotes "linear", "T" denotes "triangular", and "P" denotes "planar". For original data, see Table S3.

Among the 49 formation enthalpies, the AE maxima are $\Delta(\Delta H_f^*)_{max}$ = 0.279 eV/atom (PBE) and 0.232 eV/atom (optB88-vdW) for linear CO. The |AE| maxima are$|\Delta(\Delta H_f^*)|_{max}$ = 0.358 eV/atom (PBE) and 0.297 eV/atom (optB88-vdW) for linear

NNN. The AE minima are $\Delta(\Delta H_{\mathrm{f}}^{*})_{\min} = -0.358$ eV/atom (PBE) and $-0.297$ eV (optB88-vdW) for linear NNN. The |AE| minima are $|\Delta(\Delta H_{\mathrm{f}}^{*})|_{\min} = 0.0034$ eV/atom (PBE) for NO and 0.0027 eV/atom (optB88-vdW) for linear HCN. The MAEs (as well as $s$ and $u_{0.95}$ values) of these 49 formation enthalpies are $\Delta(\Delta H_{\mathrm{f}}^{*})_{\mathrm{MAE}} = 0.109$ ($s = 0.142, u_{0.95} = \pm 0.285$) eV/atom (PBE) and 0.097 ($s = 0.124, u_{0.95} = \pm 0.249$) eV/atom (optB88-vdW). The quite small MAE and $u_{0.95}$ values indicate that both PBE and optB88-vdW values for these formation enthalpies are overall satisfactory. In addition, the optB88-vdW result has the smaller MAE and $u_{0.95}$ values than the PBE result and therefore is even better than the PBE result. All above errors are plotted in Fig. 3 and listed in Table III for summary.

As already mentioned above, PE and MAPE can be very sensitive to small AEs when the analyzed data have relatively small magnitudes. For example, the experimental value of $\Delta H_{\mathrm{f}}^{*}$ for triangular HOO is 0.0523 eV/atom and the PAW PBE value is $-0.0012$ eV/atom. Then, the AE is $-0.0535$ eV/atom with a quite small magnitude of 0.0535 eV/atom, but the corresponding PE is $-102.44\%$ with a huge magnitude > 100%. This implies that PE or MAPE is not very proper for rating the DFT values in the case for which the rated data have relatively small magnitudes and/or the corresponding experimental data perhaps have significant uncertainties, as some of those in the list of the 49 formation enthalpies. Thus, we do not use PE and MAPE to rate our DFT values for formation enthalpies, although we plot them in Fig. 3 and list them in Table III for information only.

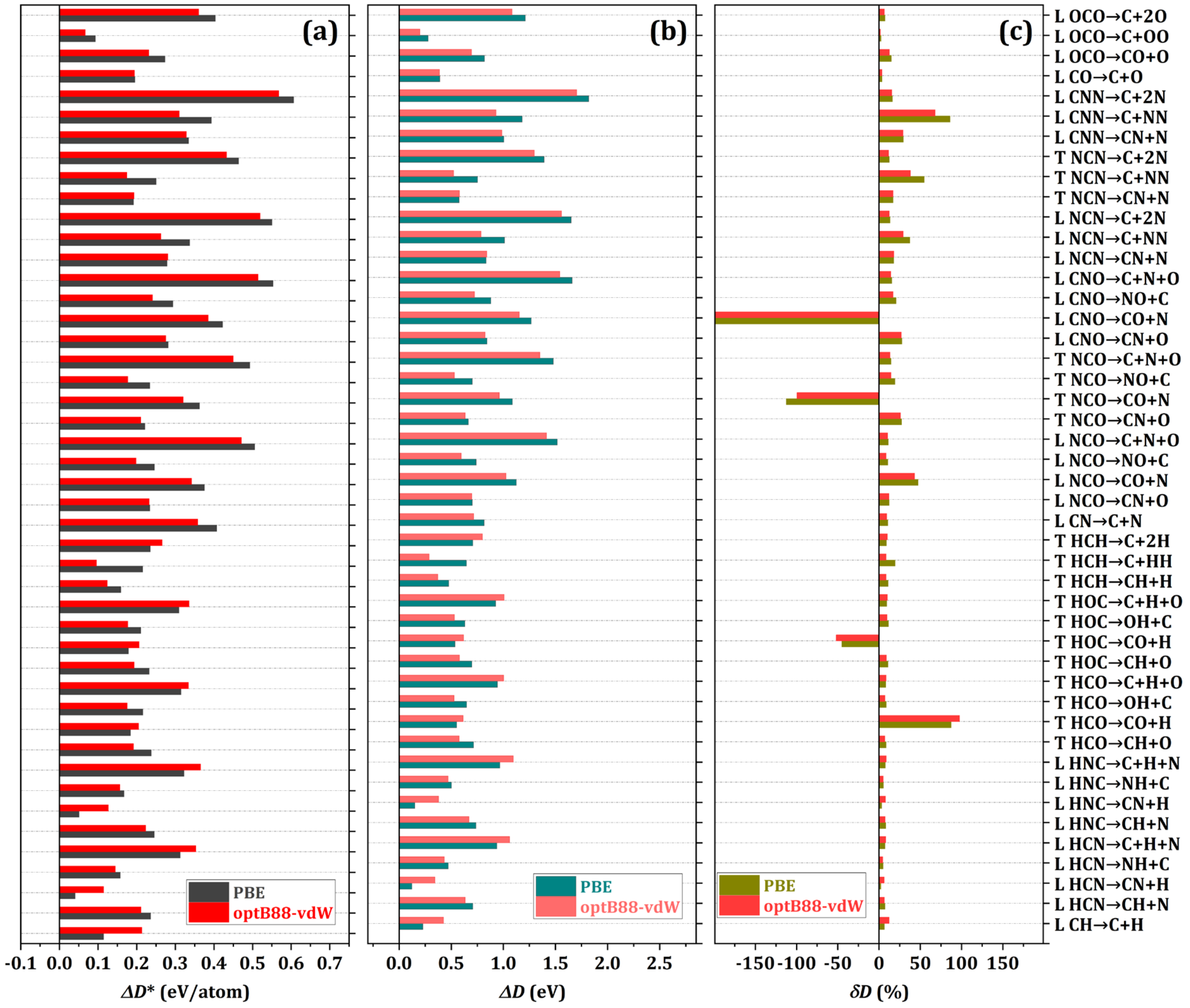

**FIG. 4.** (a, b) AEs and (c) PEs of 138 dissociation energies (including 51 atomization energies and all bond dissociation energies). On the horizontal axis, "L" denotes "linear", "T" denotes "triangular", and "P" denotes "planar". For original data, see Table S4. Part 1/3

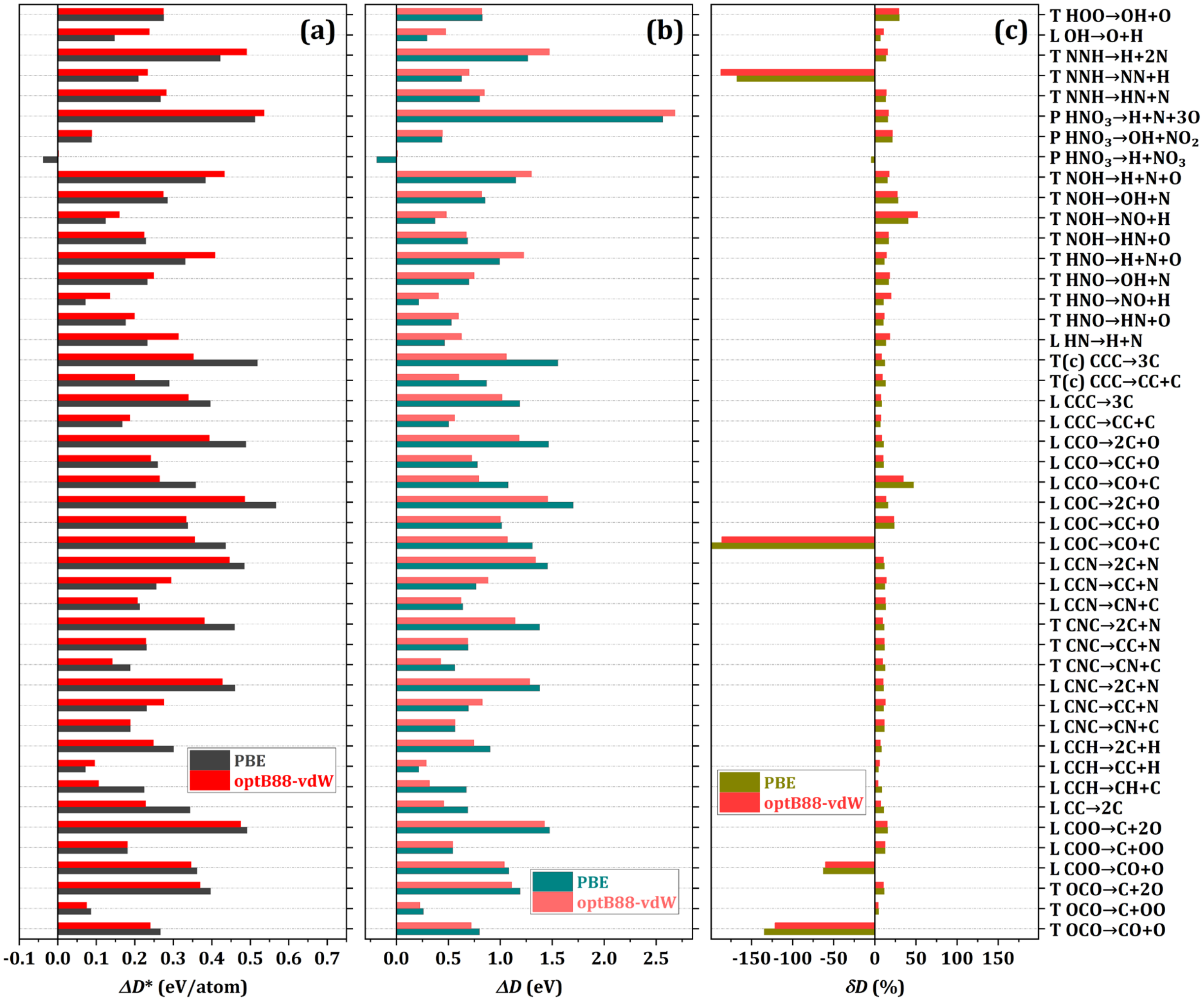

**FIG. 4.** *Continued*. Part 2/3.

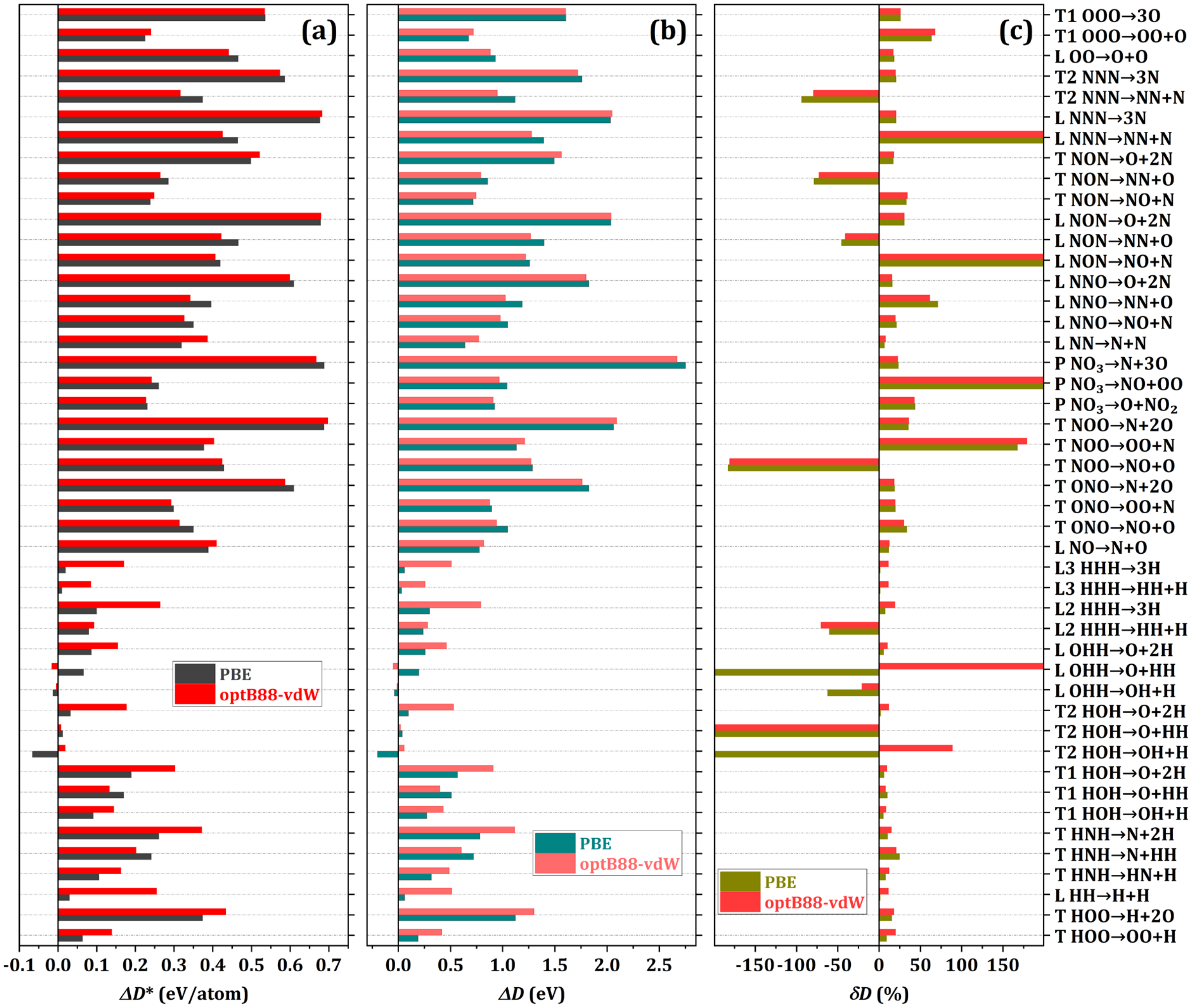

**FIG. 4.** *Continued.* Part 3/3

Among the 138 dissociation energies, the AE or |AE| maxima are $\Delta D^*_{\max} = |\Delta D^*|_{\max} = 0.688$ eV/atom (PBE) for the atomization energy of planar $NO_3$ and 0.697 eV/atom (optB88-vdW) for the atomization energy of triangular NOO. The AE minima are $\Delta D^*_{\min} = -0.066$ eV/atom (PBE) for dissociating O and HH from the triangular vdW complex HOH and $-0.016$ eV (optB88-vdW) for dissociating O and HH from the linear vdW complex OHH. The |AE| minima are $|\Delta D^*|_{\min} = 0.010$ eV/atom (PBE) for dissociating H and HH from the linear vdW complex HHH and 0.002 eV/atom (optB88-vdW) for dissociating H and $NO_3$ from the planar $HNO_3$. The MAEs (as well as $s$ and $u_{0.95}$ values) of these 138 dissociation energies are $\Delta D^*_{\mathrm{MAE}} = 0.292$ ($s = 0.334, u_{0.95} = \pm 0.660$) eV/atom (PBE) and 0.288 ($s = 0.324, u_{0.95} = \pm 0.641$) eV/atom (optB88-vdW). The quite small MAE and $u_{0.95}$ values indicate that both PBE and optB88-vdW values for these dissociation energies are overall satisfactory. Like the above description for formation enthalpies, some of 138 bond dissociation energies have relatively small magnitudes and/or the corresponding experimental data perhaps have significant uncertainties. Thus, we also do not use PE and MAPE to rate our DFT values for bond dissociation energies, although we list them in Table III for information only. Here, we also note that, among the $\Delta D^*$ values for 138 isomers plotted in Fig. 4, there are only 3 and 2 isomers with negative $\Delta D^*$ values and these negative $\Delta D^*$ values have relatively small $|\Delta D^*|$ values from PBE and optB88-vdW calculations, respectively. This behavior contrasts $\Delta H^*_{\mathrm{f}}$, for which the ratios of numbers with negative and positive values are much lower than those for the $\Delta D^*$ values.

In the literature, the error analyses for atomization energies or cohesive energies by testing various types of solids or molecules are available during developing DFT methods. Although these atomization energies or cohesive energies for solids or molecules do not exactly correspond to the formation enthalpies and dissociation energies listed in Table III, making a comparison is still informative. Below, we may as well list the MAEs and MAPEs for atomization energies or cohesive energies of various types of solids or molecules from similar or different PBE or optB88-vdW calculations in the literature. These MAEs ($\Delta E^*_{\mathrm{MAE}}$ and $\Delta E_{\mathrm{MAE}}$) and MAPEs ($\delta E_{\mathrm{MAE}}$) with corresponding $s$ and $u_{0.95}$ values are regenerated by using the original data in the literature. Note that all $\Delta E^*_{\mathrm{MAE}}$ values below always correspond to the atomization or cohesive energies *per atom*. For information, we also list $\Delta E_{\mathrm{MAE}}$ corresponding to the total atomization energies *per chemical formula* or *per primitive cell*.

$\Delta E^*_{\mathrm{MAE}} = 0.143$ ($s = 0.198, u_{0.95} = \pm 0.396$) eV/atom [$\Delta E_{\mathrm{MAE}} = 0.372$ ($s = 0.478, u_{0.95} = \pm 0.957$) eV] and $\delta E_{\mathrm{MAPE}} = 6.569\%$ ($s = 10.327\%, u_{0.95} = \pm 20.695\%$) for atomization energies of 55 tested molecules from the PAW PBE calculations of Paier *et al.* [134].

$\Delta E^*_{\mathrm{MAE}} = 0.18$ ($s = 0.25, u_{0.95} = \pm 0.53$) eV/atom [$\Delta E_{\mathrm{MAE}} = 0.22$ ($s = 0.29, u_{0.95} = \pm 0.60$) eV] and $\delta E_{\mathrm{MAPE}} = 5.09\%$ ($s = 7.20\%, u_{0.95} = \pm 15.13\%$) for cohesive energies of 18 tested solids from the PAW PBE calculations of Wu *et al.* [130].

$\Delta E^*_{\mathrm{MAE}} = 0.153$ ($s = 0.203, u_{0.95} = \pm 0.424$) eV/atom [$\Delta E_{\mathrm{MAE}} = 0.351$ ($s = 0.436, u_{0.95} = \pm 0.912$) eV] and $\delta E_{\mathrm{MAPE}} = 7.166\%$ ($s = 11.172\%, u_{0.95} = \pm 23.382\%$) for atomization energies of 19 tested molecules from the APW+lo PBE calculations of Tran *et al.* [135].

$\Delta E^*_{\mathrm{MAE}} = 0.141$ ($s = 0.194, u_{0.95} = \pm 0.389$) eV/atom [$\Delta E_{\mathrm{MAE}} = 0.367$ ($s = 0.469, u_{0.95} = \pm 0.940$) eV] and $\delta E_{\mathrm{MAPE}} = 6.476\%$ ($s = 10.110\%, u_{0.95} = \pm 20.253\%$) for atomization energies of 56 tested molecules from the PAW PBE calculations of Schimka *et al.* [131].

$\Delta E^*_{\mathrm{MAE}} = 0.13$ ($s = 0.17, u_{0.95} = \pm 0.34$) eV/atom [$\Delta E_{\mathrm{MAE}} = 0.18$ ($s = 0.25, u_{0.95} = \pm 0.51$) eV] and $\delta E_{\mathrm{MAPE}} = 4.96\%$ ($s = 6.34\%, u_{0.95} = \pm 13.12\%$) for atomization energies of 23 tested solids from the PAW PBE calculations of Klimeš *et al.* [132].

$\Delta E^*_{\mathrm{MAE}} = 0.07$ ($s = 0.10, u_{0.95} = \pm 0.21$) eV/atom [$\Delta E_{\mathrm{MAE}} = 0.08$ ($s = 0.11, u_{0.95} = \pm 0.22$) eV] and $\delta E_{\mathrm{MAPE}} = 2.91\%$ ($s = 3.89\%, u_{0.95} = \pm 8.04\%$) for atomization energies of 23 tested solids from the PAW optB88-vdW calculations of Klimeš *et al.* [132].

$\Delta E^*_{\mathrm{MAE}} = 0.34$ ($s = 0.41, u_{0.95} = \pm 0.84$) eV/atom [$\Delta E_{\mathrm{MAE}} = 0.34$ ($s = 0.41, u_{0.95} = \pm 0.84$) eV] and $\delta E_{\mathrm{MAPE}} = 8.44\%$ ($s = 11.81\%, u_{0.95} = \pm 24.24\%$) for cohesive energies of 27 transition metals from the PAW PBE calculations of Janthon *et al.* [133].

$\Delta E^*_{\mathrm{MAE}} = 0.19$ ($s = 0.26, u_{0.95} = \pm 0.53$) eV/atom [$\Delta E_{\mathrm{MAE}} = 0.27$ ($s = 0.33, u_{0.95} = \pm 0.67$) eV] and $\delta E_{\mathrm{MAPE}} = 4.99\%$ ($s = 6.35\%, u_{0.95} = \pm 12.80\%$) for cohesive energies of 44 tested solids from the PBE calculations of Tran *et al.* by using the augmented planewave plus local orbitals (APW+lo) method [16].

$\Delta E^*_{\mathrm{MAE}} = 0.13$ ($s = 0.19, u_{0.95} = \pm 0.37$) eV/atom [$\Delta E_{\mathrm{MAE}} = 0.16$ ($s = 0.21, u_{0.95} = \pm 0.42$) eV] and $\delta E_{\mathrm{MAPE}} = 3.82\%$ ($s = 5.31\%, u_{0.95} = \pm 10.71\%$) for cohesive energies of 44 tested solids from the APW+lo optB88-vdW calculations of Tran *et al.* [16].

In contrast to the above list, the results $\Delta(\Delta H^*_{\mathrm{f}})_{\mathrm{MAE}} = 0.109$ ($s = 0.142, u_{0.95} = \pm 0.285$) eV/atom (PBE) and 0.097 ($s = 0.124, u_{0.95} = \pm 0.249$) eV/atom (optB88-vdW) for 49 formation enthalpies as well as $\Delta D^*_{\mathrm{MAE}} = 0.292$ ($s = 0.334, u_{0.95} = \pm 0.660$) eV/atom (PBE) and 0.288 ($s = 0.324, u_{0.95} = \pm 0.641$) eV/atom (optB88-vdW) for 138 dissociation energies from our PAW DFT calculations are comparably good.

## V. SUMMARY

We have performed the PAW DFT calculations for the geometric structures, formation enthalpies, and dissociation energies of 82 molecules or isomers consisting of C, H, N, and/or O. We use PBE and optB88-vdW functionals typically without and with dispersion corrections, respectively. For a convenient comparison and indexing, all results are tabulated (summarized in Tables I and II) together with the corresponding data from previous experiments and AOBM calculations if available in the literature. Relative to available experimental values, the AEs and PEs of 42 bond lengths, 15 bond angles, 49 formation enthalpies, and 138 dissociation energies from our DFT calculations are illustrated (plotted in Figs. 1, 2, 3, and 4, respectively). The corresponding MAEs and MAPEs for these DFT values are also obtained (summarized in Table III) and compared with the previous error analyses from similar or different planewave DFT methods for various solids and other molecules.

As evaluated in the work, the PAW DFT results for these small molecules is especially informative before calculating various molecular adsorption properties on large-size solid materials often involving the supercells containing hundreds to thousands of atoms, for which using a planewave DFT method is computationally much more efficient than using various existent AOBMs. The data presented in this paper also provide very useful and convenient information for the further development of DFT methods.

## SUPPLEMENTARY MATERIAL

See supplementary material for formulation of formation enthalpies and dissociation energies as well as Tables S1, S2, S3, and S4, providing the original data for Figs 1, 2, 3, and 4, respectively.

## ACKNOWLEDGEMENTS

The author thanks Prof. James W. Evans and Prof. Theresa L. Windus for their comments and suggestions. The author also thanks Dr. Branko Ruscic for providing the information of ATcT and the uncertainty analysis for DFT datasets. This work was supported by the U. S. Department of Energy (DOE), Office of Science, Basic Energy Sciences, Chemical Sciences, Geosciences, and Biosciences Division. Research was performed at Ames National Laboratory, which is operated by Iowa State University under contract No. DE-AC02-07CH11358. DFT calculations were performed with a grant of computer time at the National Energy Research Scientific Computing Centre (NERSC). NERSC is a DOE Office of Science User Facility supported by the Office of Science of the U. S. DOE under Contract No. DE-AC02-05CH11231. The calculations also partly used the Extreme Science and Engineering Discovery Environment (XSEDE), which is supported by National Science Foundation under Grant No. ACI-1548562.

## DATA AVAILABILITY

The data that support the findings of this study are available within the article and its supplementary material and from the corresponding author upon reasonable request.

Supplementary Material

for "**An evaluation for geometries, formation enthalpies, and dissociation energies of diatomic and triatomic (C, H, N, O), NO3, and HNO3 molecules from the PAW DFT method with PBE and optB88-vdW functionals**"

Yong Han
Division of Chemical and Biological Sciences, Ames National Laboratory, Ames, Iowa 50011, USA.
Department of Physics and Astronomy, Iowa State University, Ames, Iowa 50011, USA.
*Email: y27h@ameslab.gov

## S1. Formulation of formation enthalpies and dissociation energies

To avoid any confusion, it is necessary to formulate the formation enthalpies and dissociation energies calculated in this work. Experimentally, the standard formation enthalpy of a chemical compound is the change of enthalpy to form the compound from its constituents at the standard states, for which the standard pressure is 100 kPa (101.325 kPa prior to 1982) [32] and the temperature is usually 298.15 K but also 0 K from extrapolation of experimental values at higher temperatures. In the standard DFT calculations, the temperature is always 0 K and the pressure is zero. Under low-pressure condition, the pressure effects can be ignored and thus the formation energies from the DFT calculations [103] can compare with the experimental values of formation enthalpies.

Consider forming a molecule with the chemical formula of $\mathrm{C}_j\mathrm{H}_k\mathrm{N}_m\mathrm{O}_n$ containing $j$ C atoms, $k$ H atoms, $m$ N atoms, and $n$ O atoms by the reaction

$$2j\mathrm{C} + k\mathrm{H}_2 + m\mathrm{N}_2 + n\mathrm{O}_2 \rightarrow 2\mathrm{C}_j\mathrm{H}_k\mathrm{N}_m\mathrm{O}_n. \tag{S1}$$

At 0 K, the formation enthalpy $\Delta H_\mathrm{f}$ is reduced to the formation energy $\Delta E_\mathrm{f}$ [103] and can be defined as

$$\Delta H_\mathrm{f} = \Delta E_\mathrm{f} = E_{\mathrm{C}_j\mathrm{H}_k\mathrm{N}_m\mathrm{O}_n} - jE \; - \frac{k}{2}E_{\mathrm{H}_2} - \frac{m}{2}E_{\mathrm{N}_2} - \frac{n}{2}E_{\mathrm{O}_2}, \tag{S2}$$

where $E_{\mathrm{C}_j\mathrm{H}_k\mathrm{N}_m\mathrm{O}_n}$ is the total energy of the molecule, $E$ is the energy per C atom in bulk graphite, $E_{\mathrm{H}_2}$ is the energy of a molecule $H_2$ in gas phase, $E_{\mathrm{N}_2}$ is the energy of a molecule $N_2$ in gas phase, and $E_{\mathrm{O}_2}$ is the energy of a molecule $O_2$ in gas phase. $\Delta H_\mathrm{f} > 0$ indicates that reaction (1) is endothermic (i.e., the product is more unfavorable than the reactants thermodynamically), and $\Delta H_\mathrm{f} < 0$ indicates that reaction (1) is exothermic (i.e., the product is more favorable than the reactants thermodynamically). For comparison purposes, the average over all atoms,

$$\Delta H_\mathrm{f}^* \equiv \frac{\Delta H_\mathrm{f}}{N} = \frac{\Delta H_\mathrm{f}}{j+k+m+n}, \tag{S3}$$

is also used in literature [136], where $N = j + k + m + n$ is the total number of atoms in the molecule $\mathrm{C}_j\mathrm{H}_k\mathrm{N}_m\mathrm{O}_n$.

For dissociation of a compound, we can first define a general dissociation energy

$$D_{\mathrm{p1+p2+p3+\cdots}} = E_{\mathrm{p1}} + E_{\mathrm{p2}} + E_{\mathrm{p3}} + \cdots - E_{\mathrm{p1+p2+p3+\cdots}}, \quad \text{(S4)}$$

and the average over all atoms,

$$D^{*}_{\mathrm{p1+p2+p3+\cdots}} \equiv \frac{D_{\mathrm{p1+p2+p3+\cdots}}}{N}, \quad \text{(S5)}$$

where p$i$ labels a product after dissociation, $E_{\mathrm{p}i}$ is the corresponding total energy of the product p$i$, and $N$ is the total number of atoms in the molecule before dissociation. $D_{\mathrm{p1+p2+p3+\cdots}} > 0$ indicates that the dissociation process is endothermic, and $D_{\mathrm{p1+p2+p3+\cdots}} < 0$ indicates that the dissociation process is exothermic. Specifically, if any p$i$ does not involve any recombination of atoms, i.e., the energy change is only from breaking bonds, then $D_{\mathrm{p1+p2+p3+\cdots}}$ is called the "bond dissociation energy". For example, for a most stable $H_2O$ molecule, $D_{\mathrm{H_2+O}} = E_{\mathrm{H_2}} + E_{\mathrm{O}} - E_{\mathrm{H_2O}}$ is not the bond dissociation energy because the dissociation reaction $H_2O \longrightarrow H_2 + O$ corresponds to breaking two O-H bonds plus a recombination of two H atoms into $H_2$, but $D_{\mathrm{OH+H}} = E_{\mathrm{OH}} + E_{\mathrm{H}} - E_{\mathrm{H_2O}}$ is the bond dissociation energy because the dissociation reaction $H_2O \longrightarrow OH + H$ corresponds to only breaking one O-H bond without any recombination of atoms. Also, if all products p$i$ are isolated single atoms in vacuum, $D_{\mathrm{p1+p2+p3+\cdots}}$ is called the "atomization energy", e.g., for the molecule $\mathrm{C}_j\mathrm{H}_k\mathrm{N}_m\mathrm{O}_n$, the atomization energy is

$$D_{j\mathrm{C}+k\mathrm{H}+m\mathrm{N}+n\mathrm{O}} = jE_{\mathrm{C}} + kE_{\mathrm{H}} + mE_{\mathrm{N}} + nE_{\mathrm{O}} - E_{\mathrm{C}_j\mathrm{H}_k\mathrm{N}_m\mathrm{O}_n}, \quad \text{(S6)}$$

where $E_{\mathrm{C}_j\mathrm{H}_k\mathrm{N}_m\mathrm{O}_n}$ is the total energy of the molecule before dissociation, $E_{\mathrm{C}}$ is the energy of one single C atom in vacuum, $E_{\mathrm{H}}$ is the energy of one single H atom in vacuum, $E_{\mathrm{N}}$ is the energy of one single N atom in vacuum, and $E_{\mathrm{O}}$ is the energy of one single O atom in vacuum. Correspondingly, the average over all atoms

$$D^{*}_{j\mathrm{C}+k\mathrm{H}+m\mathrm{N}+n\mathrm{O}} \equiv \frac{D_{j\mathrm{C}+k\mathrm{H}+m\mathrm{N}+n\mathrm{O}}}{N} = \frac{D_{j\mathrm{C}+k\mathrm{H}+m\mathrm{N}+n\mathrm{O}}}{j+k+m+n}, \quad \text{(S7)}$$

where $N$ is the total number of atoms in the molecule $\mathrm{C}_j\mathrm{H}_k\mathrm{N}_m\mathrm{O}_n$ before dissociation. Obviously, for a diatomic molecule, the above general dissociation energy, bond dissociation energy, and atomization energy are identical with no distinction.

In the above all equations, the energies $E_{\mathrm{C}_j\mathrm{H}_k\mathrm{N}_m\mathrm{O}_n}$, $E$ , $E_{\mathrm{H_2}}$, $E_{\mathrm{N_2}}$, $E_{\mathrm{O_2}}$, $E_{\mathrm{p}i}$, $E_{\mathrm{C}}$, $E_{\mathrm{H}}$, $E_{\mathrm{N}}$, and $E_{\mathrm{O}}$ can be directly obtained from the DFT calculations, and then the formation enthalpies and dissociation energies can be readily calculated from these equations. For more thermochemical concepts and relevant aspects, the work from Ruscic and Bross [35] is recommended.

## S2. DFT and experimental data for Figures

**TABLE S1.** DFT and experimental data for 42 bond lengths in Fig. 1. "PBE" and "optB88" denote our PAW PBE and optB88-vdW calculations, respectively. "Exp." denotes the selected values based on available experimental data in Table I. In the first column, "L" denotes "linear", "T" denotes "triangular", and "P" denotes "planar"; the bond of two atoms in a molecule is indicated in a parenthesis. $l$ is the bond length, listed in Table I. $\Delta l$ and $\delta l$ are the absolute and percentage errors relative to the experimental values by using Eqs. (1) and (4), respectively.

| Molecule and bond | PBE $l$ (Å) | optB88 $l$ (Å) | Exp. $l$ (Å) | PBE $\Delta l$ (Å) | optB88 $\Delta l$ (Å) | PBE $\delta l$ (%) | optB88 $\delta l$ (%) |
|---|---|---|---|---|---|---|---|
| L CH (C-H) | 1.13691448 | 1.12995479 | 1.11978600 | 0.01712848 | 0.01016879 | 1.52962056 | 0.90810098 |
| L HCN (C-H) | 1.07490936 | 1.06992917 | 1.06549000 | 0.00941936 | 0.00443917 | 0.88404039 | 0.41663216 |
| L HCN (C-N) | 1.16107445 | 1.15606992 | 1.15321000 | 0.00786445 | 0.00285992 | 0.68196200 | 0.24799612 |
| L HNC (N-H) | 1.00495238 | 1.00194864 | 0.99400000 | 0.01095238 | 0.00794864 | 1.10184869 | 0.79966167 |
| L HNC (N-C) | 1.17739641 | 1.17279626 | 1.16890000 | 0.00849641 | 0.00389625 | 0.72687202 | 0.33332663 |
| T HCO (C-H) | 1.13370652 | 1.12940706 | 1.11910000 | 0.01460652 | 0.01030706 | 1.30520224 | 0.92101341 |
| T HCO (C-O) | 1.18817237 | 1.18598664 | 1.17540000 | 0.01277237 | 0.01058664 | 1.08664069 | 0.90068393 |
| T HCH (C-H) | 1.08485196 | 1.08164467 | 1.07530000 | 0.00955196 | 0.00634467 | 0.88830635 | 0.59003757 |
| L CN (C-N) | 1.17680126 | 1.17326891 | 1.17180000 | 0.00500126 | 0.00146891 | 0.42680132 | 0.12535516 |
| L NCO (C-N) | 1.23428379 | 1.22921756 | 1.20000000 | 0.03428379 | 0.02921756 | 2.85698233 | 2.43479633 |
| L NCO (C-O) | 1.19320045 | 1.19141508 | 1.20600000 | -0.01279955 | -0.01458492 | -1.06132297 | -1.20936335 |
| L CO (C-O) | 1.14310723 | 1.13932280 | 1.12832300 | 0.01478423 | 0.01099980 | 1.31028331 | 0.97488057 |
| L OCO (C-O) | 1.17644149 | 1.17386410 | 1.16200000 | 0.01444149 | 0.01186410 | 1.24281291 | 1.02100712 |
| L CC (C-C) | 1.31416137 | 1.30927409 | 1.24250000 | 0.07166137 | 0.06677409 | 5.76751429 | 5.37417185 |
| L CCH (C-C) | 1.21130840 | 1.20653041 | 1.21652000 | -0.00521160 | -0.00998959 | -0.42840222 | -0.82116132 |
| L CCH (C-H) | 1.07201427 | 1.06751411 | 1.04653000 | 0.02548427 | 0.02098411 | 2.43512112 | 2.00511329 |
| L CNC (N-C) | 1.25305366 | 1.24819807 | 1.24500000 | 0.00805366 | 0.00319807 | 0.64688042 | 0.25687307 |
| L CCC (C-C) | 1.30183579 | 1.29477597 | 1.29471000 | 0.00712579 | 0.00006597 | 0.55037732 | 0.00509544 |
| L HN (H-N) | 1.05062020 | 1.04854018 | 1.03620000 | 0.01442020 | 0.01234018 | 1.39164262 | 1.19090727 |
| T HNO (N-H) | 1.07875335 | 1.07565811 | 1.09026000 | -0.01150665 | -0.01460189 | -1.05540457 | -1.33930357 |
| T HNO (N-O) | 1.21681670 | 1.21710621 | 1.20900000 | 0.00781670 | 0.00810621 | 0.64654266 | 0.67048910 |
| P HNOOO (N-$O_I$) | 1.44570519 | 1.45489016 | 1.40600000 | 0.03970519 | 0.04889016 | 2.82398222 | 3.47725156 |
| P HNOOO (N-$O_{II}$) | 1.22309513 | 1.22328618 | 1.21100000 | 0.01209513 | 0.01228618 | 0.99877242 | 1.01454806 |
| P HNOOO (N-$O_{III}$) | 1.20877837 | 1.20874893 | 1.19900000 | 0.00977837 | 0.00974893 | 0.81554412 | 0.81308866 |
| P HNOOO (H-$O_I$) | 0.98328154 | 0.98258196 | 0.96400000 | 0.01928154 | 0.01858196 | 2.00015936 | 1.92758971 |
| L HO (H-O) | 0.98683572 | 0.98623662 | 0.96966000 | 0.01717572 | 0.01657662 | 1.77131389 | 1.70952864 |
| T HOO (O-H) | 0.99015913 | 0.98876141 | 0.97070000 | 0.01945913 | 0.01806141 | 2.00464939 | 1.86065844 |
| T HOO (O-O) | 1.34514436 | 1.35195563 | 1.33054000 | 0.01460436 | 0.02141563 | 1.09762644 | 1.60954418 |
| L HH (H-H) | 0.75000311 | 0.74421851 | 0.74144000 | 0.00856311 | 0.00277851 | 1.15493014 | 0.37474462 |
| T HNH (N-H) | 1.03475312 | 1.03738007 | 1.02400000 | 0.01075312 | 0.01338007 | 1.05010975 | 1.30664793 |
| T HOH (O-H) | 0.97185858 | 0.97097499 | 0.95781000 | 0.01404858 | 0.01316499 | 1.46673940 | 1.37448844 |
| L NO (N-O) | 1.16873149 | 1.16764992 | 1.15077000 | 0.01796149 | 0.01687992 | 1.56082372 | 1.46683709 |
| T ONO (N-O) | 1.21184450 | 1.21280586 | 1.19300000 | 0.01884450 | 0.01980586 | 1.57958948 | 1.66017291 |
| P NOOO (N-$O_I$) | 1.25076505 | 1.25227993 | 1.24000000 | 0.01076505 | 0.01227993 | 0.86814952 | 0.99031685 |
| P NOOO (N-$O_{II}$) | 1.25095144 | 1.25236388 | 1.24000000 | 0.01095144 | 0.01236388 | 0.88318065 | 0.99708702 |
| P NOOO (N-$O_{III}$) | 1.25088042 | 1.25240112 | 1.24000000 | 0.01088042 | 0.01240112 | 0.87745347 | 1.00009016 |
| L NN (N-N) | 1.11294431 | 1.10870580 | 1.09768000 | 0.01526431 | 0.01102580 | 1.39059708 | 1.00446401 |
| L NNO (N-N) | 1.14303696 | 1.13951667 | 1.12729200 | 0.01574496 | 0.01222467 | 1.39670616 | 1.08442807 |
| L NNO (N-O) | 1.19771079 | 1.19968798 | 1.18508900 | 0.01262179 | 0.01459898 | 1.06505032 | 1.23188887 |
| L NNN (N-N) | 1.18883831 | 1.18661634 | 1.18115000 | 0.00768831 | 0.00546634 | 0.65091690 | 0.46279795 |
| L OO (O-O) | 1.23265402 | 1.23603023 | 1.20752000 | 0.02513402 | 0.02851023 | 2.08145786 | 2.36105677 |
| T OOO (O-O) | 1.28437809 | 1.28938328 | 1.27276000 | 0.01161809 | 0.01662328 | 0.91282628 | 1.30608104 |

**TABLE S2.** DFT and experimental data for 15 bond angles in Fig. 2. "PBE" and "optB88" denote our PAW PBE and optB88-vdW calculations, respectively. "Exp." denotes the selected values based on available experimental data in Table I. In the first column, "T" denotes "triangular" and "P" denotes "planar"; the bond angle of three atoms in a molecule is indicated in a parenthesis. $\Delta\theta$ and $\delta\theta$ are the absolute and percentage errors relative to the experimental value of the bond angle $\theta$ (listed in Table I) by using Eqs. (1) and (4), respectively.

| Molecule and bond angle | PBE $\theta$ (°) | optB88 $\theta$ (°) | Exp. $\theta$ (°) | PBE $\Delta\theta$ (°) | optB88 $\Delta\theta$ (°) | PBE $\delta\theta$ (%) | optB88 $\delta\theta$ (%) |
|---|---|---|---|---|---|---|---|
| T HCO (H-C-O) | 123.91205206 | 123.88214291 | 124.43000000 | -0.51794794 | -0.54785709 | -0.41625648 | -0.44029341 |
| T HCH (H-C-H) | 135.06250720 | 135.04375287 | 133.93080000 | 1.13170720 | 1.11295287 | 0.84499398 | 0.83099098 |
| T HNO (H-N-O) | 108.49165754 | 108.56885822 | 108.04700000 | 0.44465754 | 0.52185822 | 0.41154085 | 0.48299187 |
| P HNOOO ($O_I$-N-$O_{II}$) | 115.47244710 | 115.43281360 | 115.88000000 | -0.40755290 | -0.44718640 | -0.35170254 | -0.38590473 |
| P HNOOO ($O_I$-N-$O_{III}$) | 113.55835260 | 113.49290460 | 113.85000000 | -0.29164740 | -0.35709540 | -0.25616812 | -0.31365428 |
| P HNOOO ($O_{II}$-N-$O_{III}$) | 130.96920030 | 131.07428200 | 130.27000000 | 0.69920030 | 0.80428200 | 0.53673163 | 0.61739618 |
| P HNOOO (H-$O_I$-N) | 101.96825880 | 102.02320360 | 102.15000000 | -0.18174120 | -0.12679640 | -0.17791601 | -0.12412766 |
| T HOO (H-O-O) | 105.01288446 | 104.85225819 | 104.29000000 | 0.72288446 | 0.56225819 | 0.69314839 | 0.53912953 |
| T HNH (H-N-H) | 102.89355581 | 102.46321828 | 103.40000000 | -0.50644419 | -0.93678172 | -0.48979129 | -0.90597845 |
| T HOH (H-O-H) | 104.51116045 | 104.81690033 | 104.47760000 | 0.03356045 | 0.33930033 | 0.03212215 | 0.32475893 |
| T ONO (O-N-O) | 133.86015921 | 133.40281144 | 134.10000000 | -0.23984079 | -0.69718856 | -0.17885219 | -0.51990198 |
| P NOOO ($O_I$-N-$O_{II}$) | 120.03869150 | 120.00169180 | 120.00000000 | 0.03869150 | 0.00169180 | 0.03224292 | 0.00140983 |
| P NOOO ($O_I$-N-$O_{III}$) | 119.99636550 | 120.03488520 | 120.00000000 | -0.00363450 | 0.03488520 | -0.00302875 | 0.02907100 |
| P NOOO ($O_{II}$-N-$O_{III}$) | 119.96494300 | 119.96342290 | 120.00000000 | -0.03505700 | -0.03657710 | -0.02921417 | -0.03048092 |
| T OOO (O-O-O) | 118.19115792 | 117.98988230 | 116.75420000 | 1.43695792 | 1.23568230 | 1.23075480 | 1.05836218 |

**TABLE S3.** DFT and experimental data for formation enthalpies of 49 isomers in Fig. 3. “PBE” and “optB88” denote our PAW PBE and optB88-vdW calculations, respectively. “Exp.” denotes the selected values based on available experimental data in Table I. In the first column, “L” denotes “linear”, “T” denotes “triangular”, and “P” denotes “planar”. $\Delta H_\mathrm{f}$ denotes the formation enthalpy for the molecule (or isomer), listed in Table I. $\Delta(\Delta H_\mathrm{f}^*)$ [or $\Delta(\Delta H_\mathrm{f})$] and $\delta(\Delta H_\mathrm{f})$ are the absolute and percentage errors relative to the experimental value of the molecule by using Eqs. (1) and (4), respectively.

| Molecule | PBE $\Delta H_\mathrm{f}$ (eV) | optB88 $\Delta H_\mathrm{f}$ (eV) | Exp. $\Delta H_\mathrm{f}$ (eV) | PBE $\Delta(\Delta H_\mathrm{f})$ (eV) | optB88 $\Delta(\Delta H_\mathrm{f})$ (eV) | PBE $\Delta(\Delta H_\mathrm{f}^*)$ (eV/atom) | optB88 $\Delta(\Delta H_\mathrm{f}^*)$ (eV/atom) | PBE $\delta(\Delta H_\mathrm{f})$ (%) | optB88 $\delta(\Delta H_\mathrm{f})$ (%) |
|---|---|---|---|---|---|---|---|---|---|
| L CH | 6.42909884 | 6.38377449 | 6.14432253 | 0.28477631 | 0.23945196 | 0.14238816 | 0.11972598 | 4.63478786 | 3.89712554 |
| L HCN | 1.23957399 | 1.33594722 | 1.34401776 | -0.10444377 | -0.00807055 | -0.03481459 | -0.00269018 | -7.77101118 | -0.60047908 |
| L HNC | 1.85690199 | 1.94579985 | 1.99056163 | -0.13365964 | -0.04476179 | -0.04455321 | -0.01492060 | -6.71466978 | -2.24870130 |
| T HCO | 0.46495776 | 0.53396214 | 0.42888384 | 0.03607392 | 0.10507829 | 0.01202464 | 0.03502610 | 8.41111713 | 24.50040825 |
| T HOC | 2.30362775 | 2.35198673 | 2.25163758 | 0.05199016 | 0.10034914 | 0.01733005 | 0.03344971 | 2.30899336 | 4.45671821 |
| T HCH | 3.88987897 | 4.17623322 | 4.05298911 | -0.16311014 | 0.12324411 | -0.05437005 | 0.04108137 | -4.02444053 | 3.04082026 |
| L CN | 4.51499246 | 4.60640196 | 4.52638749 | -0.01139503 | 0.08001447 | -0.00569752 | 0.04000724 | -0.25174672 | 1.76773356 |
| L NCO | 1.06742555 | 1.13940081 | 1.31564039 | -0.24821485 | -0.17623958 | -0.08273828 | -0.05874653 | -18.86646587 | -13.39572587 |
| T NCO | 4.45456960 | 4.55326711 | 4.66495777 | -0.21038818 | -0.11169066 | -0.07012939 | -0.03723022 | -4.50996960 | -2.39424809 |
| L CNO | 3.64502023 | 3.73136601 | 4.03481018 | -0.38978995 | -0.30344417 | -0.12992998 | -0.10114806 | -9.66067634 | -7.52065534 |
| L NCN | 4.14900736 | 4.30053794 | 4.67677304 | -0.52776568 | -0.37623510 | -0.17592189 | -0.12541170 | -11.28482557 | -8.04475857 |
| T NCN | 5.73302946 | 5.88741197 | 6.00194856 | -0.26891910 | -0.11453659 | -0.08963970 | -0.03817886 | -4.48052987 | -1.90832339 |
| L CNN | 5.30762479 | 5.48384653 | 6.00402141 | -0.69639662 | -0.52017488 | -0.23213221 | -0.17339163 | -11.59883641 | -8.66377460 |
| L CO | -0.62114207 | -0.71507314 | -1.17949534 | 0.55835327 | 0.46442220 | 0.27917664 | 0.23221110 | -47.33831946 | -39.37465322 |
| L OCO | -3.86994994 | -3.86340063 | -4.07429804 | 0.20434810 | 0.21089741 | 0.06811603 | 0.07029914 | -5.01554138 | -5.17628834 |
| T OCO | 2.19754829 | 2.15756569 | 1.97232052 | 0.22522777 | 0.18524517 | 0.07507592 | 0.06174839 | 11.41943070 | 9.39224497 |
| L OCO | 3.04777224 | 2.97489724 | 3.10720804 | -0.05943580 | -0.13231080 | -0.01981193 | -0.04410360 | -1.91283629 | -4.25818938 |
| L CC | 8.77985785 | 8.86477971 | 8.49875294 | 0.28110491 | 0.36602677 | 0.14055246 | 0.18301339 | 3.30760186 | 4.30682917 |
| L CCH | 5.93845302 | 6.17529586 | 5.84296066 | 0.09549236 | 0.33233519 | 0.03183079 | 0.11077840 | 1.63431468 | 5.68778763 |
| L CNC | 6.87322282 | 6.89181090 | 6.96686206 | -0.09363924 | -0.07505116 | -0.03121308 | -0.02501705 | -1.34406620 | -1.07725921 |
| T CNC | 7.37512096 | 7.53087878 | 7.46538343 | -0.09026248 | 0.06549535 | -0.03008749 | 0.02183178 | -1.20908027 | 0.87732060 |
| L CCN | 6.92937168 | 6.96568999 | 7.09641543 | -0.16704376 | -0.13072544 | -0.05568125 | -0.04357515 | -2.35391741 | -1.84213347 |
| L CCO | 3.87602600 | 3.99056243 | 3.90857337 | -0.03254737 | 0.08198906 | -0.01084912 | 0.02732969 | -0.83271746 | 2.09767220 |
| L COC | 6.50136962 | 6.57493808 | 6.76786809 | -0.26649847 | -0.19293001 | -0.08883282 | -0.06431000 | -3.93770183 | -2.85067621 |
| L CCC | 8.70551620 | 8.65766895 | 8.44221585 | 0.26330035 | 0.21545310 | 0.08776678 | 0.07181770 | 3.11885358 | 2.55209184 |
| T CCC | 9.21207752 | 9.49124988 | 9.31540557 | -0.10332805 | 0.17584431 | -0.03444268 | 0.05861477 | -1.10921684 | 1.88767211 |
| L HN | 3.60223384 | 3.73246261 | 3.71807810 | -0.11584426 | 0.01438451 | -0.05792213 | 0.00719225 | -3.11570262 | 0.38688021 |
| T HNO | 0.96039057 | 0.99707962 | 1.13944781 | -0.17905724 | -0.14236818 | -0.05968575 | -0.04745606 | -15.71438683 | -12.49448902 |
| T NOH | 1.92155299 | 2.04135107 | 2.25795979 | -0.33640680 | -0.21660872 | -0.11213560 | -0.07220291 | -14.89870631 | -9.59311692 |
| P $HNO_3$ | -2.10680788 | -2.00463082 | -1.29004064 | -0.81676724 | -0.71459018 | -0.16335345 | -0.14291804 | 63.31329479 | 55.39284240 |
| T HNN | 2.01554166 | 2.16786137 | 2.61324695 | -0.59770530 | -0.44538558 | -0.19923510 | -0.14846186 | -22.87213215 | -17.04337877 |
| L HO | 0.58643538 | 0.60729350 | 0.38641107 | 0.20002431 | 0.22088244 | 0.10001215 | 0.11044122 | 51.76464320 | 57.16255539 |
| T HOO | -0.00374171 | -0.00475450 | 0.15691504 | -0.16065676 | -0.16166954 | -0.05355225 | -0.05388985 | -102.38454831 | -103.02998038 |
| T HNH | 1.55454882 | 1.73946108 | 1.95801782 | -0.40346901 | -0.21855674 | -0.13448967 | -0.07285225 | -20.60599263 | -11.16214270 |
| T1 HOH | -2.51962122 | -2.43275865 | -2.47600329 | -0.04361793 | 0.04324464 | -0.01453931 | 0.01441488 | 1.76162640 | -1.74655028 |
| T2 HOH | 2.99092661 | 2.98218990 | 2.56204746 | 0.42887915 | 0.42014244 | 0.14295972 | 0.14004748 | 16.73970361 | 16.39869861 |
| L HOH | 2.83307038 | 3.05475939 | 2.56619317 | 0.26687721 | 0.48856622 | 0.08895907 | 0.16285541 | 10.39973204 | 19.03855990 |
| L2 HHH | 2.42776063 | 2.61224404 | 2.63770663 | -0.20994599 | -0.02546258 | -0.06998200 | -0.00848753 | -7.95941407 | -0.96533035 |
| L3 HHH | 2.26871785 | 2.49365093 | 2.23764582 | 0.03107204 | 0.25600511 | 0.01035735 | 0.08533504 | 1.38860386 | 11.44082339 |
| L NO | 0.94619713 | 0.94757151 | 0.93939667 | 0.00680045 | 0.00817484 | 0.00340023 | 0.00408742 | 0.72391697 | 0.87022204 |
| T ONO | -0.19487725 | -0.10960816 | 0.38221354 | -0.57709079 | -0.49182170 | -0.19236360 | -0.16394057 | -150.98648699 | -128.67720517 |
| T NOO | 3.38878366 | 3.37830699 | 4.20167492 | -0.81289126 | -0.82336793 | -0.27096375 | -0.27445598 | -19.34683857 | -19.59618354 |
| P $NO_3$ | -0.21293888 | -0.13645417 | 0.82292301 | -1.03586190 | -0.95937718 | -0.25896547 | -0.23984430 | -125.87591819 | -116.58164473 |
| L NNO | 0.16832708 | 0.30754591 | 0.89156557 | -0.72323849 | -0.58401966 | -0.24107950 | -0.19467322 | -81.12005601 | -65.50495883 |
| L NON | 4.72217488 | 4.82848096 | 5.65370910 | -0.93153422 | -0.82522814 | -0.31051141 | -0.27507605 | -16.47651482 | -14.59622565 |
| T NON | 3.25482955 | 3.29423574 | 3.64511364 | -0.39028409 | -0.35087790 | -0.13009470 | -0.11695930 | -10.70704858 | -9.62597968 |
| L NNN | 3.61229683 | 3.79732415 | 4.68713731 | -1.07484048 | -0.88981316 | -0.35828016 | -0.29660439 | -22.93170465 | -18.98414961 |
| T2 NNN | 5.27107342 | 5.50893183 | 6.07242559 | -0.80135218 | -0.56349376 | -0.26711739 | -0.18783125 | -13.19657466 | -9.27954988 |
| T OOO | 1.28733181 | 1.21761068 | 1.49666272 | -0.20933091 | -0.27905204 | -0.06977697 | -0.09301735 | -13.98651217 | -18.64495183 |

**TABLE S4.** DFT and experimental data for 138 dissociation energies in Fig. 4. “PBE” and “optB88” denote our PAW PBE and optB88-vdW calculations, respectively. “Exp.” denotes the selected values based on available experimental data in Table I. In the first column, “L” denotes “linear”, “T” denotes “triangular”, and “P” denotes “planar”. $D$ denotes the dissociation energies, listed in Table I. $\Delta D^*$ (or $\Delta D$) and $\delta D$ are the absolute and percentage errors relative to the experimental value of the molecule by using Eqs. (1) and (4), respectively.

| Bonds | PBE $D$ (eV) | optB88 $D$ (eV) | Exp. $D$ (eV) | PBE $\Delta D$ (eV) | optB88 $\Delta D$ (eV) | PBE $\Delta D^*$ (eV/atom) | optB88 $\Delta D^*$ (eV/atom) | PBE $\delta D$ (%) | optB88 $\delta D$ (%) |
|---|---|---|---|---|---|---|---|---|---|
| L CH→C+H | 3.69706045 | 3.89485734 | 3.46790536 | 0.22915509 | 0.42695198 | 0.11457755 | 0.21347599 | 6.60788202 | 0.22915509 |
| L HCN→CH+N | 10.38595792 | 10.31113341 | 9.67753315 | 0.70842477 | 0.63360026 | 0.23614159 | 0.21120009 | 7.32030322 | 0.70842477 |
| L HCN→CN+H | 5.54435493 | 5.76476977 | 5.42144581 | 0.12290912 | 0.34332396 | 0.04096971 | 0.11444132 | 2.26709110 | 0.12290912 |
| L HCN→NH+C | 10.21988268 | 10.18083220 | 9.74728468 | 0.47259800 | 0.43354752 | 0.15753267 | 0.14451584 | 4.84850922 | 0.47259800 |
| L HCN→C+H+N | 14.08301837 | 14.20599075 | 13.14541778 | 0.93760059 | 1.06057297 | 0.31253353 | 0.35352432 | 7.13252793 | 0.93760059 |
| L HNC→CH+N | 9.76862992 | 9.70128078 | 9.03101001 | 0.73761991 | 0.67027077 | 0.24587330 | 0.22342359 | 8.16763476 | 0.73761991 |
| L HNC→CN+H | 4.92702693 | 5.15491714 | 4.77492267 | 0.15210426 | 0.37999447 | 0.05070142 | 0.12666482 | 3.18548104 | 0.15210426 |
| L HNC→NH+C | 9.60255468 | 9.57097957 | 9.10076154 | 0.50179314 | 0.47021803 | 0.16726438 | 0.15673934 | 5.51374888 | 0.50179314 |
| L HNC→C+H+N | 13.46569037 | 13.59613812 | 12.49889463 | 0.96679574 | 1.09724349 | 0.32226525 | 0.36574783 | 7.73504989 | 0.96679574 |
| T HCO→CH+O | 8.98804085 | 8.84959592 | 8.27379647 | 0.71424438 | 0.57579945 | 0.23808146 | 0.19193315 | 8.63260761 | 0.71424438 |
| T HCO→CO+H | 1.18283663 | 1.24527975 | 0.63065544 | 0.55218119 | 0.61462431 | 0.18406040 | 0.20487477 | 87.55671431 | 0.55218119 |
| T HCO→OH+C | 7.97870044 | 7.85764818 | 7.33075157 | 0.64794887 | 0.52689661 | 0.21598296 | 0.17563220 | 8.83877817 | 0.64794887 |
| T HCO→C+H+O | 12.68510130 | 12.74445326 | 11.74169146 | 0.94340984 | 1.00276180 | 0.31446995 | 0.33425393 | 8.03470135 | 0.94340984 |
| T HOC→CH+O | 7.14937087 | 7.03157133 | 6.45103236 | 0.69833851 | 0.58053897 | 0.23277950 | 0.19351299 | 10.82522097 | 0.69833851 |
| T HOC→CO+H | -0.65583335 | -0.57274484 | -1.19209830 | 0.53626495 | 0.61935346 | 0.17875498 | 0.20645115 | -44.98496036 | 0.53626495 |
| T HOC→OH+C | 6.14003046 | 6.03962359 | 5.50798747 | 0.63204299 | 0.53163612 | 0.21068100 | 0.17721204 | 11.47502600 | 0.63204299 |
| T HOC→C+H+O | 10.84643132 | 10.92642867 | 9.91902063 | 0.92741069 | 1.00740804 | 0.30913690 | 0.33580268 | 9.34982114 | 0.92741069 |
| T HCH→CH+H | 4.80815634 | 4.70185630 | 4.33039915 | 0.47775719 | 0.37145715 | 0.15925240 | 0.12381905 | 11.03263639 | 0.47775719 |
| T HCH→C+HH | 3.96734386 | 3.60808359 | 3.32019378 | 0.64715008 | 0.28788981 | 0.21571669 | 0.09596327 | 19.49133447 | 0.64715008 |
| T HCH→C+2H | 8.50521679 | 8.59671364 | 7.79826305 | 0.70695374 | 0.79845059 | 0.23565125 | 0.26615020 | 9.06552831 | 0.70695374 |
| L CN→C+N | 8.53866344 | 8.44122098 | 7.72397196 | 0.81469148 | 0.71724902 | 0.40734574 | 0.35862451 | 10.54757167 | 0.81469148 |
| L NCO→CN+O | 6.47146668 | 6.46678471 | 5.76906342 | 0.70240326 | 0.69772129 | 0.23413442 | 0.23257376 | 12.17534304 | 0.70240326 |
| L NCO→CO+N | 3.50786545 | 3.40883218 | 2.38202010 | 1.12584535 | 1.02681208 | 0.37528178 | 0.34227069 | 47.26430970 | 1.12584535 |
| L NCO→NO+C | 7.73599441 | 7.59248751 | 6.99691844 | 0.73907597 | 0.59556907 | 0.24635866 | 0.19852302 | 10.56287809 | 0.73907597 |
| L NCO→C+N+O | 15.01013012 | 14.90800569 | 13.49313902 | 1.51699110 | 1.41486667 | 0.50566370 | 0.47162222 | 11.24268485 | 1.51699110 |
| T NCO→CN+O | 3.08432263 | 3.05291841 | 2.42005696 | 0.66426567 | 0.63286145 | 0.22142189 | 0.21095382 | 27.44834832 | 0.66426567 |
| T NCO→CO+N | 0.12072140 | -0.00503412 | -0.96698636 | 1.08770776 | 0.96195224 | 0.36256925 | 0.32065075 | -112.48429193 | 1.08770776 |
| T NCO→NO+C | 4.34885036 | 4.17862121 | 3.64718649 | 0.70166387 | 0.53143472 | 0.23388796 | 0.17714491 | 19.23849711 | 0.70166387 |
| T NCO→C+N+O | 11.62298607 | 11.49413939 | 10.14351071 | 1.47947536 | 1.35062868 | 0.49315845 | 0.45020956 | 14.58543693 | 1.47947536 |
| L CNO→CN+O | 3.89387200 | 3.87481951 | 3.04916813 | 0.84470387 | 0.82565138 | 0.28156796 | 0.27521713 | 27.70276450 | 0.84470387 |
| L CNO→CO+N | 0.93027077 | 0.81686698 | -0.33787519 | 1.26814596 | 1.15474217 | 0.42271532 | 0.38491406 | -375.32970616 | 1.26814596 |
| L CNO→NO+C | 5.15839973 | 5.00052231 | 4.27733409 | 0.88106564 | 0.72318822 | 0.29368855 | 0.24106274 | 20.59847617 | 0.88106564 |
| L CNO→C+N+O | 12.43253544 | 12.31604049 | 10.77365831 | 1.65887713 | 1.54238218 | 0.55295904 | 0.51412739 | 15.39752872 | 1.65887713 |
| L NCN→CN+N | 5.56241816 | 5.56917015 | 4.72683246 | 0.83558570 | 0.84233769 | 0.27852857 | 0.28077923 | 17.67749766 | 0.83558570 |
| L NCN→C+NN | 3.70821547 | 3.48377887 | 2.69636839 | 1.01184708 | 0.78741048 | 0.33728236 | 0.26247016 | 37.52629198 | 1.01184708 |
| L NCN→C+2N | 14.10108160 | 14.01039113 | 12.45080442 | 1.65027718 | 1.55958671 | 0.55009239 | 0.51986224 | 13.25438197 | 1.65027718 |
| T NCN→CN+N | 3.97839606 | 3.98229612 | 3.40258973 | 0.57580633 | 0.57970639 | 0.19193544 | 0.19323546 | 16.92259067 | 0.57580633 |
| T NCN→C+NN | 2.12419337 | 1.89690484 | 1.37222930 | 0.75196407 | 0.52467554 | 0.25065469 | 0.17489185 | 54.79871813 | 0.75196407 |
| T NCN→C+2N | 12.51705950 | 12.42351710 | 11.12604348 | 1.39101602 | 1.29747362 | 0.46367201 | 0.43249121 | 12.50234216 | 1.39101602 |
| L CNN→CN+N | 4.40380073 | 4.38586156 | 3.39948045 | 1.00432028 | 0.98638111 | 0.33477343 | 0.32879370 | 29.54334635 | 1.00432028 |
| L CNN→C+NN | 2.54959804 | 2.30047028 | 1.36912002 | 1.18047802 | 0.93135026 | 0.39349267 | 0.31045009 | 86.22166061 | 1.18047802 |
| L CNN→C+2N | 12.94246417 | 12.82708254 | 11.12293420 | 1.81952997 | 1.70414834 | 0.60650999 | 0.56804945 | 16.35836321 | 1.81952997 |
| L CO→C+O | 11.50226467 | 11.49917351 | 11.11103601 | 0.39122866 | 0.38813750 | 0.19561433 | 0.19406875 | 3.52108171 | 0.39122866 |
| L OCO→CO+O | 6.27270764 | 6.14811105 | 5.45315012 | 0.81955752 | 0.69496093 | 0.27318584 | 0.23165364 | 15.02906590 | 0.81955752 |
| L OCO→C+OO | 11.72717277 | 11.64771744 | 11.44748094 | 0.27969183 | 0.20023650 | 0.09323061 | 0.06674550 | 2.44326098 | 0.27969183 |
| L OCO→C+2O | 17.77497231 | 17.64728456 | 16.56418613 | 1.21078618 | 1.08309843 | 0.40359539 | 0.36103281 | 7.30966298 | 1.21078618 |
| T OCO→CO+O | 0.20520941 | 0.12714473 | -0.59387265 | 0.79908206 | 0.72101738 | 0.26636069 | 0.24033913 | -134.55444691 | 0.79908206 |
| T OCO→C+OO | 5.65967454 | 5.62675112 | 5.40082092 | 0.25885362 | 0.22593020 | 0.08628454 | 0.07531007 | 4.79285698 | 0.25885362 |
| T OCO→C+2O | 11.70747408 | 11.62631824 | 10.51766085 | 1.18981323 | 1.10865739 | 0.39660441 | 0.36955246 | 11.31252709 | 1.18981323 |
| L COO→CO+O | -0.64501454 | -0.69018682 | -1.72876018 | 1.08374564 | 1.03857336 | 0.36124855 | 0.34619112 | -62.68918338 | 1.08374564 |
| L COO→C+OO | 4.80945059 | 4.80941957 | 4.26593339 | 0.54351720 | 0.54348618 | 0.18117240 | 0.18116206 | 12.74087403 | 0.54351720 |
| L COO→C+2O | 10.85725013 | 10.80898669 | 9.38277332 | 1.47447681 | 1.42621337 | 0.49149227 | 0.47540446 | 15.71472272 | 1.47447681 |
| L CC→2C | 6.93458781 | 6.70385391 | 6.24762321 | 0.68696460 | 0.45623070 | 0.34348230 | 0.22811535 | 10.99561516 | 0.68696460 |
| L CCH→CH+C | 8.34786865 | 7.99279545 | 7.67453440 | 0.67333425 | 0.31826105 | 0.22444475 | 0.10608702 | 8.77361701 | 0.67333425 |
| L CCH→CC+H | 5.11034129 | 5.18379888 | 4.89483727 | 0.21550402 | 0.28896161 | 0.07183467 | 0.09632054 | 4.40267990 | 0.21550402 |
| L CCH→2C+H | 12.04492910 | 11.88765279 | 11.14241902 | 0.90251008 | 0.74523377 | 0.30083669 | 0.24841126 | 8.09976789 | 0.90251008 |
| L CNC→CN+C | 5.49899246 | 5.49890787 | 4.93339236 | 0.56560010 | 0.56551551 | 0.18853337 | 0.18850517 | 11.46472980 | 0.56560010 |
| L CNC→CC+N | 7.10306809 | 7.23627494 | 6.40926436 | 0.69380373 | 0.82701058 | 0.23126791 | 0.27567019 | 10.82501354 | 0.69380373 |
| L CNC→2C+N | 14.03765590 | 13.94012885 | 12.65684610 | 1.38080980 | 1.28328275 | 0.46026993 | 0.42776092 | 10.90958825 | 1.38080980 |
| T CNC→CN+C | 4.99709433 | 4.85983999 | 4.43383456 | 0.56325977 | 0.42600543 | 0.18775326 | 0.14200181 | 12.70367136 | 0.56325977 |
| T CNC→CC+N | 6.60116996 | 6.59720706 | 5.90970656 | 0.69146340 | 0.68750050 | 0.23048780 | 0.22916683 | 11.70046931 | 0.69146340 |
| T CNC→2C+N | 13.53575777 | 13.30106097 | 12.15728831 | 1.37846946 | 1.14377266 | 0.45948982 | 0.38125755 | 11.33862609 | 1.37846946 |
| L CCN→CN+C | 5.44284361 | 5.42502878 | 4.80280256 | 0.64004105 | 0.62222622 | 0.21334702 | 0.20740874 | 13.32640773 | 0.64004105 |
| L CCN→CC+N | 7.04691924 | 7.16239585 | 6.27971098 | 0.76720826 | 0.88268487 | 0.25573609 | 0.29422829 | 12.21725422 | 0.76720826 |
| L CCN→2C+N | 13.98150705 | 13.86624976 | 12.52729273 | 1.45421432 | 1.33895703 | 0.48473811 | 0.44631901 | 11.60836860 | 1.45421432 |
| L COC→CO+C | 0.73471114 | 0.49430559 | -0.57418054 | 1.30889168 | 1.06848613 | 0.43629723 | 0.35616204 | -227.95821003 | 1.30889168 |
| L COC→CC+O | 5.30238800 | 5.28962519 | 4.28873478 | 1.01365322 | 1.00089041 | 0.33788441 | 0.33363014 | 23.63525066 | 1.01365322 |
| L COC→2C+O | 12.23697581 | 11.99347910 | 10.53631653 | 1.70065928 | 1.45716257 | 0.56688643 | 0.48572086 | 16.14092811 | 1.70065928 |
| L CCO→CO+C | 3.36005476 | 3.07868124 | 2.28511417 | 1.07494059 | 0.79356707 | 0.35831353 | 0.26452236 | 47.04100121 | 1.07494059 |
| L CCO→CC+O | 7.92773162 | 7.87400084 | 7.14854771 | 0.77918391 | 0.72545313 | 0.25972797 | 0.24181771 | 10.89989102 | 0.77918391 |
| L CCO→2C+O | 14.86231943 | 14.57785475 | 13.39612946 | 1.46618997 | 1.18172529 | 0.48872999 | 0.39390843 | 10.94487759 | 1.46618997 |
| L CCC→CC+C | 7.93156448 | 7.99142757 | 7.42973035 | 0.50183413 | 0.56169722 | 0.16727804 | 0.18723241 | 6.75440576 | 0.50183413 |
| L CCC→3C | 14.86615229 | 14.69528148 | 13.67741574 | 1.18873655 | 1.01786574 | 0.39624552 | 0.33928858 | 8.69123653 | 1.18873655 |
| T(c) CCC→CC+C | 7.42500316 | 7.15784664 | 6.55643698 | 0.86856618 | 0.60140966 | 0.28952206 | 0.20046989 | 13.24753334 | 0.86856618 |
| T(c) CCC→3C | 14.35959097 | 13.86170055 | 12.80401873 | 1.55557224 | 1.05768182 | 0.51852408 | 0.35256061 | 12.14909373 | 1.55557224 |
| L HN→H+N | 3.86313569 | 4.02515855 | 3.39813309 | 0.46500260 | 0.62702546 | 0.23250130 | 0.31351273 | 13.68406078 | 0.46500260 |
| T HNO→HN+O | 5.66574304 | 5.73516654 | 5.13694661 | 0.52879643 | 0.59821993 | 0.17626548 | 0.19940664 | 10.29398332 | 0.52879643 |
| T HNO→NO+H | 2.25474302 | 2.44480691 | 2.03894204 | 0.21580098 | 0.40586487 | 0.07193366 | 0.13528829 | 10.58396829 | 0.21580098 |
| T HNO→OH+N | 4.82247787 | 4.87352001 | 4.12415018 | 0.69832769 | 0.74936983 | 0.23277590 | 0.24978994 | 16.93264449 | 0.69832769 |
| T HNO→H+N+O | 9.52887873 | 9.76032509 | 8.53507970 | 0.99379903 | 1.22524539 | 0.33126634 | 0.40841513 | 11.64369941 | 0.99379903 |
| T NOH→HN+O | 4.70458062 | 4.69089510 | 4.01843463 | 0.68614599 | 0.67246047 | 0.22871533 | 0.22415349 | 17.07495709 | 0.68614599 |
| T NOH→NO+H | 1.29358060 | 1.40053547 | 0.92045079 | 0.37312981 | 0.48008468 | 0.12437660 | 0.16002823 | 40.53772528 | 0.37312981 |
| T NOH→OH+N | 3.86131545 | 3.82924857 | 3.00563820 | 0.85567725 | 0.82361037 | 0.28522575 | 0.27453679 | 28.46907022 | 0.85567725 |
| T NOH→H+N+O | 8.56771631 | 8.71605365 | 7.41656772 | 1.15114859 | 1.29948593 | 0.38371620 | 0.43316198 | 15.52131160 | 1.15114859 |
| P $HNO_3$→H+$NO_3$ | 4.16349287 | 4.36269567 | 4.35195683 | -0.18846396 | 0.01073884 | -0.03769279 | 0.00214777 | -4.33055671 | -0.18846396 |
| P $HNO_3$→OH+$NO_2$ | 2.49836600 | 2.50231617 | 2.05865488 | 0.43971112 | 0.44366129 | 0.08794222 | 0.08873226 | 21.35914679 | 0.43971112 |
| P $HNO_3$→H+N+3O | 18.64387672 | 18.76160266 | 16.08130444 | 2.56257228 | 2.68029822 | 0.51251446 | 0.53605964 | 15.93510208 | 2.56257228 |
| T NNH→HN+N | 6.78312525 | 6.82790736 | 5.98204916 | 0.80107609 | 0.84585820 | 0.26702536 | 0.28195273 | 13.39133244 | 0.80107609 |
| T NNH→NN+H | 0.25339481 | 0.32645365 | -0.37425378 | 0.62764859 | 0.70070743 | 0.20921620 | 0.23356914 | -167.70668070 | 0.62764859 |
| T NNH→H+2N | 10.64626094 | 10.85306591 | 9.38018225 | 1.26607869 | 1.47288366 | 0.42202623 | 0.49096122 | 13.49737834 | 1.26607869 |
| L OH→O+H | 4.70640086 | 4.88680508 | 4.41098134 | 0.29541952 | 0.47582374 | 0.14770976 | 0.23791187 | 6.69736489 | 0.29541952 |
| T HOO→OH+O | 3.61407686 | 3.61183156 | 2.78788489 | 0.82619197 | 0.82394667 | 0.27539732 | 0.27464889 | 29.63508166 | 0.82619197 |
| T HOO→OO+H | 2.27267818 | 2.49906952 | 2.08218177 | 0.19049641 | 0.41688775 | 0.06349880 | 0.13896258 | 9.14888452 | 0.19049641 |
| T HOO→H+2O | 8.32047772 | 8.49863664 | 7.19881442 | 1.12166330 | 1.29982222 | 0.37388777 | 0.43327407 | 15.58122264 | 1.12166330 |
| L HH→H+H | 4.53787293 | 4.98863005 | 4.47806988 | 0.05980305 | 0.51056017 | 0.02990152 | 0.25528008 | 1.33546478 | 0.05980305 |
| T HNH→HN+H | 4.31662149 | 4.48731655 | 3.99905345 | 0.31756804 | 0.48826310 | 0.10585601 | 0.16275437 | 7.94108024 | 0.31756804 |
| T HNH→N+HH | 3.64188425 | 3.52384505 | 2.91920019 | 0.72268406 | 0.60464486 | 0.24089469 | 0.20154829 | 24.75623497 | 0.72268406 |
| T HNH→N+2H | 8.17975718 | 8.51247510 | 7.39718654 | 0.78257064 | 1.11528856 | 0.26085688 | 0.37176285 | 10.57930115 | 0.78257064 |
| T1 HOH→OH+H | 5.37499306 | 5.53436718 | 5.10144588 | 0.27354718 | 0.43292130 | 0.09118239 | 0.14430710 | 5.36215000 | 0.27354718 |
| T1 HOH→O+HH | 5.54352099 | 5.43254221 | 5.03436107 | 0.50915992 | 0.39818114 | 0.16971997 | 0.13272705 | 10.11369489 | 0.50915992 |
| T1 HOH→O+2H | 10.08139392 | 10.42117226 | 9.51243033 | 0.56896359 | 0.90874193 | 0.18965453 | 0.30291398 | 5.98126417 | 0.56896359 |
| T2 HOH→OH+H | -0.13555477 | 0.11941863 | 0.06322204 | -0.19877681 | 0.05619659 | -0.06625894 | 0.01873220 | -314.41060663 | -0.19877681 |
| T2 HOH→O+HH | 0.03297316 | 0.01759366 | -0.00414571 | 0.03711887 | 0.02173937 | 0.01237296 | 0.00724646 | -895.35657344 | 0.03711887 |
| T2 HOH→O+2H | 4.57084609 | 5.00622371 | 4.47425521 | 0.09659088 | 0.53196850 | 0.03219696 | 0.17732283 | 2.15881470 | 0.09659088 |
| L OHH→OH+H | 0.02230146 | 0.04684914 | 0.05907634 | -0.03677488 | -0.01222720 | -0.01225829 | -0.00407573 | -62.24975833 | -0.03677488 |
| L OHH→O+HH | 0.19082939 | -0.05497583 | -0.00725499 | 0.19808438 | -0.04772084 | 0.06602813 | -0.01590695 | -2730.31958186 | 0.19808438 |
| L OHH→O+2H | 4.72870232 | 4.93365422 | 4.47010950 | 0.25859282 | 0.46354472 | 0.08619761 | 0.15451491 | 5.78493250 | 0.25859282 |
| L2 HHH→HH+H | -0.15882417 | -0.11792902 | -0.39798795 | 0.23916378 | 0.28005893 | 0.07972126 | 0.09335298 | -60.09322190 | 0.23916378 |
| L2 HHH→3H | 4.37904876 | 4.87070103 | 4.07937654 | 0.29967222 | 0.79132449 | 0.09989074 | 0.26377483 | 7.34602998 | 0.29967222 |
| L3 HHH→HH+H | 2.26871785 | 2.49365093 | 2.23764582 | 0.03107204 | 0.25600511 | 0.01035735 | 0.08533504 | 1.38860386 | 0.03107204 |
| L3 HHH→3H | 4.53809154 | 4.98929415 | 4.47943735 | 0.05865419 | 0.50985680 | 0.01955140 | 0.16995227 | 1.30940987 | 0.05865419 |
| L NO→N+O | 7.27413571 | 7.31551818 | 6.49616876 | 0.77796695 | 0.81934942 | 0.38898348 | 0.40967471 | 11.97578115 | 0.77796695 |
| T ONO→NO+O | 4.16497415 | 4.05696323 | 3.11554382 | 1.04943033 | 0.94141941 | 0.34981011 | 0.31380647 | 33.68369677 | 1.04943033 |
| T ONO→OO+N | 5.39131032 | 5.37291429 | 4.49500448 | 0.89630584 | 0.87790981 | 0.29876861 | 0.29263660 | 19.94004335 | 0.89630584 |
| T ONO→N+2O | 11.43910986 | 11.37248141 | 9.61170967 | 1.82740019 | 1.76077174 | 0.60913340 | 0.58692391 | 19.01222831 | 1.82740019 |
| T NOO→NO+O | 0.58131324 | 0.56904808 | -0.70373391 | 1.28504715 | 1.27278199 | 0.42834905 | 0.42426066 | -182.60412523 | 1.28504715 |
| T NOO→OO+N | 1.80764941 | 1.88499914 | 0.67575038 | 1.13189903 | 1.20924876 | 0.37729968 | 0.40308292 | 167.50253633 | 1.13189903 |
| T NOO→N+2O | 7.85544895 | 7.88456626 | 5.79259031 | 2.06285864 | 2.09197595 | 0.68761955 | 0.69732532 | 35.61202378 | 2.06285864 |
| P $NO_3$→O+$NO_2$ | 3.04196140 | 3.02662957 | 2.11762758 | 0.92433382 | 0.90900199 | 0.23108346 | 0.22725050 | 43.64949882 | 0.92433382 |
| P $NO_3$→NO+OO | 1.15913601 | 1.08402568 | 0.11649439 | 1.04264162 | 0.96753129 | 0.26066040 | 0.24188282 | 895.01443862 | 1.04264162 |
| P $NO_3$→N+3O | 14.48107126 | 14.39911098 | 11.72934761 | 2.75172365 | 2.66976337 | 0.68793091 | 0.66744084 | 23.46015941 | 2.75172365 |
| L NN→N+N | 10.39286613 | 10.52661226 | 9.75442567 | 0.63844046 | 0.77218659 | 0.31922023 | 0.38609330 | 6.54513640 | 0.63844046 |
| L NNO→NO+N | 5.97430311 | 5.90333173 | 4.92499730 | 1.04930581 | 0.97833443 | 0.34976860 | 0.32611148 | 21.30571346 | 1.04930581 |
| L NNO→NN+O | 2.85557269 | 2.69223765 | 1.66678185 | 1.18879084 | 1.02545580 | 0.39626361 | 0.34181860 | 71.32252170 | 1.18879084 |
| L NNO→O+2N | 13.24843882 | 13.21884991 | 11.42121788 | 1.82722094 | 1.79763203 | 0.60907365 | 0.59921068 | 15.99847726 | 1.82722094 |
| L NON→NO+N | 1.42045531 | 1.38239668 | 0.16271903 | 1.25773628 | 1.21967765 | 0.41924543 | 0.40655922 | 772.94969651 | 1.25773628 |
| L NON→NN+O | -1.69827511 | -1.82869740 | -3.09580735 | 1.39753224 | 1.26710995 | 0.46584408 | 0.42236998 | -45.14273919 | 1.39753224 |
| L NON→O+2N | 8.69459102 | 8.69791486 | 6.65904325 | 2.03554777 | 2.03887161 | 0.67851592 | 0.67962387 | 30.56817156 | 2.03554777 |
| T NON→NO+N | 2.88780064 | 2.91664190 | 2.17131449 | 0.71648615 | 0.74532741 | 0.23882872 | 0.24844247 | 32.99780614 | 0.71648615 |
| T NON→NN+O | -0.23092978 | -0.29445218 | -1.08721189 | 0.85628211 | 0.79275971 | 0.28542737 | 0.26425324 | -78.75945041 | 0.85628211 |
| T NON→O+2N | 10.16193635 | 10.23216008 | 8.66763871 | 1.49429764 | 1.56452137 | 0.49809921 | 0.52150712 | 17.23996218 | 1.49429764 |
| L NNN→NN+N | 1.58413624 | 1.46598198 | 0.19008071 | 1.39405553 | 1.27590127 | 0.46468518 | 0.42530042 | 733.40191519 | 1.39405553 |
| L NNN→3N | 11.97700237 | 11.99259424 | 9.94451674 | 2.03248563 | 2.04807750 | 0.67749521 | 0.68269250 | 20.43825446 | 2.03248563 |
| T2 NNN→NN+N | -0.07464035 | -0.24562570 | -1.19500029 | 1.12035994 | 0.94937459 | 0.37345331 | 0.31645820 | -93.75394713 | 1.12035994 |
| T2 NNN→3N | 10.31822578 | 10.28098656 | 8.55985031 | 1.75837547 | 1.72113625 | 0.58612516 | 0.57371208 | 20.54212875 | 1.75837547 |
| L OO→O+O | 6.04779954 | 5.99956712 | 5.11670830 | 0.93109124 | 0.88285882 | 0.46554562 | 0.44142941 | 18.19707479 | 0.93109124 |
| T1 OOO→OO+O | 1.73656796 | 1.78217288 | 1.06168469 | 0.67488327 | 0.72048819 | 0.22496109 | 0.24016273 | 63.56720362 | 0.67488327 |
| T1 OOO→3O | 7.78436750 | 7.78174000 | 6.17840025 | 1.60596725 | 1.60333975 | 0.53532242 | 0.53444658 | 25.99325370 | 1.60596725 |